\documentclass[10pt,letterpaper,twocolumn]{article}
\usepackage[margin=2cm]{geometry}

\usepackage{graphicx}
\usepackage{array}
\usepackage{subfig}
\usepackage{algorithm,algorithmic}
\usepackage{color}
\usepackage{url}
\usepackage{hyperref}
\usepackage{listings}
\usepackage[none]{hyphenat}

\begin{document}

\date{August 15, 2017 (first draft)}

\title{A Collective Knowledge workflow for collaborative research into\\multi-objective autotuning and machine learning techniques}

\author{Grigori Fursin$^{1,2}$, Anton Lokhmotov$^1$, Dmitry Savenko$^3$, and Eben Upton$^{4,5}$ \\
\\
$^1$~dividiti, UK ; $^2$~cTuning foundation, France ; $^3$~Xored, Russia \\
$^4$~Raspberry Pi foundation, UK ; $^5$~Broadcom, UK\\
}

\maketitle

\begin{abstract}
 Developing efficient software and hardware has never been harder whether it is
for a tiny IoT device or an Exascale supercomputer.
Apart from the ever growing design and optimization complexity, there exist even
more fundamental problems such as lack of interdisciplinary knowledge required
for effective software/hardware co-design, and a growing technology transfer
gap between academia and industry.

We introduce our new educational initiative to tackle these problems by
developing Collective Knowledge (CK), a unified experimental framework for
computer systems research and development.
We use CK to teach the community how to make their research artifacts and
experimental workflows portable, reproducible, customizable and reusable while
enabling sustainable R\&D and facilitating technology transfer.
We also demonstrate how to redesign multi-objective autotuning and machine learning
as a portable and extensible CK workflow.
Such workflows enable researchers to experiment with different applications,
data sets and tools; crowdsource experimentation across diverse platforms;
share experimental results, models, visualizations; gradually expose more
design and optimization choices using a simple JSON API; and ultimately build
upon each other's findings.

As the first practical step, we have implemented customizable 
compiler autotuning, crowdsourced optimization 
of diverse workloads across Raspberry Pi 3 devices,
reduced the execution time and code size by up to 40\%,
and applied machine learning to predict optimizations.
We hope such approach will help teach students how to build 
upon each others' work to enable efficient and self-optimizing
software/hardware/model stack for emerging workloads.

\end{abstract}

{\bf Keywords:}
{\it\small 
 collective knowledge,
customizable research workflows,
portable package manager,
reusable artifacts,
collaborative optimization,
customizable autotuning,
machine learning,
crowd-fuzzing,
software/hardware co-design competitions, 
technology transfer,
Raspberry Pi

}

{\bf Live CK repository:} \newline
\textit{\href{https://github.com/ctuning/ck-rpi-optimization-results}{github.com/ctuning/ck-rpi-optimization-results}} (0.9GB)

{\bf Interactive report:} \newline
\textit{\href{http://cKnowledge.org/rpi-crowd-tuning}{cKnowledge.org/rpi-crowd-tuning}}

{\bf Archives of CK repositories at FigShare:} \newline
\textit{\href{https://doi.org/10.6084/m9.figshare.5789007.v2}{doi.org/10.6084/m9.figshare.5789007.v2}}



\section{Introduction} 
\label{introduction} 
\begin{figure*}[htb]
     \centering
      \includegraphics[width=5.2in]
      {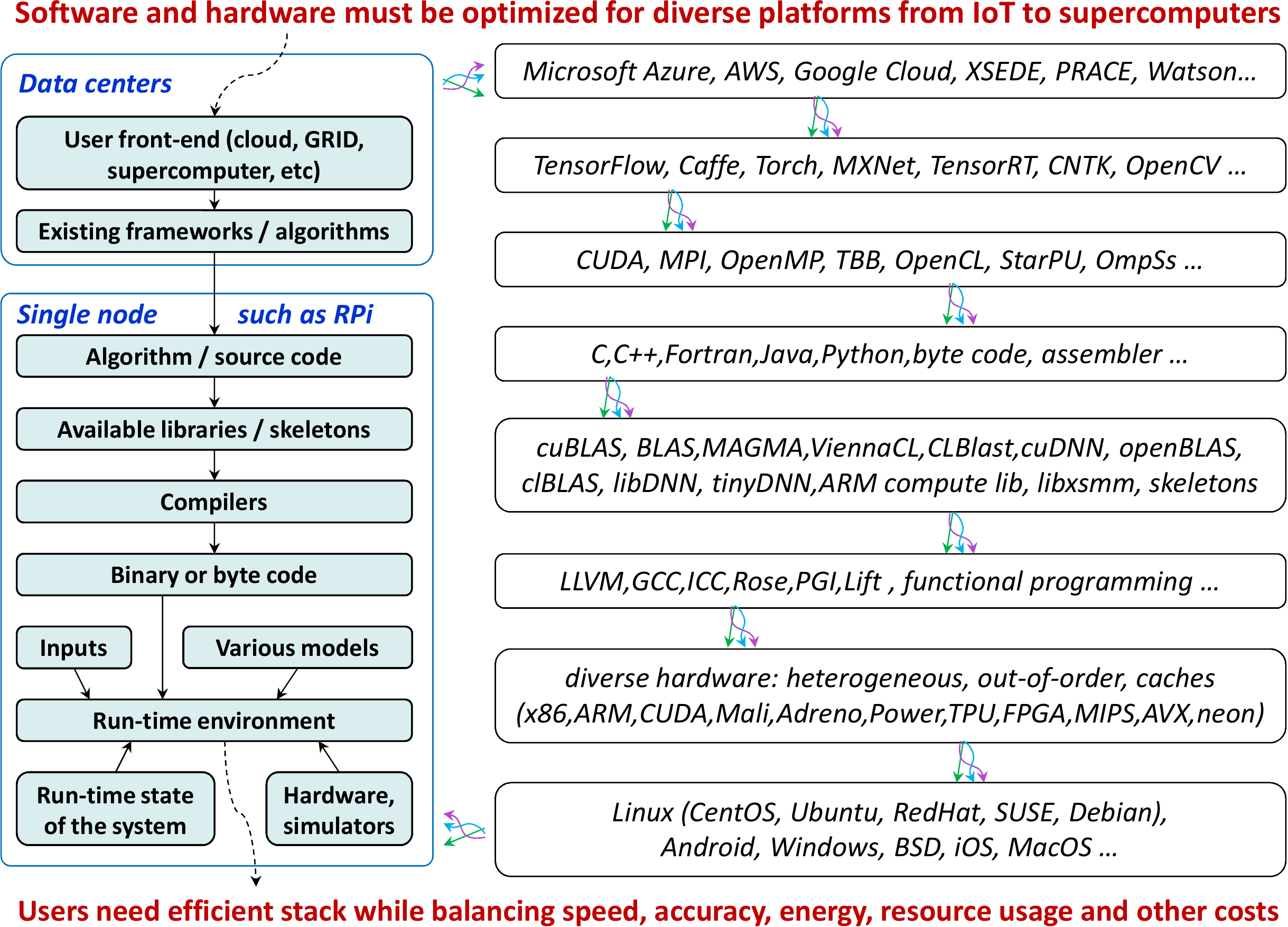} 

     \caption{
       Too many design and optimization choices at all levels of the continuously changing software and hardware stack
       make it extremely challenging and time consuming to design efficient computer systems for realistic workloads.
     }

     \label{fig:introduction}
   \end{figure*}

Many recent international roadmaps for computer systems research
appeal to reinvent computing~\cite{hipeac_roadmap2017,Dongarra:2011:IES:1943326.1943339,prace}.
Indeed, developing, benchmarking, optimizing and co-designing hardware and software
has never been harder, no matter if it is for embedded and IoT devices,
or data centers and Exascale supercomputers.
This is caused by both physical limitations of existing technologies
and an unmanageable complexity of continuously changing computer systems
which already have too many design and optimization choices and objectives
to consider at all software and hardware levels~\cite{fursin:hal-01054763},
as conceptually shown in Figure~\ref{fig:introduction}.
That is why most of these roadmaps now agree with our vision
that such problems should be solved in a close collaboration
between industry, academia and end-users~\cite{Fur2009,cm:29db2248aba45e59:cd11e3a188574d80}.


However, after we initiated artifact evaluation (AE)~\cite{ctuning-ae1,childers2016artifact}
at several premier ACM and IEEE conferences to reproduce and validate experimental results 
from published papers, we noticed an even more
fundamental problem: a growing technology transfer gap between academic
research and industrial development.
After evaluating more than 100 artifacts from the leading computer systems conferences
in the past 4 years, we noticed that only a small fraction of research artifacts 
could be easily customized, ported to other environments and hardware, reused, 
and built upon.
We have grown to believe that is this due to a lack of a common workflow framework
that could simplify implementation and sharing of artifacts and workflows as
portable, customizable and reusable components with some common API and meta information
vital for open science~\cite{new_pub_model}.

At the same time, companies are always under pressure and rarely have time to
dig into numerous academic artifacts shared as CSV/Excel files
and ``black box'' VM and Docker images, or adapt numerous ad-hoc scripts 
to realistic and ever changing workloads, software and hardware.
That is why promising techniques may remain in academia for decades while just
being incrementally improved, put on the shelf when leading students graduate,
and ``reinvented'' from time to time.

Autotuning is one such example: this very popular technique has been actively
researched since the 1990s to automatically explore large optimization spaces
and improve efficiency of computer systems~\cite{atlas, fftw, CSS99, VE00,
FOK02, Tapus:2002:AHT:762761.762771, vista, spiral, LCYP04, la2004, PE2006,
Shende:2006:TPP:1125980.1125982, 1742-6596-125-1-012089,
DBLP:conf/ipps/HartonoNS09, 29db2248aba45e59:a31e374796869125, tnld10,
openbenchmarking, Ren:2010:GPC:1849301.1849332, Grauer-Gray2012-hn,
DBLP:conf/cgo/GreweWO13, Khan:2013:SAC:2400682.2400690, ansel:pact:2014,
DBLP:conf/sc/TsaiLKD16, DBLP:conf/supercomputer/AbdelfattahHTD16}.
Every year, dozens of autotuning papers get published to optimize some components
of computer systems, improve and speed up exploration and co-design strategies,
and enable run-time adaptation.
Yet, when trying to make autotuning practical (in particular, by applying
machine learning) we faced numerous challenges with integrating such published
techniques into real, complex and continuously evolving software and hardware
stack~\cite{Fur2009,fursin:hal-01054763,cm:29db2248aba45e59:cd11e3a188574d80,new_pub_model}.

Eventually, these problems motivated us to develop a common experimental
framework and methodology similar to physics and other natural sciences to
collaboratively improve autotuning and other techniques.
As part of this educational initiative, we implemented an extensible, portable
and technology-agnostic workflow for autotuning using the open-source
Collective Knowledge framework (CK)~\cite{ck,ck-date16}.
Such workflows help researchers to reuse already shared applications, kernels,
data sets and tools, or add their own ones using a common JSON API and
meta-description~\cite{json-org}. 
Moreover, such workflows can automatically
adapt compilation and execution to a given environment on a given device
using integrated cross-platform package manager.

Our approach takes advantage of a powerful and holistic top-down methodology
successfully used in physics and other sciences when learning complex systems.
The key idea is to let novice researchers first master simple compiler flag
autotuning scenarios while learning interdisciplinary techniques including
machine learning and statistical analysis.
Researchers can then gradually increase complexity to enable automatic and
collaborative co-design of the whole SW/HW stack by exposing more design and
optimization choices, multiple optimization objectives (execution time, code
size, power consumption, memory usage, platform cost, accuracy, etc.),
crowdsource autotuning across diverse devices provided by volunteers similar to
SETI@home~\cite{Anderson:2002:SEP:581571.581573}, 
continuously exchange and discuss optimization results, and
eventually build upon each other's results.

We use our approach to optimize diverse kernels and real workloads such as {\tt
zlib} in terms of speed and code size by crowdsourcing compiler flag autotuning
across Raspberry Pi3 devices using the default GCC 4.9.2 and the latest GCC
7.1.0 compilers.
We have been able to achieve up to 50\% reductions in code size and from 15\%
to 8 times speed ups across different workloads over the ``-O3'' baseline.
Our CK workflow and all related artifacts are available at GitHub 
to allow researchers to compare and improve various exploration strategies 
(particularly based on machine learning algorithms such as KNN, GA, SVM,
deep learning, though further documentation of APIs is still 
required)~\cite{29db2248aba45e59:a31e374796869125,cm:29db2248aba45e59:cd11e3a188574d80}.
We have also shared all experimental results in our open repository of
optimization knowledge~\cite{live-ck-repo,live-ck-repo-rpi-gcc710} to be
validated and reproduced by the community.

We hope that our approach will serve as a practical foundation
for open, reproducible and sustainable computer systems research
by connecting students, scientists, end-users, hardware designers
and software developers to learn together how to co-design
the next generation of efficient and self-optimizing computer systems,
particularly via reproducible competitions such as ReQuEST~\cite{request}.

This technical report is organized as follows.
Section~\ref{sec:converting} introduces the Collective Knowledge framework (CK)
and the concept of sharing artifacts as portable, customizable and reusable components.
Section~\ref{sec:autotuning} describes how to implement a customizable,
multi-dimensional and multi-objective autotuning as a CK workflow.
Section~\ref{sec:flag_autotuning} shows how to optimize compiler flags using
our universal CK autotuner.
Section~\ref{sec:crowdtuning} presents a snapshot of the latest optimization
results from collaborative tuning of GCC flags for numerous shared workloads
across Raspberry Pi3 devices.
Section~\ref{sec:collaborative} shows optimization results of zlib and other
realistic workloads for GCC 4.9.2 and GCC 7.1.0 across Raspberry Pi3 devices.
Section~\ref{sec:crowdfuzzing} describes how implement and crowdsource fuzzing
of compilers and systems for various bugs using our customizable CK autotuning workflow.
Section~\ref{sec:crowdmodeling} shows how to predict optimizations via CK for previously
unseen programs using machine learning.
Section~\ref{sec:features} demonstrates how to select and autotune models
and features to improve optimization predictions while reducing complexity.
Section~\ref{sec:datasets} shows how to enable efficient, input-aware and adaptive 
libraries and programs via CK.
Section~\ref{sec:competitions} presents CK as an open platform to 
support reproducible and Pareto-efficient co-design competitions 
of the whole software/hardware/model stack for emerging workloads 
such as deep learning and quantum computing.
We present future work in Section~\ref{sec:conclusions}.
We also included Artifact Appendix to allow students try our framework, 
participate in collaborative autotuning, gradually document APIs and
improve experimental workflows.


\section{Converting ad-hoc artifacts to portable and reusable components with JSON API}
\label{sec:converting}
Artifact sharing and reproducible experimentation are key
for our collaborative approach to machine-learning based optimization
and co-design of computer systems, which was first prototyped
during the EU-funded MILEPOST project~\cite{Fur2009,milepost,29db2248aba45e59:a31e374796869125}.
Indeed, it is difficult, if not impossible, and time consuming to build useful predictive models
without large and diverse training sets (programs, data sets),
and without crowdsourcing design and optimization space exploration
across diverse hardware~\cite{new_pub_model,cm:29db2248aba45e59:cd11e3a188574d80}.%

While we have been actively promoting artifact sharing for the past 10 years
since the MILEPOST project~\cite{Fur2009,ctuning1}, it is still relatively rare in the community
systems community.
We have begun to understand possible reasons for that through our Artifact
Evaluation initiative~\cite{ctuning-ae1,childers2016artifact} 
at PPoPP, CGO, PACT, SuperComputing and other leading ACM and IEEE conferences
which has attracted over a hundred of artifacts in the past few years.

   \begin{figure*}[htbp]
     \centering
      \includegraphics[width=6.5in]
      {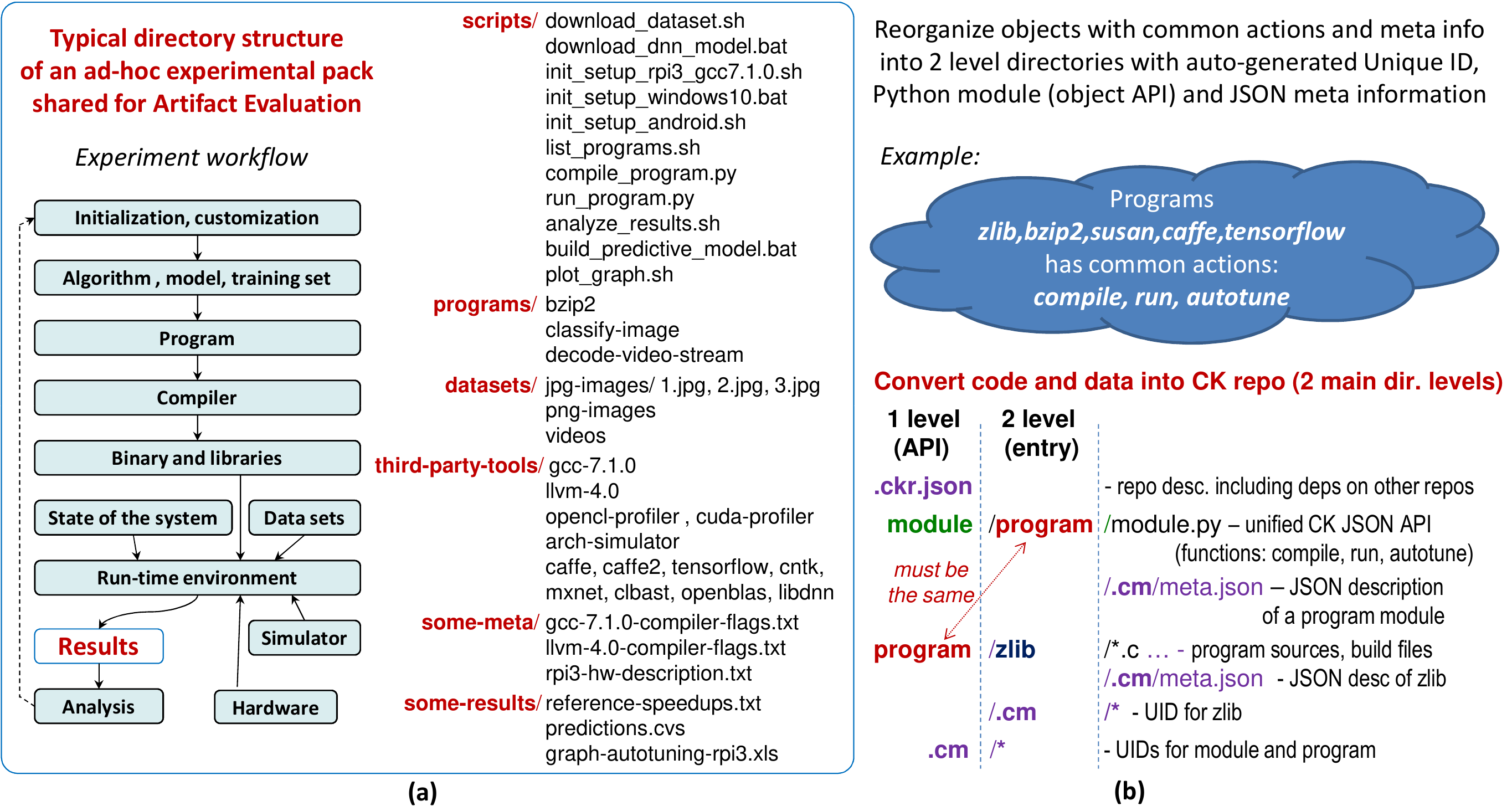} 
     \caption{
       Reorganizing ad-hoc experimental packs into reusable, customizable and discoverable components
       with JSON API and meta information using the Collective Knowledge framework.
     }
     \label{fig:convert-to-ck}
   \end{figure*}

Unfortunately, nearly all the artifacts have been shared simply as zip
archives, GitHub/GitLab/Bitbucket repositories, or VM/Docker images, with many
ad-hoc scripts to prepare, run and visualize experiments, as shown in
Figure~\ref{fig:convert-to-ck}a.
While a good step towards reproducibility, such ad-hoc artifacts are hard to
reuse and customize as they do not provide a common API and meta information.

Some popular and useful services such as Zenodo~\cite{zenodo} and
FigShare~\cite{figshare} allow researchers to upload individual artifacts to
specific websites while assigning DOI~\cite{doi} and providing some meta
information.
This helps the community to discover the artifacts, but does not necessarily
make them easy to reuse.

After an ACM workshop on reproducible research methodologies (TRUST'14)~\cite{trust2014}
and a Dagstuhl Perspective workshop on Artifact Evaluation~\cite{childers2016artifact},
we concluded that compute systems research lacked a common experimental
framework in contrast with other sciences~\cite{doi:10.1093/bioinformatics/bth361}.

Together with our fellow researchers, we also assembled the following wish-list
for such a framework:

\begin{itemize}

\item it should be able to help researchers quickly organize their local code
and data into discoverable and reusable components with a unique ID, common API
and unified meta information, rather than being forced to upload them to the
web from the start;

\item it should be open-source with a permissive license to simplify technology
transfer;

\item it should be portable, simple to install and use from the command line;

\item it should allow to assemble experimental workflows by simply plugging in
shared components;

\item it should support native non-virtualized execution of such workflows,
i.e.\ not only via Virtual Machine~\cite{Smith:2005:AVM:1069588.1069632} and
Docker~\cite{docker}, critical for empirical program optimization and hardware
co-design experiments;

\item it should be able to adapt to continuously evolving software environments
and support different versions of tools such as rapidly evolving
compilers and libraries;

\item it should include a local web server to simplify crowdsourcing of
experiments and visualization of results in workgroups.

\end{itemize}

Since there was no available open-source framework with all these features,
we decided to develop such a framework, Collective Knowledge (CK)~\cite{ck,ck-date16}, 
from scratch with initial support from the EU-funded TETRACOM project~\cite{tetracom}.
CK is implemented as a small and portable Python module with a command line
front-end to assist users in converting their local objects (code and data)
into searchable, reusable and shareable directory entries with user-friendly
aliases and auto-generated Unique ID, JSON API and JSON meta
information~\cite{json-org}, as described in~\cite{ck-date16,ck-concepts} and 
conceptually shown in Figure~\ref{fig:convert-to-ck}b.

The user first creates a new local CK repository as follows:
\begin{flushleft}
\texttt{\$ ck add repo:new-ck-repo}
\end{flushleft}

Initially, it is just an empty directory:
\begin{flushleft}
\texttt{\$ ck find repo:new-ck-repo} \newline
\texttt{\$ ls `ck find repo:new-ck-repo`}
\end{flushleft}

Now, the user starts adding research artifacts as CK components with extensible APIs.
For example, after noticing that we always perform 3 common actions on all our benchmarks
during our experiments, "compile", "run" and "autotune", we want to provide a common
API for these actions and benchmarks, rather than writing ad-hoc scripts.
The user can provide such an API with actions by adding a new CK module to a CK repository as follows:
\begin{flushleft}
\texttt{\$ ck add new-ck-repo:module:program}
\end{flushleft}
CK will then create two levels of directories \textit{module} and \textit{program} in the \textit{new-ck-repo}
and will add a dummy \textit{module.py} where common object actions can be implemented later.
CK will also create a sub-directory \textit{.cm} (collective meta) 
with an automatically generated Unique ID of this module and various pre-defined 
descriptions in JSON format (date and time of module creation, author, license, etc)
to document provenance of the CK artifacts.

Users can now create holders (directories) for such objects sharing common CK module and an API as follows:
\begin{flushleft}
\texttt{\$ ck add new-ck-repo:program:new-benchmark}
\end{flushleft}
CK will again create two levels of directories: 
the first one specifying used CK module (\textit{program}) 
and the second one with alias \textit{new-benchmark} 
to keep objects.
CK will also create three files in an internal \textit{.cm} directory:

\begin{itemize}

\item \textbf{meta.json} - an empty JSON file which can be gradually extended 
to describe a given object (such as added program in our example);

\item \textbf{info.json} - a JSON file with the date and time of the last modification
as well as license, copyright and author information to keep attribution of all updates
for open research;

\item \textbf{desc.json} - an empty JSON file to describe types of keys in \textbf{meta.json}
(useful for automatic type checking) and their value ranges (useful for autotuning
as we will show later in this report).

\end{itemize}

Users can then find a path to a newly created object holder (CK entry) using
the \textit{ck find program:new-benchmark} command and then copy all files and
sub-directories related to the given object using standard OS shell commands.

   \begin{figure*}[htbp]
     \centering
      \includegraphics[width=6in]
      {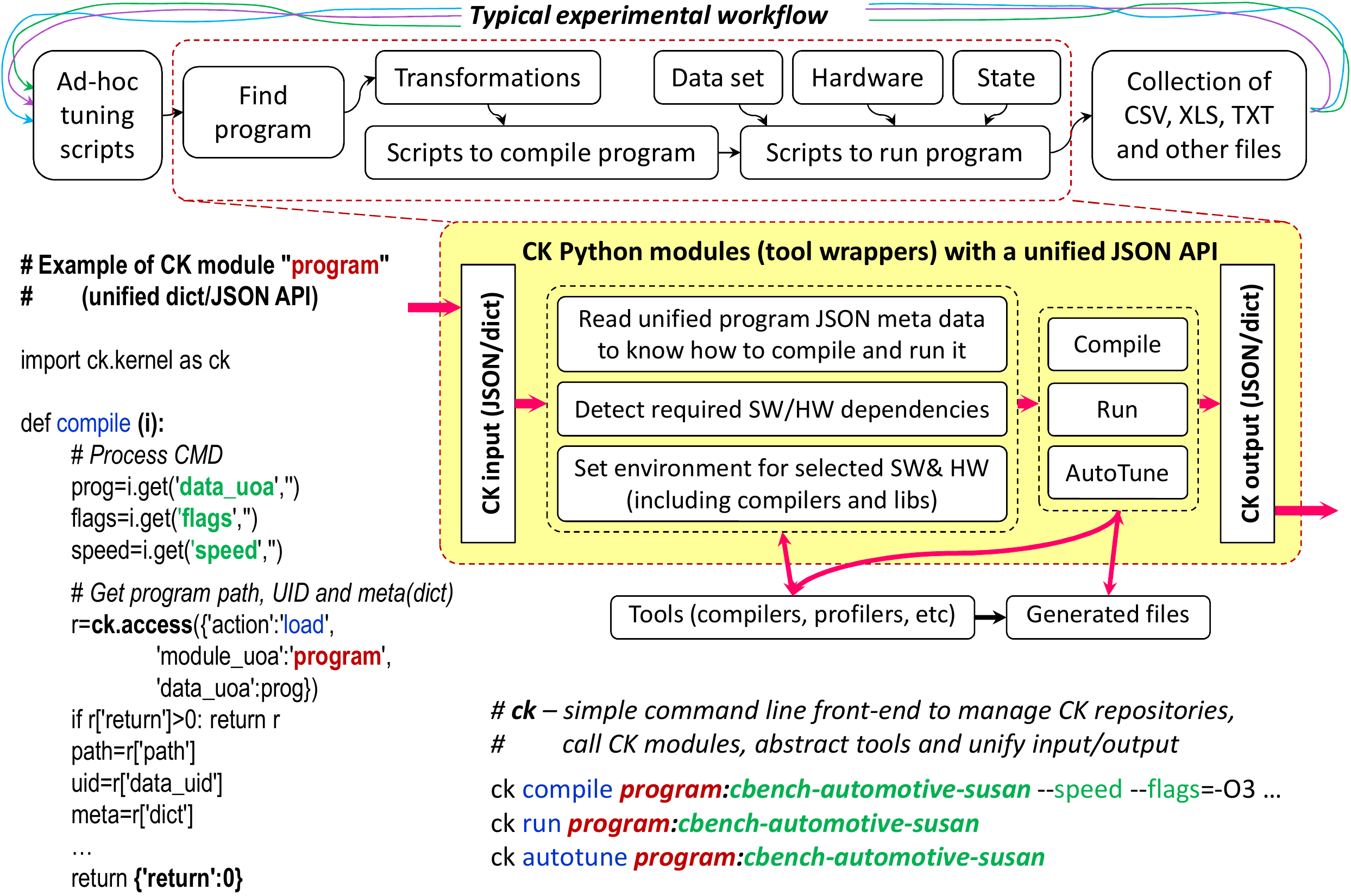} 
     \caption{
        Converting ad-hoc scripts, tools and workflows to CK Python modules
        and standardized directories with actions, unified JSON API, and JSON meta information.
     }
     \label{fig:ck-workflows}
   \end{figure*}

\begin{figure*}[htbp]
     \centering
      \includegraphics[width=4.8in]
      {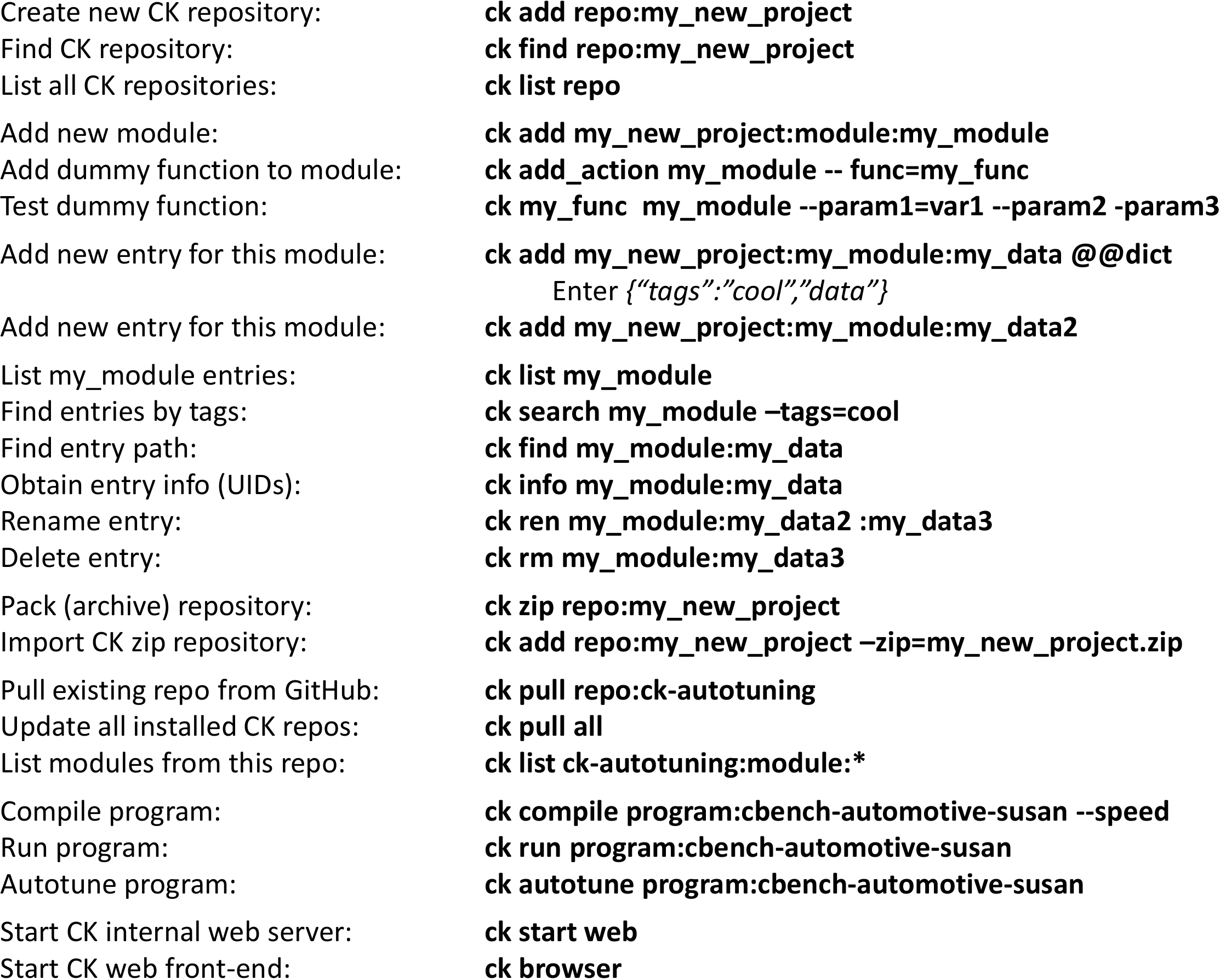} 
     \caption{
        Main CK commands to create new or pull existing repositories, add modules, manage entries, perform actions, and use a local web server.
     }
     \label{fig:ck-commands}
   \end{figure*}

This allows to get rid of ad-hoc scripts by implementing actions inside
reusable CK Python modules as shown in Figure~\ref{fig:ck-workflows}.
For example, the user can add an action to a given module such as \textit{compile program}
as follows:
\begin{flushleft}
\texttt{\$ ck add\_action module:program --func=compile}
\end{flushleft}
CK will create a dummy function body with an input dictionary \textit{i}
inside \textit{module.py} in the CK \textit{module:program} entry.
Whenever this function is invoked via CK using the following format:
\begin{flushleft}
\texttt{\$ ck compile program:some\_entry --param1=val1}
\end{flushleft}
the command line will be converted to \textit{i} dictionary
and printed to the console to help novice users understand the CK API.
The user can now substitute this dummy function with a specific action on a specific entry
(some program in our example based on its meta information)
as conceptually shown in Figure~\ref{fig:ck-workflows}.
The above example shows how to call CK functions from Python modules rather than from the command line
using the \textit{ck.access} function.
It also demonstrates how to find a path to a given \textit{program} entry,
load its meta information and unique ID.
For the reader's convenience, Figure~\ref{fig:ck-commands} lists several important CK commands.

This functionality should be enough to start implementing unified compilation and execution
of shared programs.
For example, the \textit{program} module can read instructions about how to compile and run
a given program from the JSON meta data of related entries, prepare and execute portable sub-scripts,
collect various statistics, and embed them to the output dictionary in a unified way.
This can be also gradually extended to include extra tools into compilation and execution workflow
such as code instrumentation and profiling.

\begin{figure*}[ht]
     \centering
      \includegraphics[width=5.6in]
      {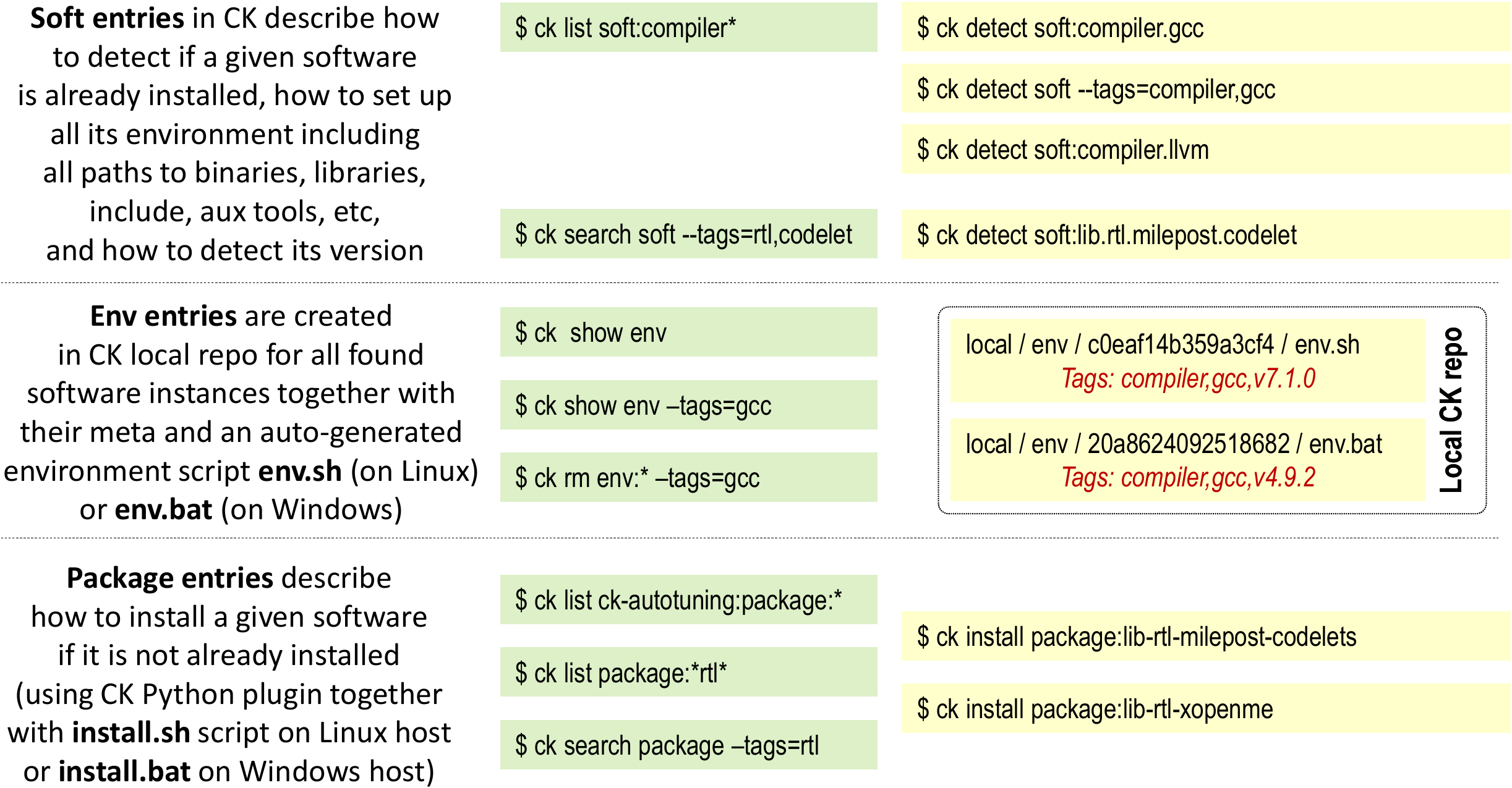} 
     \caption{
   CK modules implementing portable package manager with JSON API to enable cross-platform CK workflows.
   The community shares CK entries with Python scripts and JSON meta information via Git repositories
   to describe how to detect, build and install any software. This approach also simplify
   co-existence of multiple versions of the same tool.
     }
     \label{fig:portable-package-manager}
   \end{figure*}

Here we immediately face another problem common for computer systems research:
how to support multiple versions of various and continuously evolving tools and libraries?
However, since we no longer hardwire calls to specific tools directly in scripts
but invoke them from higher-level CK modules, we can detect all required tools
and set up their environment before execution.
To support this concept even better, we have developed a cross-platform package manager
as a \href{https://github.com/ctuning/ck-env}{ck-env} repository~\cite{ck-env} with several CK modules
including \textit{soft}, \textit{env}, \textit{package}, \textit{os} and \textit{platform}.
These modules allow the community to describe various operating systems (Linux, Windows, MacOS, Android);
detect platform features (\textit{ck detect platform}); detect multiple-versions of
already installed software (\textit{ck detect soft:compiler.gcc});
prepare CK entries with their environments for a given OS and platform
using \textit{env} module (\textit{ck show env}) thus allowing easy co-existence of multiple versions of a given tool;
install missing software using \textit{package} modules;
describe software dependencies using simple tags
in a program meta description (such as \textit{compiler,gcc} or \textit{lib,caffe}),
and ask the user to select an appropriate version during program compilation when multiple software
versions are registered in the CK as shown in Figure~\ref{fig:portable-package-manager}.

Such approach extends the concept of package managers including
Spack~\cite{Gamblin:2015:SPM:2807591.2807623} and EasyBuild~\cite{DBLP:conf/sc/HosteTGW12}
by integrating them directly with experimental CK workflows while using unified CK API,
supporting any OS and platform, and allowing the community to gradually extend existing
detection or installation procedures via CK Python scripts and CK meta data.

\begin{figure}[htbp]
     \centering
      \includegraphics[width=3.4in]
      {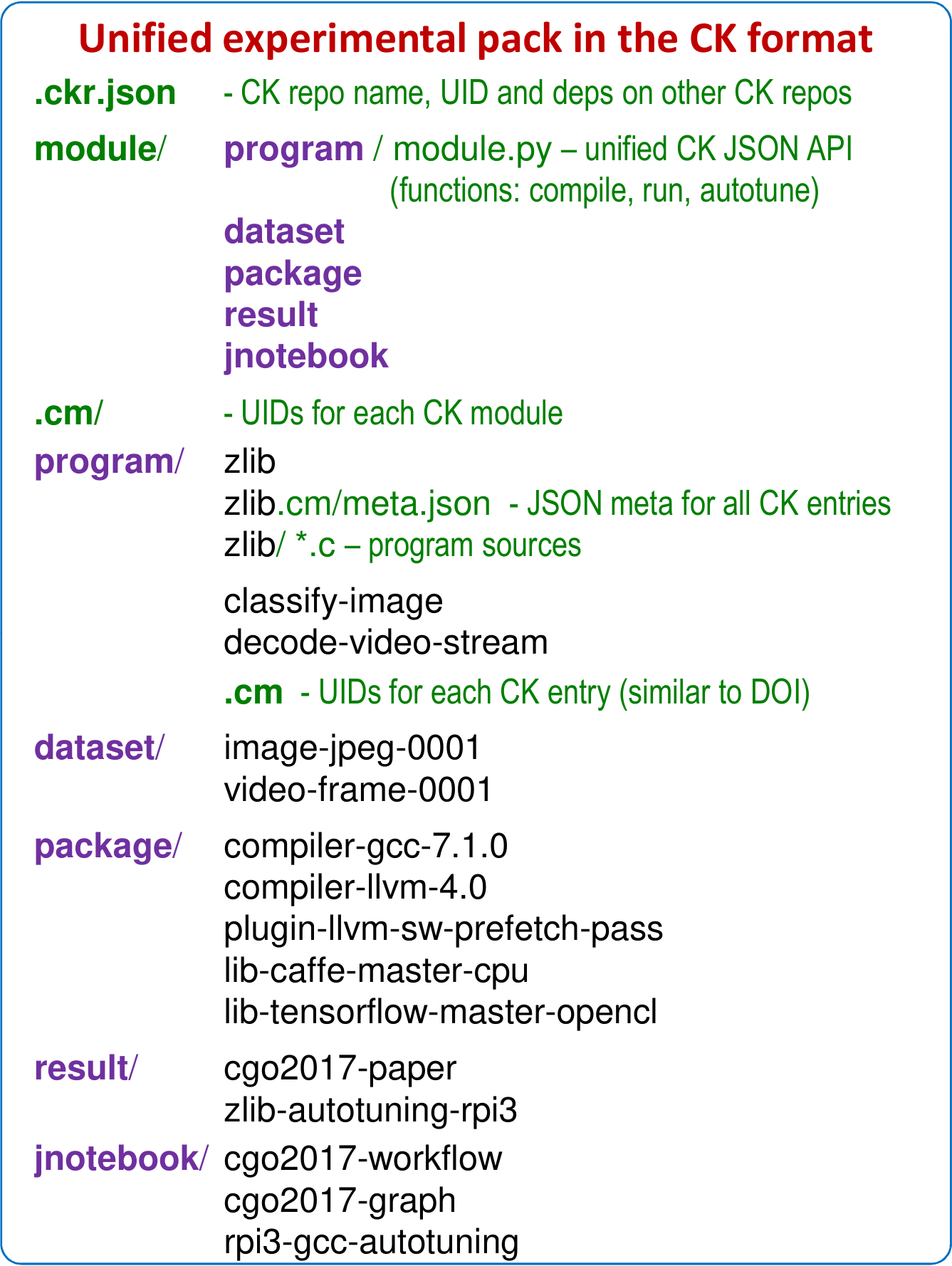} 
     \caption{
        Typical experiment pack with reusable and discoverable components 
        shared in the CK format with two level directory structure (module and data).
     }
     \label{fig:ck-repo}
   \end{figure}

Note that this CK approach encourages reuse of all such existing CK modules
from shared CK repositories rather then writing numerous ad-hoc scripts.
It should indeed be possible to substitute most of ad-hoc scripts
from public research projects (Figure~\ref{fig:convert-to-ck})
with just a few above modules and entries (Figure~\ref{fig:ck-repo}),
and then collaboratively extend them, thus dramatically improving research productivity.
For this reason, we keep track of all publicly shared modules and their repositories
in this \href{https://github.com/ctuning/ck/wiki/Shared-modules}{wiki page}.
The user will just need to add/update a \textit{.ckr.json} file
in the root directory of a given CK repository to describe a dependency
on other existing CK repositories with required modules or entries.
Since it is possible to uniquely reference any CK entry by two Unique IDs
(\textit{module UID:object UID}), we also plan to develop a simple web service
to automatically index and discover all modules similar to DOI.

The open, file-based format of CK repositories allows researchers
to continue editing entries and their meta directly using their favourite editors.
It also simplifies exchange of these entries using Git repositories, zip archives, Docker images 
and any other popular tool.
At the same time, schema-free and human readable Python dictionaries
and JSON files helps users to collaboratively extend actions, API and meta information
while keeping backward compatibility.
Such approach should let the community to gradually and collaboratively convert and
cross-link all existing ad-hoc code and data into unified components
with extensible API and meta information.
This, in turn, allows users organize their own research while reusing existing artifacts,
building upon them, improving them and continuously contributing back to Collective Knowledge
similar to Wikipedia.

We also noticed that CK can help students reduce preparation time 
for Artifact Evaluation~\cite{ctuning-ae1} at conferences while automating preparation 
and validation of experiments  since all artifacts, workflows and repositories are immediatelly ready
to be shared, ported and plugged in to research workflows.

For example, the highest ranked artifact from
the CGO'17 article~\cite{Ainsworth:2017:SPI:3049832.3049865}
was implemented and shared using the CK framework~\cite{cgo17-artifact}.
That is why CK is now used and publicly extended by leading
companies~\cite{ck-date16},
universities~\cite{Ainsworth:2017:SPI:3049832.3049865}
and organizations~\cite{ck-acm} to encourage, support and simplify
technology transfer between academia and industry.


\section{Assembling portable and customizable autotuning workflow}
\label{sec:autotuning}
Autotuning combined with various run-time adaptation,
genetic and machine learning techniques is a popular approach
in computer systems research to automatically explore multi-dimensional 
design and optimization spaces~\cite{atlas, europar97x, CGJ1997, Nis1998,
fftw, CSS99, VE00, KKO2000, FOK02, SAMP2003, Tapus:2002:AHT:762761.762771,
vista, spiral, LCYP04, la2004, FOTP2005, PE2006, HE2008, BCCP2008, JGVP2009,
Ansel:2009:PLC:1542476.1542481, Mars:2010:CAE:1772954.1772991,
DBLP:conf/cc/MooreC13, DBLP:conf/cf/ShenVSAS13, DBLP:conf/cgo/GreweWO13,
Miceli:2012:APA:2451764.2451792,
Manotas:2014:SSE:2568225.2568297,
ashouri2016cobayn}.

CK allows to unify such techniques by developing
a common, universal, portable, customizable, multi-dimensional
and multi-objective autotuning workflow as a CK module
(\textit{pipeline}%
\footnote{We use the term \textit{pipeline} similar to experiments in physics and electronics
where an output of one object is chained to an input of another one.}
from the public
\href{https://github.com/ctuning/ck-autotuning}{ck-autotuning} repository
with the \textit{autotune} function).
This allows us to abstract autotuning by decoupling it from the autotuned objects
such as \textit{"program"}.
Users just need to provide a compatible function \textit{"pipeline"} in a CK module 
which they want to be autotuned with a specific API including the following keys in both input and output:
\begin{itemize}

\item \textbf{dependencies} to describe software dependencies via portable package manager from the CK;

\item \textbf{choices} to expose various design and optimization knobs \textbf{c} such as algorithmic parameters, model topology, source-to-source transformations, compiler flags, hardware configurations, etc.;

\item \textbf{characteristics} to monitor optimized behavior \textbf{b} such as execution time, code size, compilation time, energy, memory usage, accuracy, resiliency, costs, etc.;

\item \textbf{features} to expose various object features \textbf{f} such as semantic program and data set features, hardware counters, platform properties, etc.;

\item \textbf{state} to define run-time system state \textbf{s} such as hardware frequencies, network status, cache state, etc.

\end{itemize}

   \begin{figure*}[!htbp]
     \centering
      \includegraphics[width=6.6in]
      {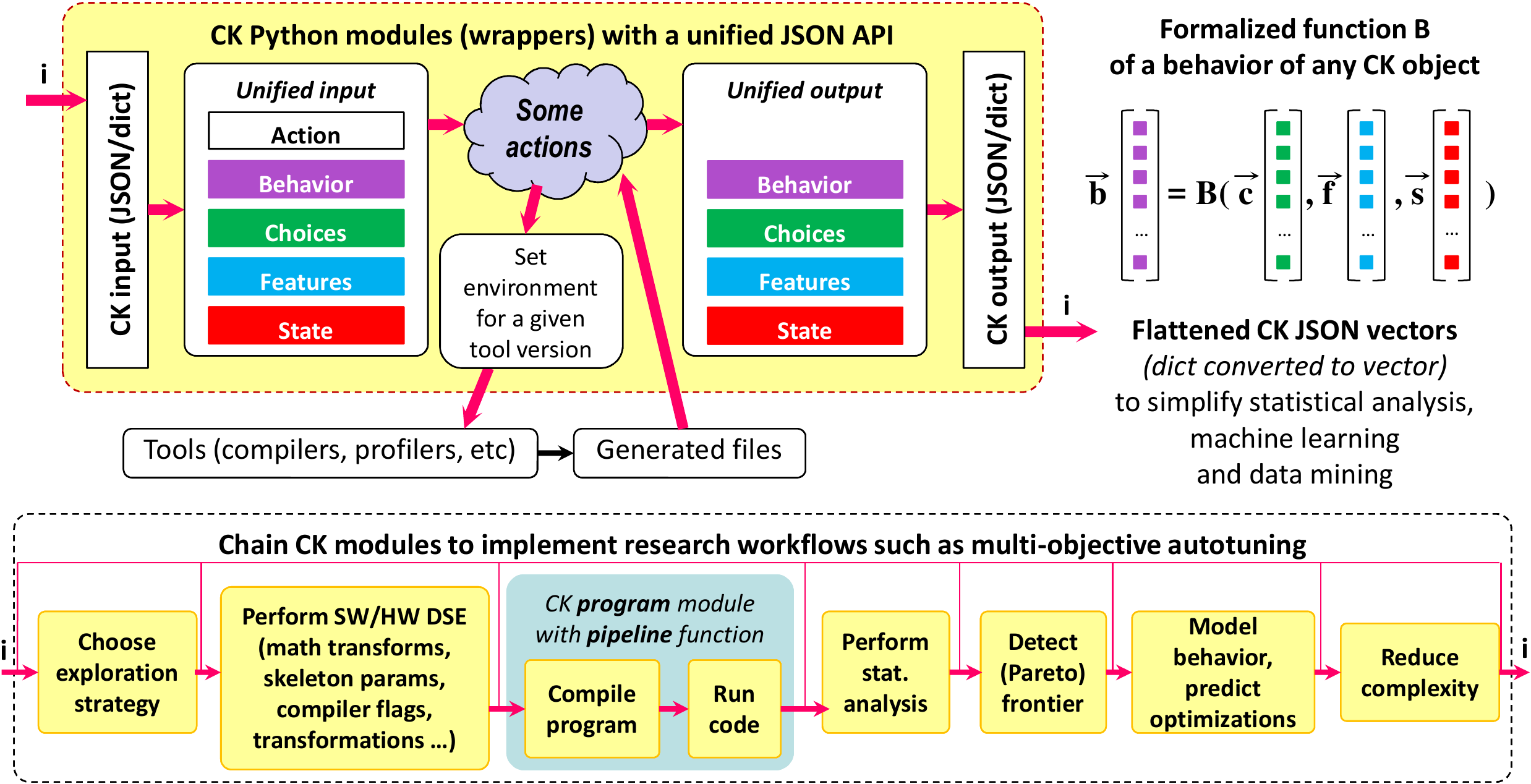} 
     \caption{
       Chaining together various CK modules with JSON API and JSON meta information
       to implement universal, portable, customizable, multi-dimensional
       and multi-objective autotuner gradually extended by the community.
     }
     \label{fig:ck-universal-autotuning-workflow}
   \end{figure*}

Autotuning can now be implemented as a universal and extensible workflow
applied to any object with a matching JSON API by chaining together
related CK modules with various exploration strategies,
program transformation tools, compilers,
program compilation and execution pipeline, architecture simulators,
statistical analysis, Pareto frontier filter and other components,
as conceptually shown in Figure~\ref{fig:ck-universal-autotuning-workflow}.
Researchers can also use unified machine learning CK modules
(wrappers to R and scikit-learn~\cite{scikit-learn})
to model the relationship between \textbf{c}, \textbf{f}, \textbf{s}
and the observed behavior~\textbf{b}, increase coverage, speed up (focus) exploration,
and predict efficient optimizations~\cite{fursin:hal-01054763,cm:29db2248aba45e59:cd11e3a188574d80}.
They can also take advantage of a universal complexity reduction module
which can automatically simplify found solutions without changing their behavior,
reduce models and features without sacrificing accuracy,
localize performance issues via differential analysis~\cite{FOTP04},
reduce programs to localize bugs, and so on.

Even more importantly, our concept of a universal autotuning workflow,
knowledge sharing and artifact reuse can help teach students
how to apply a well-established holistic and top-down
experimental methodology from natural sciences to continuously 
learn and improve the behavior of complex computer systems~\cite{ck-date16,fursin:hal-01054763}.
Researchers can continue exposing more design and optimization knobs~\textbf{c},
behavioral characteristics~\textbf{b}, static and dynamic features~\textbf{f},
and run-time state \textit{state} to optimize and model behavior
of various interconnected objects from the workflow depending on their
research interests and autotuning scenarios.

Such scenarios are also implemented as CK modules
and describe which sets of choices to select, 
how to autotune them and which multiple characteristics to trade off.
For example, existing scenarios include
"autotuning OpenCL parameters to improve execution time",
"autotuning GCC flags to balance execution time and code size",
"autotune LLVM flags to reduce execution time",
"automatically fuzzing compilers to detect bugs",
"exploring CPU and GPU frequency in terms of execution time and power consumption",
"autotuning deep learning algorithms in terms of speed, accuracy, energy, memory usage and costs",
and so on.

You can see some of the autotuning scenarios using the following commands:\newline
\begin{flushleft}
\texttt{\$ ck pull repo:ck-crowdtuning}\newline
\texttt{\$ ck search module --tags="program optimization"}\newline
\texttt{\$ ck list program}\newline
\end{flushleft}
They can then be invoked from the command line as follows:\newline
\begin{flushleft}
\texttt{\$ ck autotune program:[CK program alias] --scenario=[above CK scenario alias]}
\end{flushleft}


\section{Implementing universal compiler flag autotuning}
\label{sec:flag_autotuning}
In this section we would like to show how to 
customize our universal autotuning workflow
to tackle an old but yet unsolved problem 
of finding the most efficient 
selection of compiler flag which minimizes 
program size and execution time.

\begin{figure}[]
     \centering
      \includegraphics[width=3.4in]
      {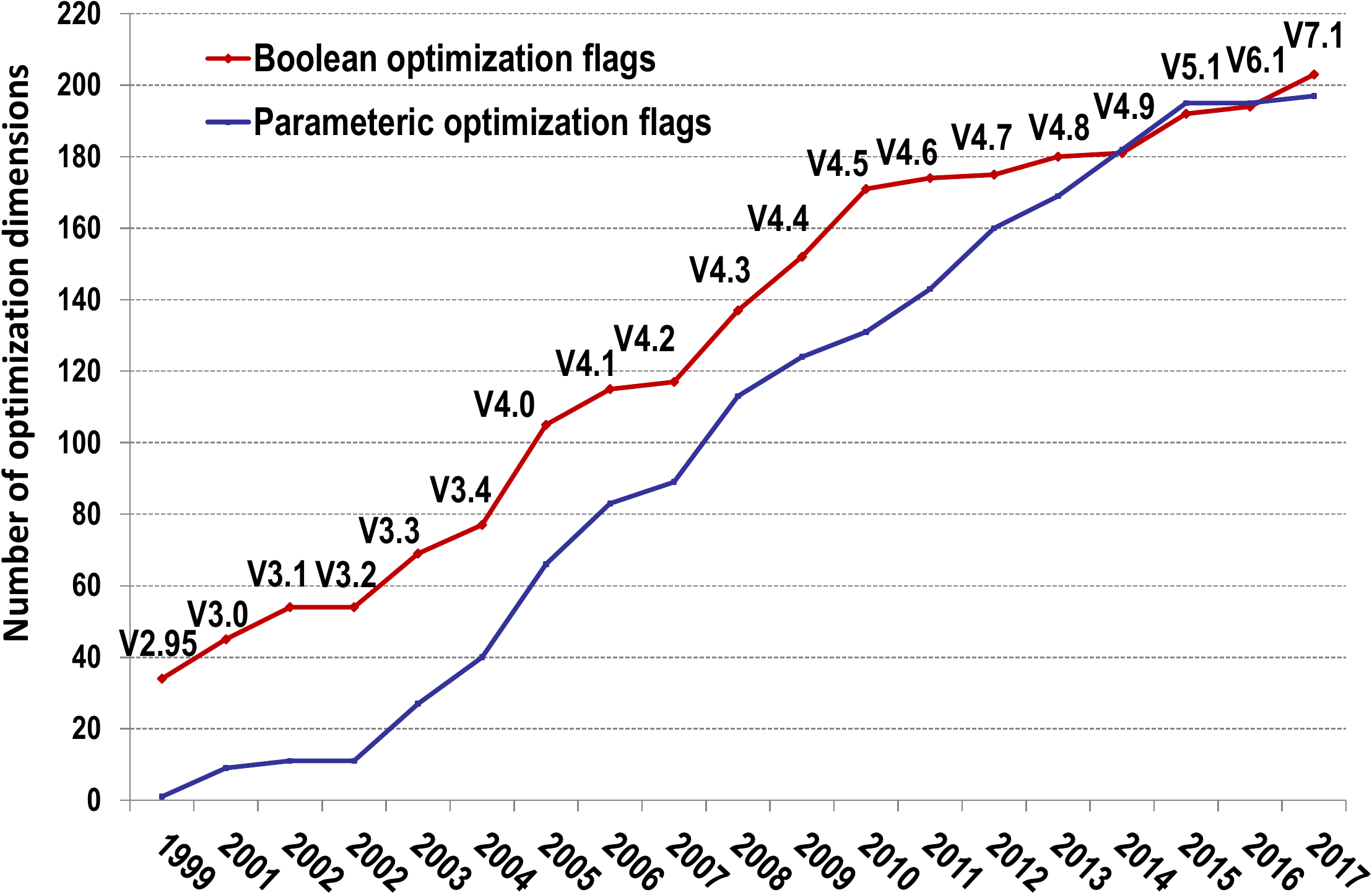} 
     \caption{
       Continuously rising number of boolean and parametric optimization flags 
       in GCC over years (obtained by automatically parsing GCC source code 
       and manual pages, therefore small variation is possible).
     }
     \label{fig:rising-compiler-flags}
   \end{figure}

Indeed, the raising complexity of ever changing hardware 
made development of compilers very challenging.
Popular GCC and LLVM compilers nowadays include hundreds 
of optimizations (Figure~\ref{fig:rising-compiler-flags}) 
and often fail to produce efficient code (execution time and code size)
on realistic workloads within a reasonable compilation time~\cite{atlas, europar97x, citeulike:1671417,
Hall:2009:CRN:1461928.1461946, fursin:hal-01054763}.
Such large design and optimization spaces mean
that hardware and compiler designers can afford to explore
only a tiny fraction of the whole optimization space 
using just few ad-hoc benchmarks and data sets on a few architectures
in a tough mission to assemble \textit{-O3}, \textit{-Os} and other
optimization levels across all supported architectures and workloads.

Our idea is to keep compiler as a simple collection of code analysis and transformation routines 
and separate it from optimization heuristics.
In such case we can use CK autotuning workflow to collaboratively optimize multiple 
shared benchmarks and realistic workloads across diverse hardware, exchange optimization results,
and continuously learn and update compiler optimization heuristics for a given hardware 
as a compiler plugin.
We will demonstrate this approach by randomly optimizing compiler flags 
for \textit{susan corners} program with aging \textit{GCC 4.9.2}, the latest \textit{GCC 7.1.0}
and compare them with~\textit{Clang 3.8.1}.
We already monitor and optimize execution time and code size of this popular image processing 
application across different compilers and platforms for many years~\cite{29db2248aba45e59:a31e374796869125}.
That is why we are interested to see if we can still improve it with the CK autotuner 
on the latest \textit{Raspberry Pi 3 (Model B)} devices (RPi3) extensively used for educational purposes.

\begin{figure}[]
     \centering
      \includegraphics[width=2.8in]
      {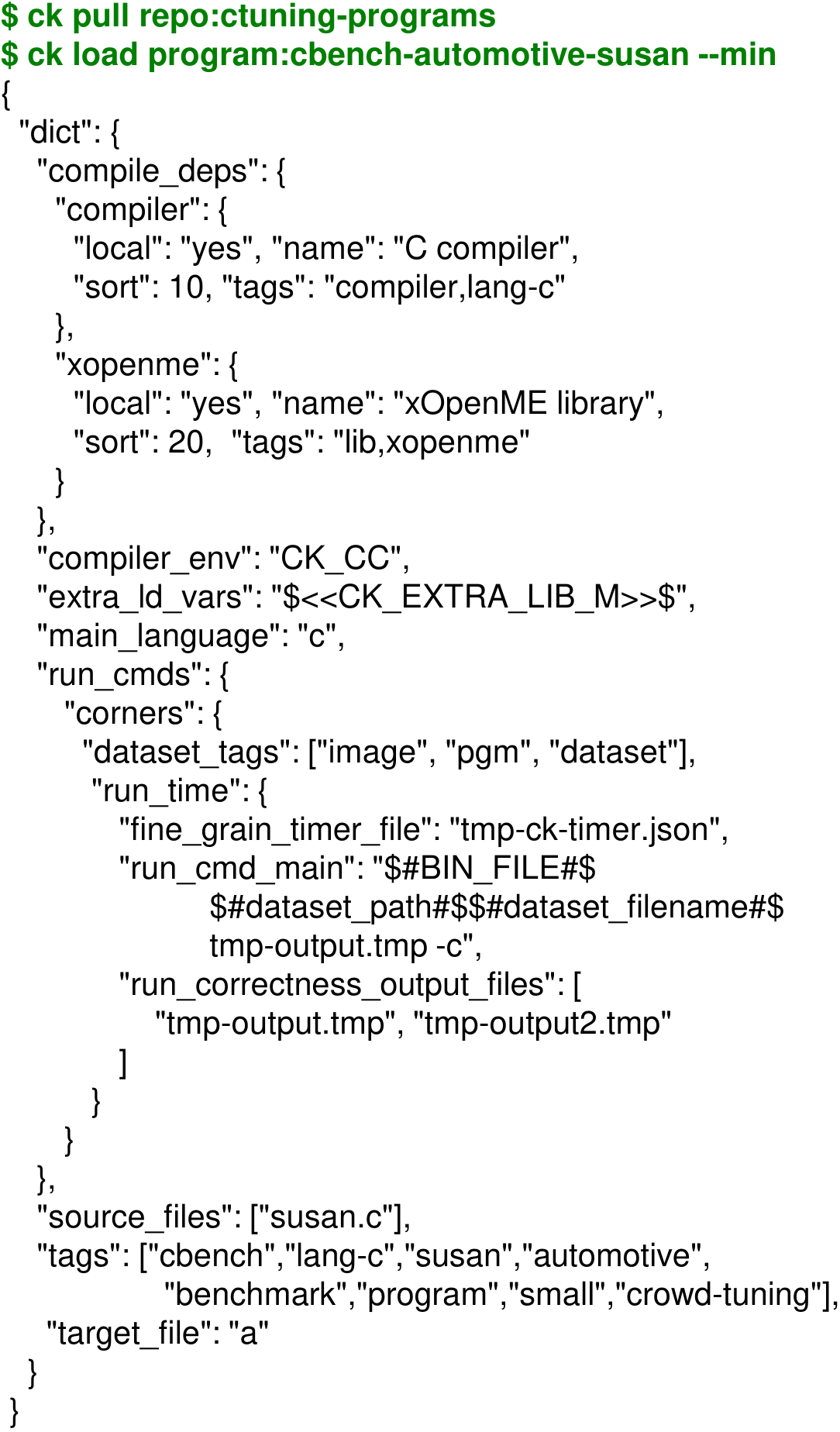} 
     \caption{
       CK JSON meta information for susan corners (image processing program) to describe software dependencies as well as how to compile and run it.
     }
     \label{fig:susan-corners-ck-json-meta}
   \end{figure}

First of all, we added \textit{susan} program with \textit{corners} algorithm 
to the \textit{ctuning-programs} repository with the JSON meta information 
describing compilation and execution 
as shown in Figure~\ref{fig:susan-corners-ck-json-meta}.

We can then test its compilation and execution by invoking the program pipeline as following:
\begin{flushleft}
\texttt{\$ ck pipeline program:cbench-automotive-susan}\newline
\end{flushleft}

CK program pipeline will first attempt to detect platform features
(OS, CPU, GPU) and embed them to the input dictionary using key \textit{features}.
Note that in case of cross-compilation for a target platform different from the host one
(Android, remote platform via SSH, etc), 
it is possible to specify such platform using CK \textit{os} entries and \textit{--target\_os=} flag.

For example, it is possible to compile and run a given CK program for Android via adb as following:
\begin{flushleft}
\texttt{\$ ck ls os}\newline
\texttt{\$ ck pipeline program:cbench-automotive-susan --target\_os=android21-arm64}\newline
\end{flushleft}

Next, CK will try to resolve software dependencies and prepare environment for compilation
by detecting already installed compilers using CK \textit{soft:compiler.*} entries 
or installing new ones if none was found using CK \textit{package:compiler.*}.
Each installed compiler for each target 
will have an associated CK entry with prepared environment 
to let computer systems researchers work with different 
versions of different tools:
\begin{flushleft}
\texttt{\$ ck show env}\newline
\texttt{\$ ck show env --target\_os=android21-arm64}\newline
\texttt{\$ ck show env --tags=compiler}\newline
\end{flushleft}

\begin{figure}[]
     \centering
      \includegraphics[width=3.0in]
      {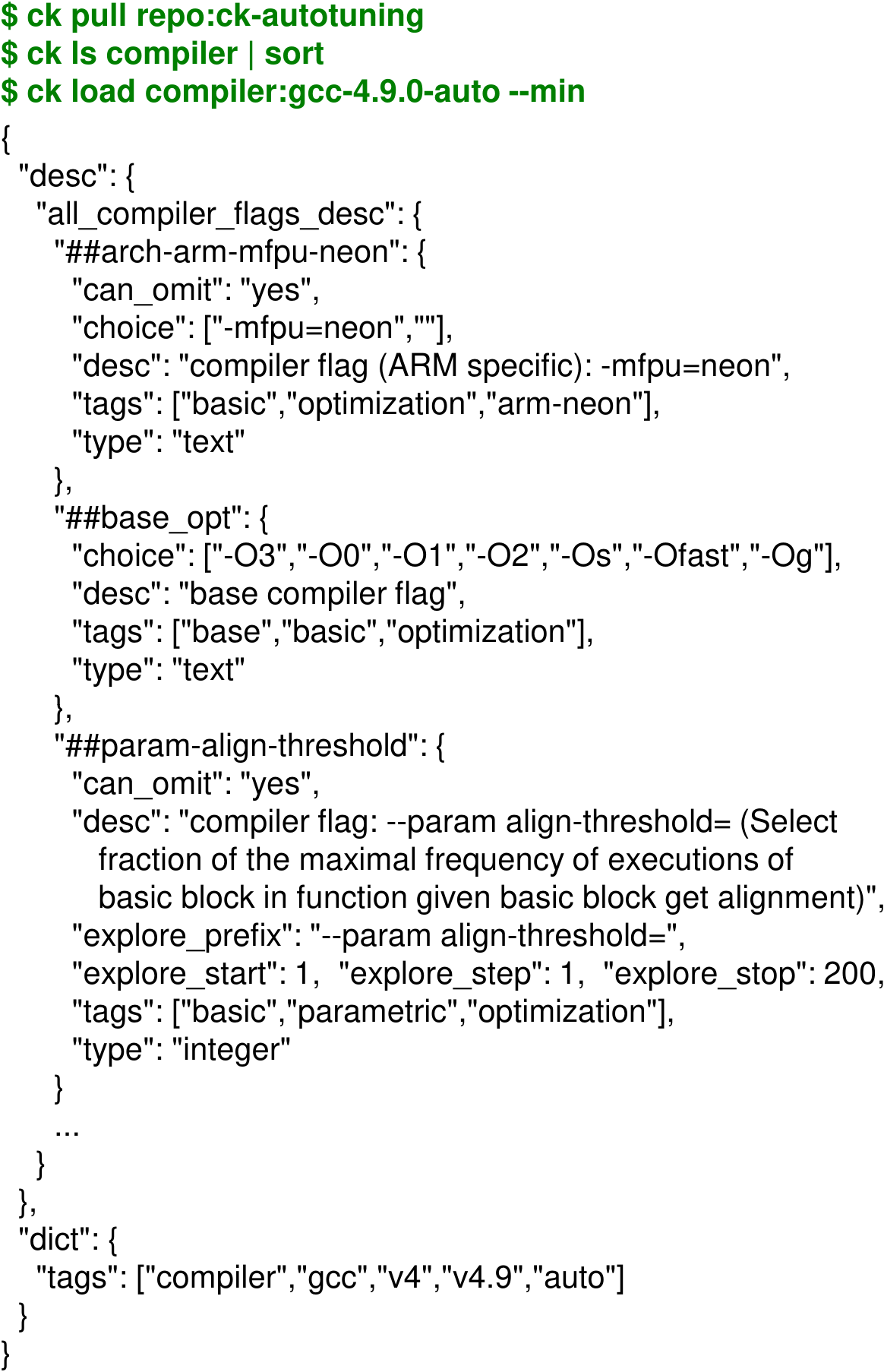} 
     \caption{
       CK JSON description of compiler flags for GCC 4.9 to enable universal autotuning.
     }
     \label{fig:ck-gcc-meta}
   \end{figure}

Automatically detected version of a selected compiler is used by CK
to find and preload all available optimization flags 
from related \textit{compiler:*} entries to the \textit{choices} key
of a pipeline input.
An example of such flags and tags in the CK JSON format 
for GCC 4.9 is shown in Figure~\ref{fig:ck-gcc-meta}.
The community can continue extending such descriptions for different compilers
including \textit{GCC, LLVM, Julia, Open64, PathScale, Java, MVCC, ICC and PGI}
using either public \href{https://github.com/ctuning/ck-autotuning}{ck-autotuning} repository
or their own ones.

Finally, CK program pipeline compiles a given program, runs it on a target platform
and fills in sub-dictionary \textit{characteristics} 
with compilation time, object and binary sizes, MD5 sum of the binary, execution time,
used energy (if supported by a used platform), and all other obtained measurements
in the common pipeline dictionary.

We are now ready to implement universal compiler flag autotuning coupled with
this program pipeline.
For a proof-of-concept, we implemented GCC compiler flags exploration strategy
which automatically generate N random combinations of compiler flags, 
compile a given program with each combination, runs it and record all results 
(inputs and outputs of a pipeline) in a reproducible form in a \textit{local} 
CK repository using \textit{experiment} module from 
the \href{https://github.com/ctuning/ck-analytics}{ck-analytics} 
repository:

\begin{flushleft}
\texttt{\$ ck pull repo:ck-crowdtuning}\newline
\texttt{\$ ck info module:experiment.tune.compiler.flags.gcc}\newline
\end{flushleft}

\begin{figure}[]
     \centering
      \includegraphics[width=3.0in]
      {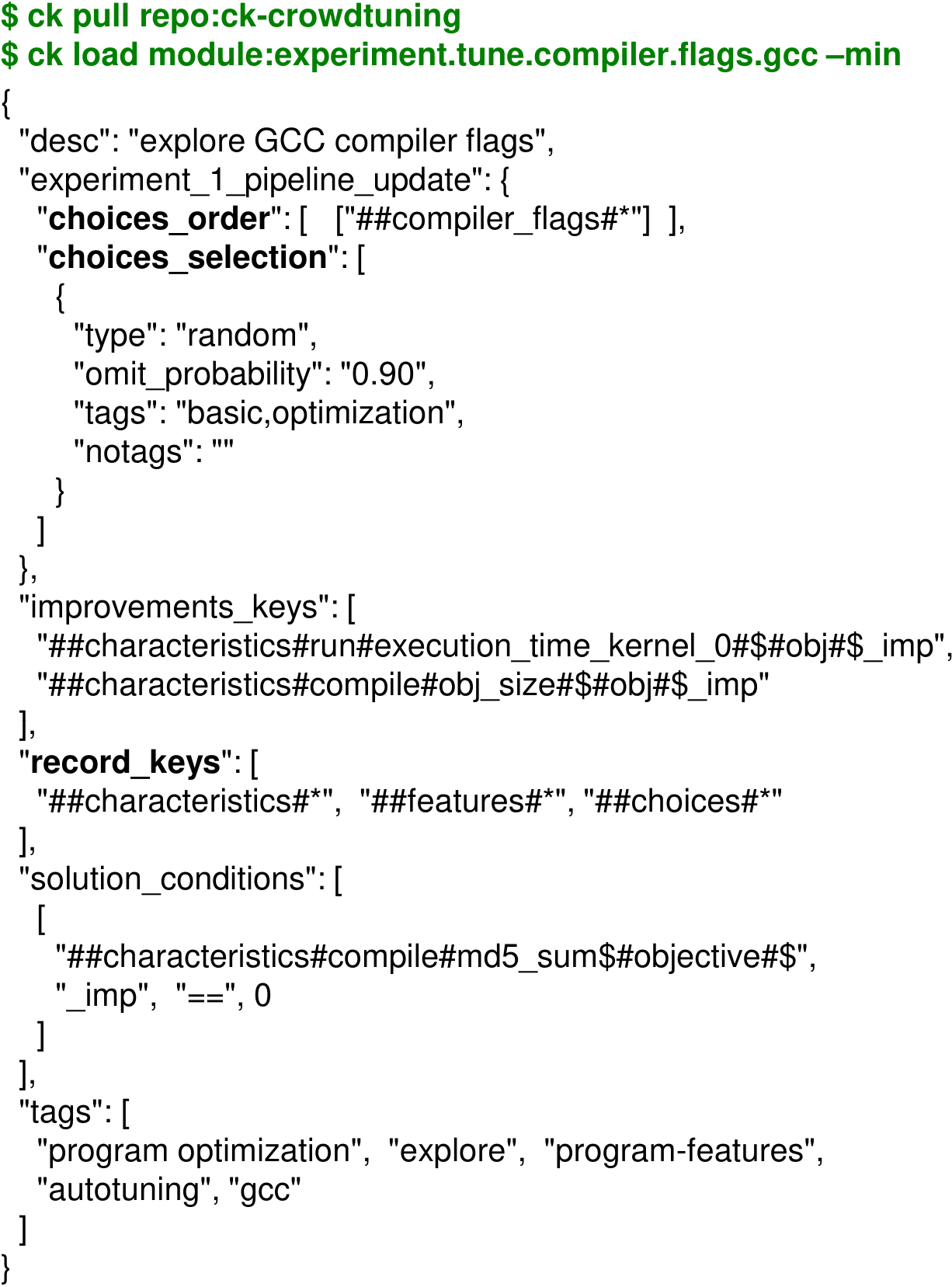} 
     \caption{
       CK JSON description of random autotuning of compiler flags applied to program pipeline.
     }
     \label{fig:ck-gcc-tuning-meta}
   \end{figure}

The JSON meta information of this module describes which keys to select
in the program pipeline, how to tune them, and which characteristics to monitor
and record as shown in Figure~\ref{fig:ck-gcc-tuning-meta}.
Note that a string starting with \emph{\#\#} is used to reference any key
in a complex, nested JSON or Python dictionary (\textit{CK flat key} \cite{fursin:hal-01054763}).
Such \emph{flat key} always starts with \emph{\#} 
followed by \emph{\#key} if it is a dictionary key or
\emph{@position\_in\_a\_list} if it is a value in a list. 
CK also supports wild cards in such flat keys 
such as \emph{"\#\#compiler\_flags\#\*"} and \emph{"\#\#characteristics\#\*}
to be able to select multiple sub-keys, dictionaries 
and lists in a given dictionary.

We can now invoke this CK experimental scenario from the command line as following:

\begin{flushleft}
\texttt{\$ ck autotune program:cbench-automotive-susan --iterations=300 --repetitions=3 
  --scenario=experiment.tune.compiler.flags.gcc
  --cmd\_key=corners --record\_uoa=tmp-susan-corners-gcc4-300-rnd}
\end{flushleft}

CK will generate 300 random combinations of compiler flags, compile \textit{susan corners} program 
with each combination, run each produced code 3 times to check variation, and record
results in the \textit{experiment:tmp-susan-corners-gcc4-300-rnd}.
We can now visualize these autotuning results using the following command line:
\begin{flushleft}
\texttt{\$ ck plot graph:tmp-susan-corners-gcc4-300-rnd}
\end{flushleft}

\begin{figure*}[!htbp]
    \centering
     \includegraphics[width=6.9in]
     {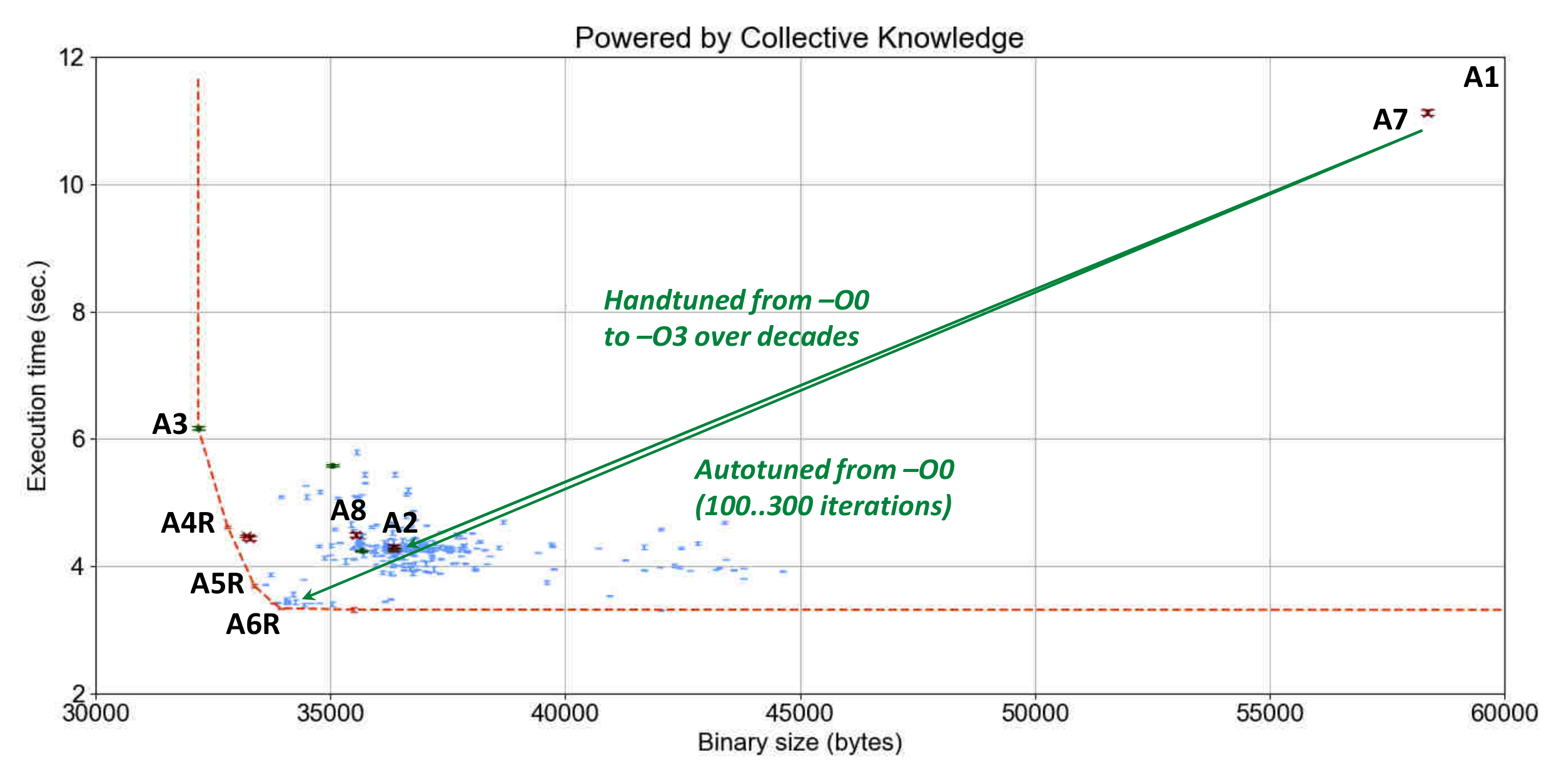} 
     \vspace{0.1in}
         \begin{tabular}{|l|l|l|l|p{3.2in}|}
     \hline
      \textbf{ID} & \textbf{Compiler} & \textbf{Time (sec.)} & \textbf{Size (bytes)} & \textbf{Flags} \\ 
     \hline
      \textbf{ \href{http://cknowledge.org/repo/web.php?wcid=experiment:f9e6ec8d198c36c3\&subpoint=7cf654adb86fb606}{A1} } &  GCC 4.9.2  &  11.7 $\pm$ 0.0  &  60560  & {\small  }\\
     \hline
      \textbf{ \href{http://cknowledge.org/repo/web.php?wcid=experiment:0b867dd820354a8b\&subpoint=5734f47e4214a783}{A2} } &  GCC 4.9.2  &  4.3 $\pm$ 0.1  &  36360  & {\small -O3 }\\
     \hline
      \textbf{ \href{http://cknowledge.org/repo/web.php?wcid=experiment:b2b26ab783304fc4\&subpoint=7d87c22a2425da10}{A3} } &  GCC 4.9.2  &  6.2 $\pm$ 0.1  &  32184  & {\small -Os }\\
     \hline
      \textbf{ \href{http://cknowledge.org/repo/web.php?wcid=experiment:98688a71f99ac30b\&subpoint=4bcd9dad6b249a79}{A4R} } &  GCC 4.9.2  &  4.2 $\pm$ 0.0  &  32448  & {\small -O3 -fno-guess-branch-probability -fno-if-conversion -fno-ivopts -fno-schedule-insns -fsingle-precision-constant --param max-unswitch-insns=5 }\\
     \hline
      \textbf{ \href{http://cknowledge.org/repo/web.php?wcid=experiment:984b2d8abc3c4415\&subpoint=78c281b4cab897a6}{A5R} } &  GCC 4.9.2  &  3.7 $\pm$ 0.1  &  33376  & {\small -O3 -fbranch-probabilities -fno-ivopts -fno-sched-dep-count-heuristic }\\
     \hline
      \textbf{ \href{http://cknowledge.org/repo/web.php?wcid=experiment:7af17ca204080b57\&subpoint=5a464ecf81b60098}{A6R} } &  GCC 4.9.2  &  3.4 $\pm$ 0.0  &  33804  & {\small -O3 -fno-inline-small-functions -fno-ivopts -fno-tree-partial-pre }\\
     \hline
      \textbf{ \href{http://cknowledge.org/repo/web.php?wcid=experiment:a32e34c31b900930\&subpoint=64ec888d2e6a0669}{A7} } &  CLANG 3.8.1  &  11.1 $\pm$ 0.1  &  58368  & {\small  }\\
     \hline
      \textbf{ \href{http://cknowledge.org/repo/web.php?wcid=experiment:99141b3313132494\&subpoint=a63f42ac837e38d0}{A8} } &  CLANG 3.8.1  &  4.5 $\pm$ 0.1  &  35552  & {\small -O3 }\\
     \hline
    \end{tabular}     
     \vspace{0.1in}
    \caption{
      Results of GCC 4.9.2 random compiler flag autotuning of susan corners program on Raspberry~Pi~3 (Model~B) 
      device using CK with a highlighted frontier (trading-off execution time and code size) 
      and best found combinations of flags on this frontier.
    }
    \label{fig:autotuning-susan-gcc4}
  \end{figure*}

Figure~\ref{fig:autotuning-susan-gcc4} shows a manually annotated graph 
with the outcome of GCC 4.9.2 random compiler flags autotuning 
applied to susan corners on an RPi3 device in terms of execution 
time with variation and code size.
Each blue point on this graph is related to one combination of random compiler flags.
The red line highlights the frontier of all autotuning results (not necessarily Pareto optimal) 
which trade off execution time and code size during multi-objective optimization.
We also plotted points when default GCC compilation is used without any flags 
or with \textit{-O3} and \textit{-Os} optimization levels.
Finally, we decided to compare optimization results with \textit{Clang 3.8.1} also available on RPi3.

Besides showing that \textit{GCC -O3} (optimization choice \textbf{A2})
and \textit{Clang -O3} (optimization choice \textbf{A8}) can produce a very similar code, 
these results confirm well that it is indeed possible to automatically obtain execution time 
and binary size of \textit{-O3} and \textit{-Os} levels in comparison with non-optimized code 
within tens to hundreds autotuning iterations (green improvement vectors with ~3.6x execution time speedup 
and ~1.6x binary size improvement).
The graph also shows that it is possible to improve best optimization level \textit{-O3} 
much further and obtain ~1.3x execution time speedup (optimization solution \textbf{A6R}
or obtain 11\% binary size improvement without sacrifying original execution time
(optimization solution \textbf{A4R}).
Such automatic squeezing of a binary size without sacrificing performance 
can be very useful for the future IoT devices.

Note that it is possible to browse all results in a user-friendly way 
via web browser using the following command:

\begin{flushleft}
\texttt{\$ ck browse experiment:tmp-susan-corners-gcc4-300-rnd}
\end{flushleft}

CK will then start internal CK web server 
available in the \href{https://github.com/ctuning/ck-web}{ck-web}
repository, will run a default web browser, and will 
open a web page with all given experimental results.
Each experiment on this page has an associated button 
with a command line to replay it via CK such as:                          

\begin{flushleft}
\texttt{\$ ck replay experiment:7b41a4ac1b3b4f2b --point=00e81f4e4abb371d}
\end{flushleft}

CK will then attempt to reproduce this experiment using the same input
and then report any differences in the output.
This simplifies validation of shared experimental results 
(optimizations, models, bugs) by the community
and possibly with a different software and hardware setup
(CK will automatically adapt the workflow to a user platform).

We also provided support to help researchers 
visualize their results as interactive graphs 
using popular D3.js library as demonstrated in this 
\href{http://cknowledge.org/repo/web.php?wcid=graph:6b6d77a51c74ec1a&subgraph=rpi3-autotuning-susan-gcc4-interactive}{link}.

Looking at above optimization results one may notice 
that one of the original optimization solutions on a frontier \textbf{A4} 
has ~40 optimization flags, while \textbf{A4R} only 7 as shown in Table~\ref{fig:autotuning-susan-gcc4-reduce}.
The natural reason is that not all randomly selected flags contribute to improvements.
That is why we developed a simple and universal complexity reduction algorithm.
It iteratively and randomly removes choices from a found solution one by one
if they do not influence monitored characteristics such as execution time and code size
in our example.

  \begin{table*}[]
    \centering
         \begin{tabular}{|l|p{6.2in}|}
     \hline
      \textbf{ID} & \textbf{Flags} \\ 
     \hline
      \textbf{ \href{http://cknowledge.org/repo/web.php?wcid=experiment:8556dbf4e51a825d\&subpoint=86ca1630895041c1}{A4} } & {\small -O3 -fira-algorithm=priority -fcaller-saves -fno-devirtualize-speculatively -fno-function-cse -fgcse-sm -fno-guess-branch-probability -fno-if-conversion -fno-inline-functions-called-once -fipa-reference -fno-ira-loop-pressure -fira-share-save-slots -fno-isolate-erroneous-paths-dereference -fno-ivopts -floop-nest-optimize -fmath-errno -fmove-loop-invariants -fsched-last-insn-heuristic -fsched2-use-superblocks -fno-schedule-insns -fno-signed-zeros -fsingle-precision-constant -fno-tree-sink -fno-unsafe-loop-optimizations --param asan-instrument-reads=1 --param gcse-unrestricted-cost=5 --param l1-cache-size=11 --param large-function-growth=33 --param loop-invariant-max-bbs-in-loop=636 --param max-completely-peel-loop-nest-depth=7 --param max-delay-slot-live-search=163 --param max-gcse-insertion-ratio=28 --param max-inline-insns-single=282 --param max-inline-recursive-depth-auto=0 --param max-jump-thread-duplication-stmts=6 --param max-last-value-rtl=4062 --param max-pipeline-region-insns=326 --param max-sched-region-blocks=17 --param max-tail-merge-iterations=2 --param max-unswitch-insns=5 --param max-vartrack-expr-depth=6 --param min-spec-prob=1 --param omega-eliminate-redundant-constraints=1 --param omega-max-keys=366 --param omega-max-wild-cards=36 --param sms-dfa-history=0 }\\
     \hline
      \textbf{ \href{http://cknowledge.org/repo/web.php?wcid=experiment:98688a71f99ac30b\&subpoint=4bcd9dad6b249a79}{A4R} } & {\small -O3 -fno-guess-branch-probability -fno-if-conversion -fno-ivopts -fno-schedule-insns -fsingle-precision-constant --param max-unswitch-insns=5 }\\
     \hline
    \end{tabular}     
    \caption{
      One of original optimization solutions found after autotuning with random selection of compiler flags (A4) 
      and reduced optimization solution (A4R) which results in the same or better execution time and code size.
    }
    \label{fig:autotuning-susan-gcc4-reduce}
  \end{table*}

Such complexity reduction (pruning) of an existing solution can be invoked as following
(flag \textit{--prune\_md5} tells CK to exclude a given choice without running code
if MD5 of a produced binary didn't change, thus considerably speeding up flag pruning):

\begin{flushleft}
\texttt{\$ck replay experiment:93974bf451f957eb --point=74e9c9f14b424ba7 --prune --prune\_md5 @prune.json}
\end{flushleft}

\begin{figure}[]
     \centering
      \includegraphics[width=3.0in]
      {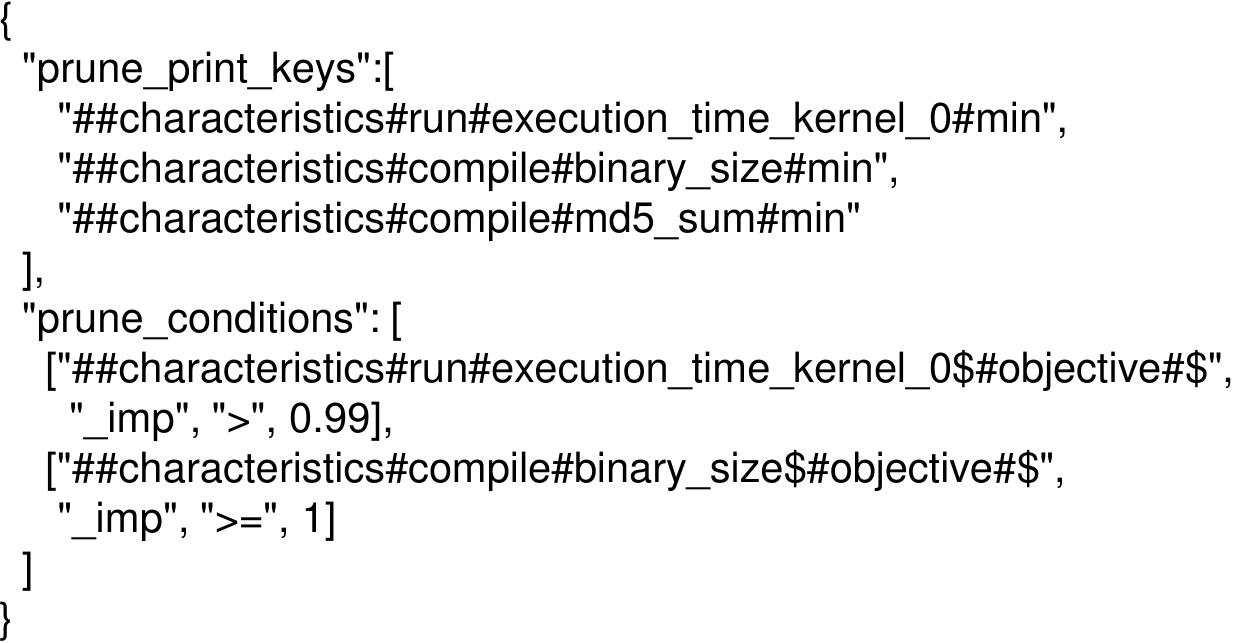} 
     \caption{
       CK JSON description of conditions on choices in a pipeline input to reduce choices from a found optimization solution.
     }
     \label{fig:ck-pruning-meta}
   \end{figure}

The \textit{'prune.json'} file describes conditions on program pipeline keys 
when a given choice should be removed as shown in Figure~\ref{fig:ck-pruning-meta}.

\begin{figure}[]
     \centering
      \includegraphics[width=3.0in]
      {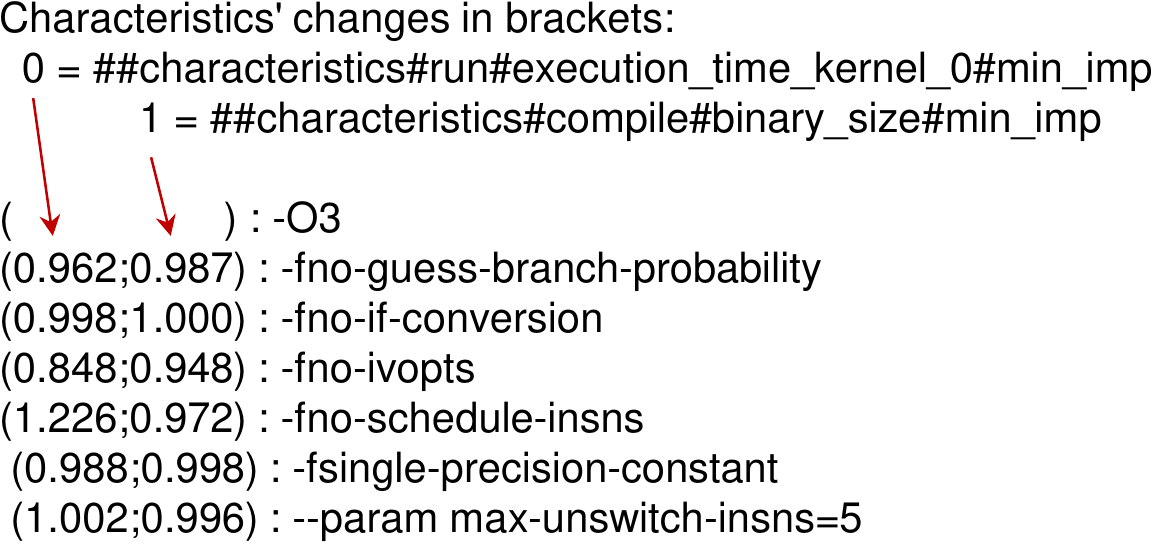} 
     \caption{
       Contribution of individual compiler flags to improvements or degradations of monitored characteristics during universal complexity reduction.
     }
     \label{fig:ck-pruning-contribution}
   \end{figure}

Such universal complexity reduction approach helps software engineers better understand
individual contribution of each flag to improvements or degradations of all monitored
characteristics such as execution time and code size as shown in Figure~\ref{fig:ck-pruning-contribution}.

Asked by compiler developers, we also provided an extension to our complexity reduction module 
to turn off explicitly all available optimization choices one by one
if they do not influence found optimization result.
Table~\ref{fig:autotuning-susan-gcc4-invert} demonstrates this approach and shows all compiler optimizations contributing to the found optimization solution.
It can help improve internal optimization heuristics, global optimization levels such as \textit{-O3},
and improve machine learning based optimization predictions.
This extension can be invoked by adding flags \textit{--prune\_invert --prune\_invert\_do\_not\_remove\_key}
when reducing complexity of a given solution such as:
\begin{flushleft}
\texttt{\$ ck replay experiment:93974bf451f957eb --point=74e9c9f14b424ba7 --prune --prune\_md5 --prune\_invert --prune\_invert\_do\_not\_remove\_key @prune.json}
\end{flushleft}

  \begin{table*}[]
   \centering
        \begin{tabular}{|l|p{6.2in}|}
     \hline
      \textbf{ID} & \textbf{Flags} \\ 
     \hline
      \textbf{ \href{http://cknowledge.org/repo/web.php?wcid=experiment:7af17ca204080b57\&subpoint=5a464ecf81b60098}{A6R} } & {\small -O3 -fno-inline-small-functions -fno-ivopts -fno-tree-partial-pre }\\
     \hline
      \textbf{ \href{http://cknowledge.org/repo/web.php?wcid=experiment:f594f90a1545babf\&subpoint=de11e85953947388}{A6RI} } & {\small \textbf{-O3} -fno-inline-small-functions -fno-ivopts -fno-tree-bit-ccp -fno-tree-partial-pre -fno-tree-pta -fno-associative-math -fno-auto-inc-dec -fno-branch-probabilities -fno-branch-target-load-optimize -fno-branch-target-load-optimize2 -fno-caller-saves -fno-check-data-deps -fno-combine-stack-adjustments -fno-conserve-stack -fno-compare-elim \textbf{-fcprop-registers} \textbf{-fcrossjumping} \textbf{-fcse-follow-jumps} -fno-cse-skip-blocks -fno-cx-limited-range -fno-data-sections \textbf{-fdce} -fno-delayed-branch -fno-devirtualize -fno-devirtualize-speculatively -fno-early-inlining -fno-ipa-sra -fno-expensive-optimizations -fno-fat-lto-objects -fno-fast-math -fno-finite-math-only -fno-float-store \textbf{-fforward-propagate} -fno-function-sections -fno-gcse-after-reload -fno-gcse-las -fno-gcse-lm -fno-graphite-identity -fno-gcse-sm -fno-hoist-adjacent-loads -fno-if-conversion \textbf{-fif-conversion2} -fno-indirect-inlining -fno-inline-functions -fno-inline-functions-called-once -fno-ipa-cp -fno-ipa-cp-clone -fno-ipa-pta \textbf{-fipa-pure-const} -fno-ipa-reference -fno-ira-hoist-pressure -fno-ira-loop-pressure -fno-ira-share-save-slots \textbf{-fira-share-spill-slots} \textbf{-fisolate-erroneous-paths-dereference} -fno-isolate-erroneous-paths-attribute -fno-keep-inline-functions -fno-keep-static-consts -fno-live-range-shrinkage -fno-loop-block -fno-loop-interchange -fno-loop-strip-mine -fno-loop-nest-optimize -fno-loop-parallelize-all -fno-lto -fno-merge-all-constants -fno-merge-constants -fno-modulo-sched -fno-modulo-sched-allow-regmoves \textbf{-fmove-loop-invariants} -fno-branch-count-reg -fno-defer-pop -fno-function-cse \textbf{-fguess-branch-probability} \textbf{-finline} \textbf{-fmath-errno} -fno-peephole \textbf{-fpeephole2} -fno-sched-interblock -fno-sched-spec -fno-signed-zeros -fno-toplevel-reorder -fno-trapping-math -fno-zero-initialized-in-bss \textbf{-fomit-frame-pointer} -fno-optimize-sibling-calls -fno-partial-inlining -fno-peel-loops -fno-predictive-commoning -fno-prefetch-loop-arrays -fno-ree -fno-rename-registers \textbf{-freorder-blocks} -fno-reorder-blocks-and-partition -fno-rerun-cse-after-loop -fno-reschedule-modulo-scheduled-loops -fno-rounding-math -fno-sched2-use-superblocks \textbf{-fsched-pressure} -fno-sched-spec-load -fno-sched-spec-load-dangerous -fno-sched-group-heuristic \textbf{-fsched-critical-path-heuristic} -fno-sched-spec-insn-heuristic -fno-sched-rank-heuristic -fno-sched-dep-count-heuristic \textbf{-fschedule-insns} \textbf{-fschedule-insns2} -fno-section-anchors -fno-selective-scheduling -fno-selective-scheduling2 -fno-sel-sched-pipelining -fno-sel-sched-pipelining-outer-loops -fno-shrink-wrap -fno-signaling-nans -fno-single-precision-constant -fno-split-ivs-in-unroller -fno-split-wide-types -fno-strict-aliasing \textbf{-fstrict-overflow} -fno-tracer -fno-tree-builtin-call-dce -fno-tree-ccp \textbf{-ftree-ch} -fno-tree-coalesce-vars -fno-tree-copy-prop \textbf{-ftree-copyrename} \textbf{-ftree-dce} \textbf{-ftree-dominator-opts} -fno-tree-dse \textbf{-ftree-forwprop} -fno-tree-fre -fno-tree-loop-if-convert -fno-tree-loop-if-convert-stores \textbf{-ftree-loop-im} -fno-tree-phiprop -fno-tree-loop-distribution -fno-tree-loop-distribute-patterns -fno-tree-loop-linear \textbf{-ftree-loop-optimize} -fno-tree-loop-vectorize -fno-tree-pre \textbf{-ftree-reassoc} -fno-tree-sink \textbf{-ftree-slsr} \textbf{-ftree-sra} -fno-tree-switch-conversion -fno-tree-tail-merge \textbf{-ftree-ter} -fno-tree-vectorize \textbf{-ftree-vrp} -fno-unit-at-a-time -fno-unroll-all-loops -fno-unroll-loops -fno-unsafe-loop-optimizations -fno-unsafe-math-optimizations -fno-unswitch-loops -fno-variable-expansion-in-unroller -fno-vect-cost-model -fno-vpt -fno-web -fno-whole-program -fno-wpa \textbf{-fexcess-precision=standard} \textbf{-ffp-contract=off} \textbf{-fira-algorithm=CB} \textbf{-fira-region=all} }\\
     \hline
    \end{tabular}     
   \caption{
     Explicitly switching off all compiler flags one by one if they do not influence the optimization result - 
     useful to understand all compiler optimizations which contributed to the found solution. 
   }
   \label{fig:autotuning-susan-gcc4-invert}
  \end{table*}

We have been analyzing already aging \textit{GCC 4.9.2} because 
it is still the default compiler for Jessy Debian distribution on RPi3.
However, we would also like to check how our universal autotuner
works with the latest \textit{GCC 7.1.0}.

Since there is no yet a standard Debian GCC 7.1.0 package available for RPi3,
we need to build it from scratch.
This is not a straightforward task since we have to pick up correct 
configuration flags which will adapt GCC build to quite outdated RPi3 libraries.
However, once we manage to do it, we can automate this process
using CK \textit{package} module. 

We created a public \href{https://github.com/ctuning/ck-dev-compilers}{ck-dev-compilers} repository
to automate building and installation of various compilers including GCC and LLVM via CK.
It is therefore possible to install GCC 7.1.0 on RPi3 as following 
(see Appendix or GitHub repository ReadMe file for more details):

\begin{flushleft}
\texttt{\$ ck pull repo:ck-dev-compilers \newline
\$ ck install package:compiler-gcc-any-src-linux-no-deps --env.PARALLEL\_BUILDS=1 --env.GCC\_COMPILE\_CFLAGS=-O0 --env.GCC\_COMPILE\_CXXFLAGS=-O0 --env.EXTRA\_CFG\_GCC=--disable-bootstrap --env.RPI3=YES --force\_version=7.1.0}
\end{flushleft}

This CK package has an \textit{install.sh} script which is customized 
using environment variables or \textit{--env} flags to build GCC for a target platform.
The JSON meta data of this CK package provides optional software dependencies 
which CK has to resolve before installation (similar to CK compilation).
If installation succeeded, you should be able to see two prepared environments
for GCC 4.9.2 and GCC 7.1.0 which now co-exist in the system.

\begin{flushleft}
\texttt{\$ ck show env --tags=gcc}
\end{flushleft}

Whenever we now invoke CK autotuning, CK software and package manager 
will detect multiple available versions of a required software dependency
and will let you choose which compiler version to use.

Let us now autotune the same \textit{susan corners} program 
by generating 300 random combinations of \textit{GCC 7.1.0} compiler flags
and record results in the \textit{experiment:tmp-susan-corners-gcc7-300-rnd}:

\begin{flushleft}
\texttt{\$ ck autotune program:cbench-automotive-susan --iterations=300 --repetitions=3 
  --scenario=experiment.tune.compiler.flags.gcc
  --cmd\_key=corners --record\_uoa=tmp-susan-corners-gcc7-300-rnd}
\end{flushleft}

\begin{figure*}[]
     \centering
      \includegraphics[width=6.9in]
      {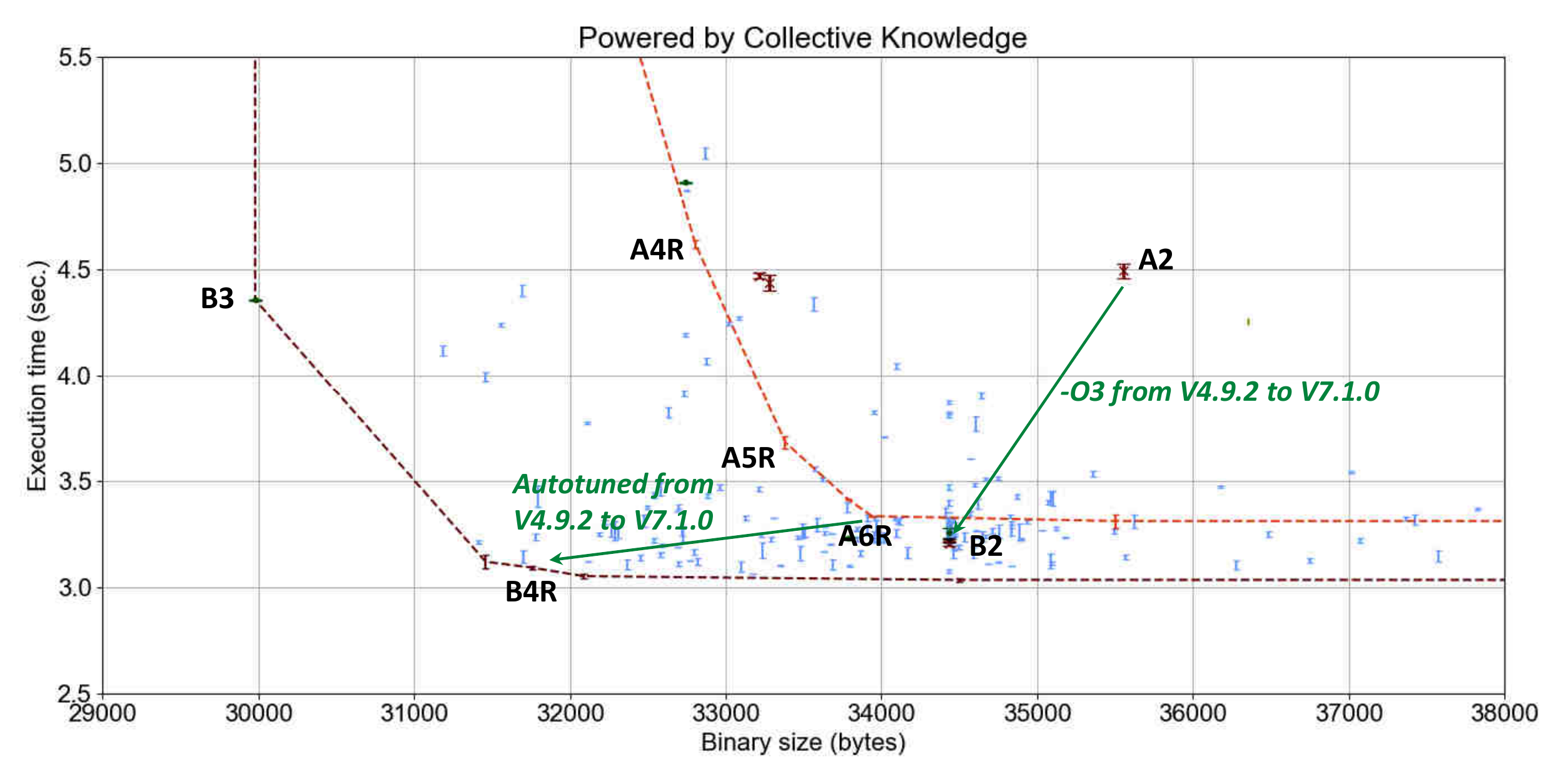} 
      \vspace{0.1in}
          \begin{tabular}{|l|l|l|l|p{3.2in}|}
     \hline
      \textbf{ID} & \textbf{Compiler} & \textbf{Time (sec.)} & \textbf{Size (bytes)} & \textbf{Flags} \\ 
     \hline
      \textbf{ \href{http://cknowledge.org/repo/web.php?wcid=experiment:0b867dd820354a8b\&subpoint=5734f47e4214a783}{A2} } &  GCC 4.9.2  &  4.3 $\pm$ 0.1  &  36360  & {\small -O3 }\\
     \hline
      \textbf{ \href{http://cknowledge.org/repo/web.php?wcid=experiment:984b2d8abc3c4415\&subpoint=78c281b4cab897a6}{A5R} } &  GCC 4.9.2  &  3.7 $\pm$ 0.1  &  33376  & {\small -O3 -fbranch-probabilities -fno-ivopts -fno-sched-dep-count-heuristic }\\
     \hline
      \textbf{ \href{http://cknowledge.org/repo/web.php?wcid=experiment:7af17ca204080b57\&subpoint=5a464ecf81b60098}{A6R} } &  GCC 4.9.2  &  3.4 $\pm$ 0.0  &  33804  & {\small -O3 -fno-inline-small-functions -fno-ivopts -fno-tree-partial-pre }\\
     \hline
      \textbf{ \href{http://cknowledge.org/repo/web.php?wcid=experiment:ea8ded6ddee6093a\&subpoint=6006d3eadf403088}{B1} } &  GCC 7.1.0  &  11.5 $\pm$ 0.0  &  58008  & {\small  }\\
     \hline
      \textbf{ \href{http://cknowledge.org/repo/web.php?wcid=experiment:00fa4e108053ac7b\&subpoint=de68bdc517447085}{B2} } &  GCC 7.1.0  &  3.2 $\pm$ 0.0  &  34432  & {\small -O3 }\\
     \hline
      \textbf{ \href{http://cknowledge.org/repo/web.php?wcid=experiment:eeba1f30493b8bd4\&subpoint=5d64cf89c67817be}{B3} } &  GCC 7.1.0  &  4.4 $\pm$ 0.0  &  29980  & {\small -Os }\\
     \hline
      \textbf{ \href{http://cknowledge.org/repo/web.php?wcid=experiment:f4f13d284194463a\&subpoint=5421ba4fdca2c4b6}{B4} } &  GCC 7.1.0  &  3.1 $\pm$ 0.1  &  31460  & {\small -O3 -fno-cx-fortran-rules -fno-devirtualize -fno-expensive-optimizations -fno-if-conversion -fira-share-save-slots -fno-ira-share-spill-slots -fno-ivopts -fno-loop-strip-mine -finline -fno-math-errno -frounding-math -fno-sched-rank-heuristic -fno-sel-sched-pipelining-outer-loops -fno-semantic-interposition -fsplit-wide-types -fno-tree-ccp -ftree-dse }\\
     \hline
      \textbf{ \href{http://cknowledge.org/repo/web.php?wcid=experiment:2b6646924c79bf81\&subpoint=e8a36aa042438771}{B4R} } &  GCC 7.1.0  &  3.1 $\pm$ 0.1  &  31420  & {\small -O3 -fno-expensive-optimizations -fno-ivopts -fno-math-errno }\\
     \hline
    \end{tabular}     
      \vspace{0.1in}
     \caption{
      Results of GCC 7.1.0 random compiler flag autotuning of susan corners program on Raspberry~Pi~3 (Model~B) 
      device using CK with a highlighted frontier (trading-off execution time and code size), 
      best combinations of flags on this frontier, and comparison with the results from GCC 4.9.2.
     }
     \label{fig:autotuning-susan-gcc7}
   \end{figure*}

Figure~\ref{fig:autotuning-susan-gcc7} shows the results of such \textit{GCC 7.1.0}
compiler flag autotuning (\textbf{B} points) and compares them 
against \textit{GCC 4.9.2} (\textbf{A} points).
Note that this graph is also available in interactive form~\href{http://cknowledge.org/repo/web.php?wcid=graph:96fd8e4c8394b1bc&subgraph=rpi3-autotuning-susan-gcc7-interactive}{online}.

It is interesting to see considerable improvement in execution time of susan corners 
when moving from GCC 4.9 to GCC 7.1 with the best optimization level \textit{-O3}.
This graph also shows that new optimization added during the past 3 years opened up
many new opportunities thus considerably expanding autotuning frontier (light red
dashed line versus dark red dashed line).
Autotuning only managed to achieve a modest improvement of a few percent over \textit{-O3}.

On the other hand, GCC \textit{-O3} and \textit{-Os} are still far from achieving
best trade-offs for execution time and code size.
For example, it is still possible to improve a program binary size 
by ~10\% (reduced solution \textbf{B4R}) without degrading best achieved 
execution time with the \textit{-O3} level (\textbf{-O3}), or improve 
execution time of \textbf{-Os} level by ~28\% while slightly degrading code size by ~5\%.

Note that for readers' convenience we added scripts to reproduce and validate 
all results from this section to the following CK entries:

\begin{flushleft}
\texttt{\$ ck pull repo:ck-rpi-optimization-results \newline
\$ ck find script:rpi3-susan*}
\end{flushleft}

These results confirm that it is difficult to manually prepare compiler optimization
heuristic which can deliver good trade offs between execution time and code size 
in such a large design and optimization spaces.
They also suggest that either susan corners or similar code 
was eventually added to the compiler regression testing suite,
or some engineer check it manually and fixed compiler heuristic.
However, there is also no guarantee that future GCC versions will still
perform well on the susan corners program.
Neither these results guarantee that GCC 7.1.0
will perform well on other realistic workloads or devices.


\section{Crowdsourcing autotuning}
\label{sec:crowdtuning}
We use our universal CK autotuning workflow to teach students and end-users 
how to automatically find good trade offs between multiple characteristics 
for any individual program, data set, compiler, environment and hardware.
At the same time, automatically tuning many realistic workloads
is very costly and can easily take from days to weeks and months~\cite{29db2248aba45e59:a31e374796869125}.

Common experimental frameworks can help tackle this problem too by 
crowdsourcing autotuning across diverse hardware provided by volunteers and combining it with online
classification, machine learning and run-time adaptation~\cite{Fur2009,JGVP2009,cm:29db2248aba45e59:cd11e3a188574d80}.
However, our previous frameworks did not cope well with "big data" problem
(cTuning framework~\cite{Fur2009,new_pub_model} based on MySQL database) 
or were too "heavy" (Collective Mind aka cTuning 3 framework~\cite{fursin:hal-01054763}).

Extensible CK workflow framework combined with our cross-platform package manager, 
internal web server and machine learning, helped solve most of the above issues.
For example, we introduced a notion of a remote repository in the CK - 
whenever such repository is accessed CK simply forward all JSON requests 
to an appropriate web server.

CK always has a default remote repository \textit{remote-ck} connected
with a public optimization repository running CK web serve 
at~\url{cKnowledge.org/repo}: 

\begin{flushleft}
\texttt{\$ ck load repo:remote-ck --min}
\end{flushleft}

For example, one can see publicly available experiments from command line as following:
\begin{flushleft}
\texttt{\$ ck list remote-ck:experiment:* | sort}
\end{flushleft}

Such organization allows one to crowdsource autotuning, i.e. distributing autotuning 
of given shared workloads in a cloud or across diverse platforms simply by using remote 
repositories instead of local ones.
On the other hand, it does not address the problem of optimizing larger applications
with multiple hot spots.
It also does not solve the "big data" problem 
when a large amount of data from multiple participants
needed for reproducibility will be continuously aggregated in a CK server.

\begin{figure}[!htbp]
     \centering
      \includegraphics[width=2.5in]
      {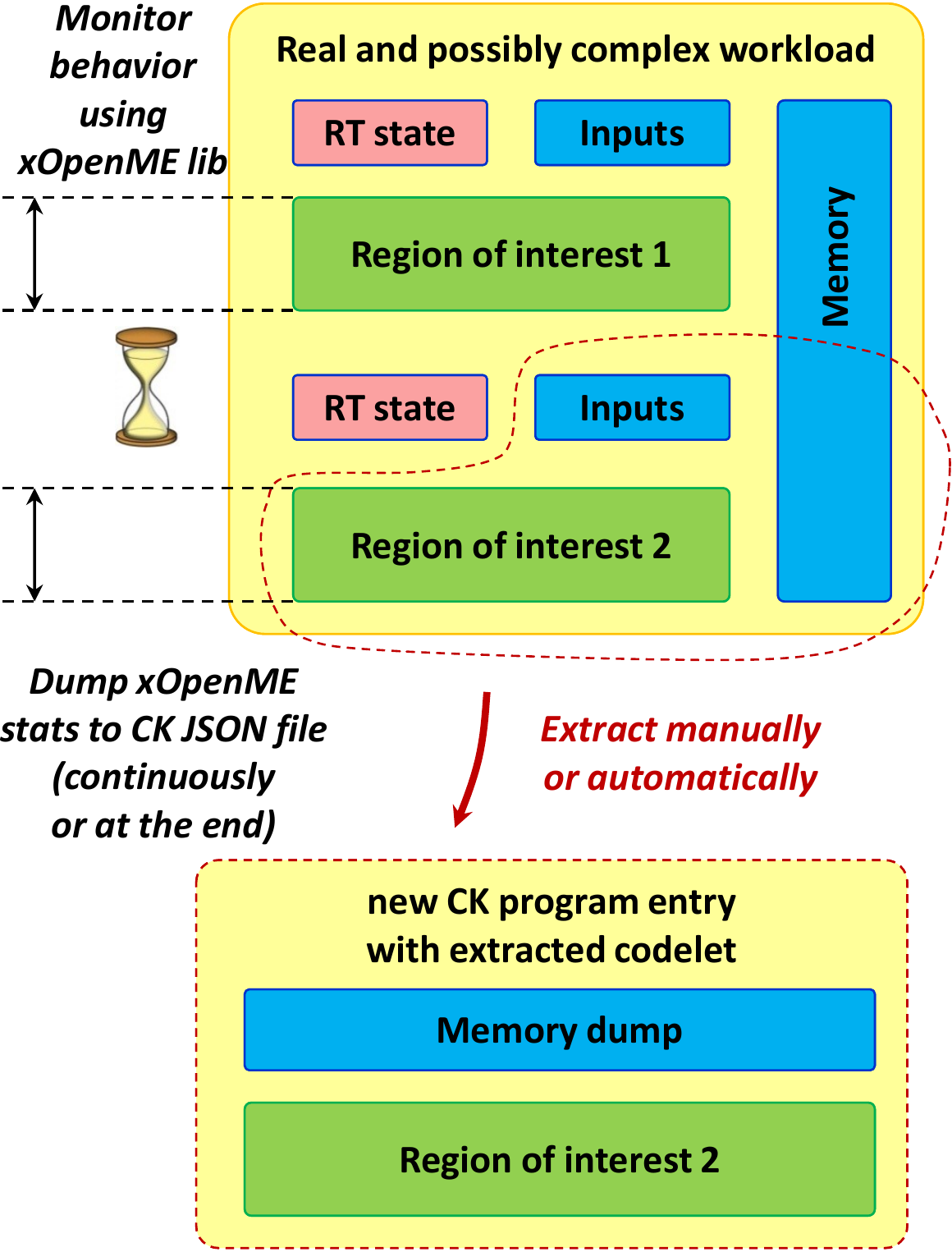} 
     \caption{
       Preparing larger applications such as Firefox and Chrome for CK-based autotuning: 
       a) instrumenting, monitoring and optimizing hot code regions using "XOpenME" library 
       b) extracting code regions from a large application with a run-time data set and register them in the CK as standalone programs (codelets)
     }
     \label{fig:ck-codelets}
   \end{figure}

However, we have been already addressing the first problem by either 
instrumenting, monitoring and optimizing hot code regions in large applications 
using our small "XOpenME" library, or even extracting such code regions 
from a large application with a run-time data set and registering them 
in the CK as standalone programs (codelets or computational species) 
as shown in Figure~\ref{fig:ck-codelets}
(~\cite{fursin:hal-01054763}).

In the MILEPOST project~\cite{29db2248aba45e59:a31e374796869125} 
we used a proprietary "codelet extractor" tool from CAPS Entreprise 
(now dissolved) to automatically extract such hot spots with their data sets 
from several real software projects and 8 popular benchmark suits 
including NAS, MiBench, SPEC2000, SPEC2006, Powerstone, UTDSP and SNU-RT.
We shared those of them with a permissive license as CK programs 
in the \href{https://github.com/ctuning/ctuning-programs}{ctuning-programs} repository
to be compatible with the presented CK autotuning workflow.
We continue adding real, open-source applications and libraries as CK program entries
(GEMM, HOG, SLAM, convolutions) or manually extracting and sharing interesting code 
regions from them with the help of the community.
Such a large collection of diverse and realistic workloads 
should help make computer systems research more applied and practical.

As many other scientists, we also faced a big data problem when continuously 
aggregating large amounts of raw optimization data during crowd-tuning
for further processing including machine learning~\cite{new_pub_model}.
We managed to solve this problem in the CK by using 
online pre-processing of raw data and online classification 
to record only the most efficient optimization solutions 
(on a frontier in case of multi-objective autotuning) 
along with unexpected behavior (bugs and numerical 
instability)~\cite{cm:29db2248aba45e59:cd11e3a188574d80}.
It is now possible to invoke crowd-tuning of GCC compiler flags (improving execution time) in the CK as following:
\begin{flushleft}
\texttt{\$ ck crowdtune program --iterations=50 --scenario=8289e0cf24346aa7}
\end{flushleft}

or

\begin{flushleft}
\texttt{\$ ck crowdsource program.optimization --iterations=50 --scenario=8289e0cf24346aa7}
\end{flushleft}

\begin{figure*}[!htbp]
     \centering
      \includegraphics[width=6.0in]
       {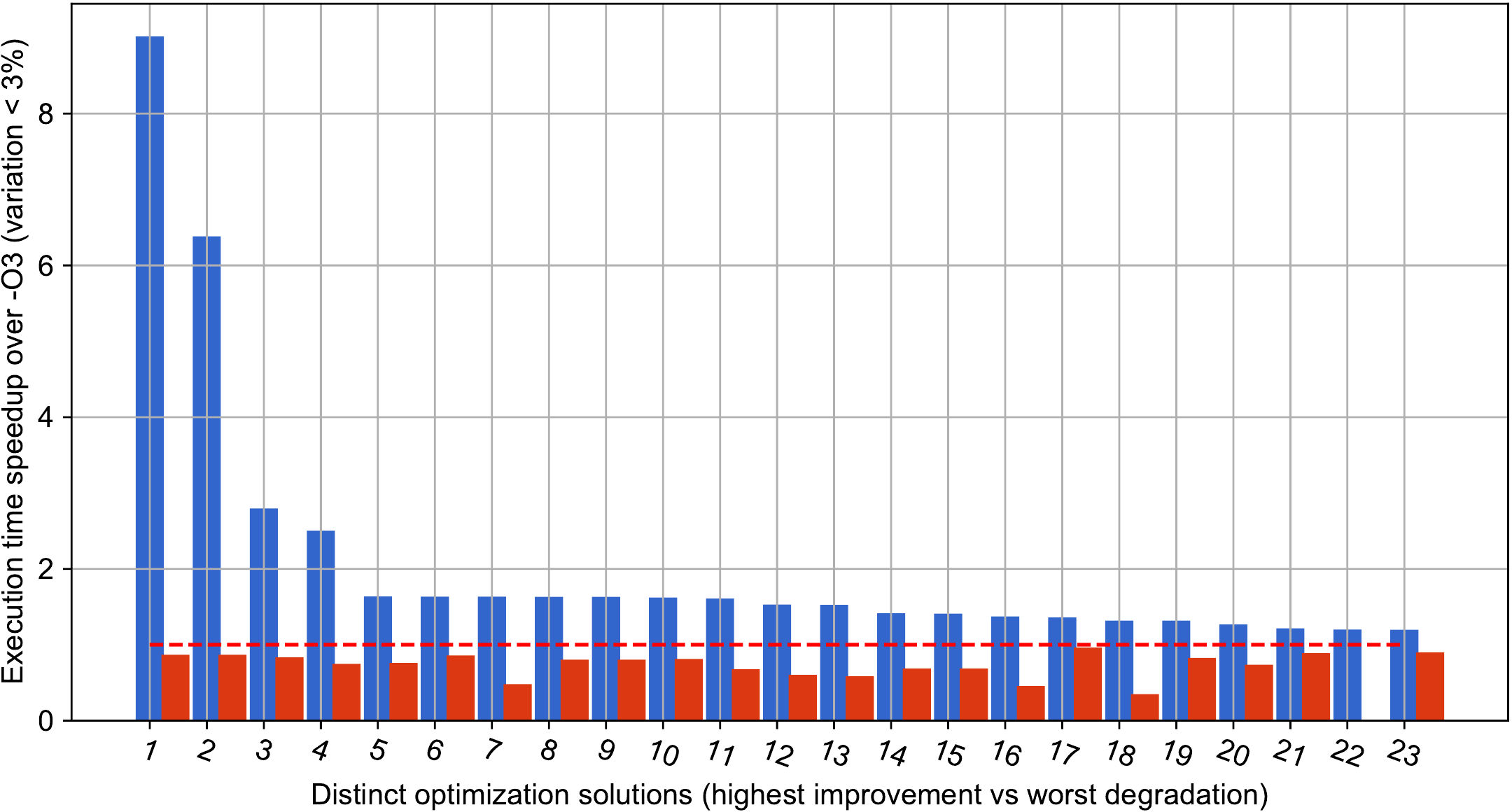} 
      \vspace{0.1in}
          \begin{tabular}{|r|p{4.5in}|p{0.5in}|p{0.5in}|}
     \hline
     \textbf{Solution} & \textbf{Pruned flags (complexity reduction)} & \textbf{Best species} & \textbf{Worst species} \\ 
     \hline
      1 & -O3 -flto & 6 & 3 \\
     \hline
      2 & -O3 -fno-inline -flto & 1 & 1 \\
     \hline
      3 & -O3 -fno-if-conversion2 -funroll-loops & 2 & 1 \\
     \hline
      4 & -O3 -fpeel-loops -ftracer & 1 & 3 \\
     \hline
      5 & -O3 -floop-nest-optimize -fno-sched-interblock -fno-tree-copy-prop -funroll-all-loops & 4 & 1 \\
     \hline
      6 & -O3 -funroll-loops & 2 & 3 \\
     \hline
      7 & -O3 -floop-strip-mine -funroll-loops & 1 & 1 \\
     \hline
      8 & -O3 -fno-inline -fno-merge-all-constants -fno-tree-ccp -funroll-all-loops & 2 & 3 \\
     \hline
      9 & -O3 -fno-tree-loop-if-convert -funroll-all-loops & 3 & 2 \\
     \hline
      10 & -O3 -fno-section-anchors -fselective-scheduling2 -fno-tree-forwprop -funroll-all-loops & 2 & 2 \\
     \hline
      11 & -O3 -fno-ivopts -funroll-loops & 4 & 1 \\
     \hline
      12 & -O3 -fno-tree-ch -funroll-all-loops & 1 & 1 \\
     \hline
      13 & -O3 -fno-move-loop-invariants -fno-tree-ch -funroll-loops & 1 & 2 \\
     \hline
      14 & -O3 -fira-algorithm=priority -fno-ivopts & 1 & 2 \\
     \hline
      15 & -O3 -fno-ivopts & 2 & 4 \\
     \hline
      16 & -O3 -fno-sched-spec -fno-tree-ch & 1 & 2 \\
     \hline
      17 & -O3 -fno-ivopts -fselective-scheduling -fwhole-program & 1 & 1 \\
     \hline
      18 & -O3 -fno-omit-frame-pointer -fno-tree-loop-optimize & 1 & 4 \\
     \hline
      19 & -O3 -fno-auto-inc-dec -ffinite-math-only & 1 & 2 \\
     \hline
      20 & -O3 -fno-guess-branch-probability -fira-loop-pressure -fno-toplevel-reorder & 1 & 5 \\
     \hline
      21 & -O3 -fselective-scheduling2 -fno-tree-pre & 2 & 2 \\
     \hline
      22 & -O3 -fgcse-sm -fno-move-loop-invariants -fno-tree-forwprop -funroll-all-loops -fno-web & 1 & 0 \\
     \hline
      23 & -O3 -fno-schedule-insns -fselective-scheduling2 & 1 & 2 \\
     \hline
    \end{tabular} 
      \vspace{0.1in}
     \caption{
      Snapshot of top performing combinations of GCC 4.9.2 compiler flags together with highest speedups and worst degradations achieved across all shared CK workloads on RPi3.
     }
     \label{fig:ck-snapshot-of-results-gcc4}
   \end{figure*}

In contrast with traditional autotuning, CK will first query \textit{remote-ck} repository
to obtain all most efficient optimization choices aka solutions (combinations of random compiler flags in our example)
for a given trade-off scenario (GCC compiler flag tuning to minimize execution time), compiler version,
platform and OS.
CK will then select a random CK program (computational species),
compiler and run it with all these top optimizations,
and then try N extra random optimizations (random combinations of GCC flags) 
to continue increasing design and optimization space coverage.
CK will then send the highest improvements of monitored characteristics 
(execution time in our example) achieved for each optimization solution as well as worst degradations
back to a public server.
If a new optimization solution if also found during random autotuning,
CK will assign it a unique ID (\textit{solution\_uid} 
and will record it in a public repository.
At the public server side, CK will merge improvements and degradations for a given
program from a participant with a global statistics while recording how many programs 
achieved the highest improvement (best species) or worst degradation (worst species) for a given optimization
as shown in Figure~\ref{fig:ck-snapshot-of-results-gcc4}.

This figure shows a snapshot of public optimization results 
with top performing combinations of GCC 4.9.2 compiler flags
on RPi3 devices which minimize execution time of shared CK workloads 
(programs and data sets) in comparison with \textit{-O3} optimization level.
It also shows the highest speedup and the worse degradation achieved
across all CK workloads for a given optimization solution, as well
as a number of workloads where this solution was the best or the worst
(online classification of all optimization solutions).
Naturally this snapshot automatically generated from the public repository 
at the time of publication may slightly differ from continuously updated 
live optimization results available at this~\href{http://cknowledge.org/repo/web.php?template=cknowledge&wcid=8289e0cf24346aa7:d24a4fde9f120e10}{link}.
These results confirm that GCC 4.9.2 misses many optimization opportunities 
not covered by \textit{-O3} optimization level.

   \begin{figure*}[!htbp]
     \centering
      \includegraphics[width=6in]
       {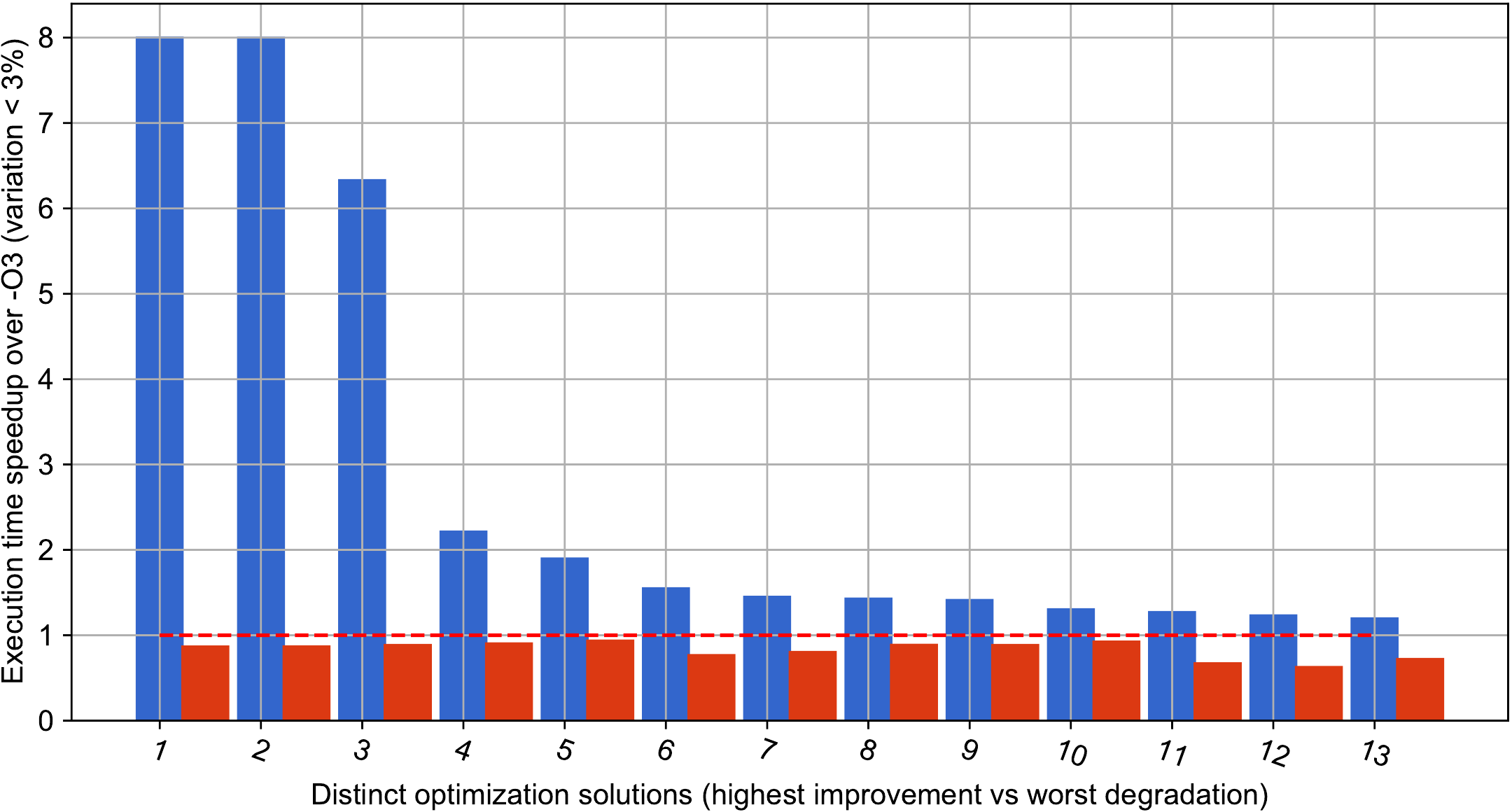} 
      \vspace{0.1in}
          \begin{tabular}{|r|p{4.5in}|p{0.5in}|p{0.5in}|}
     \hline
     \textbf{Solution} & \textbf{Pruned flags (complexity reduction)} & \textbf{Best species} & \textbf{Worst species} \\ 
     \hline
      1 & -O3 -fno-delayed-branch -flto -fno-selective-scheduling2 -fno-whole-program & 6 & 0 \\
     \hline
      2 & -O3 -flto & 4 & 1 \\
     \hline
      3 & -O3 -fno-inline -flto & 2 & 1 \\
     \hline
      4 & -O3 -fno-cprop-registers -flto -funroll-all-loops & 3 & 1 \\
     \hline
      5 & -O3 -fno-tree-fre -funroll-all-loops & 2 & 1 \\
     \hline
      6 & -O3 -fno-predictive-commoning -fno-schedule-insns -funroll-loops & 3 & 3 \\
     \hline
      7 & -O3 -funroll-loops & 3 & 0 \\
     \hline
      8 & -O3 -fno-tree-ter -funroll-all-loops & 3 & 1 \\
     \hline
      9 & -O3 -fno-merge-all-constants -fselective-scheduling2 -funroll-loops & 1 & 0 \\
     \hline
      10 & -O3 -fno-devirtualize-at-ltrans -fno-predictive-commoning -fno-tree-pre & 1 & 2 \\
     \hline
      11 & -O3 -fcheck-data-deps -fira-loop-pressure -fno-isolate-erroneous-paths-dereference -fno-sched-dep-count-heuristic -fsection-anchors -fsemantic-interposition -fno-tree-ch -fno-tree-loop-linear -fno-tree-partial-pre & 2 & 2 \\
     \hline
      12 & -O3 -fno-schedule-insns -ftracer & 2 & 3 \\
     \hline
      13 & -O3 -fno-auto-inc-dec -fguess-branch-probability -fipa-pure-const -freorder-blocks -fselective-scheduling2 -ftree-ccp -fno-tree-pre -ftree-tail-merge & 1 & 1 \\
     \hline
    \end{tabular} 
      \vspace{0.1in}
     \caption{
      Snapshot of top performing combinations of GCC 7.1.0 compiler flags together with highest speedups and worst degradations achieved across all shared CK workloads on RPi3.
     }
     \label{fig:ck-snapshot-of-results-gcc7}
   \end{figure*}

Figure~\ref{fig:ck-snapshot-of-results-gcc7} with optimization results 
for GCC 7.1.0 also confirms that this version was considerably improved 
in comparison with GCC 4.9.2
(latest live results are available in our public optimization repository
at this \href{http://cknowledge.org/repo/web.php?wcid=8289e0cf24346aa7:79bca2b76876b5c6}{link}):
there are fewer efficient optimization solutions found during crowd-tuning
14 vs 23 showing the overall improvement of the \textit{-O3} optimization level.

Nevertheless, GCC 7.1.0 still misses many optimization opportunities 
simply because our long-term experience suggests that it is infeasible 
to prepare one universal and efficient optimization heuristics 
with good multi-objective trade-offs for all continuously 
evolving programs, data sets, libraries, optimizations and platforms.
That is why we hope that our approach of combining a common workflow framework
adaptable to software and hardware changes, public repository of optimization knowledge, 
universal and collaborative autotuning across multiple hardware platforms 
(e.g.~provided by volunteers or by HPC providers), and community involvement 
should help make optimization and testing of compilers
more automatic and sustainable~\cite{cm:29db2248aba45e59:cd11e3a188574d80,ck-date16}.
Rather than spending considerable amount of time on writing their own autotuning and crowd-tuning
frameworks, students and researchers can quickly reuse shared workflows, 
reproduce and learn already existing optimizations, try to improve optimization heuristics, 
and validate their results by the community. 

Furthermore, besides using \textit{-Ox} compiler levels, academic and industrial users 
can immediately take advantage of various shared optimizations solutions automatically
found by volunteers for a given compiler and hardware via CK using \textit{solution\_uid} flag.
For example, users can test the most efficient combination of compiler flags 
which achieved the highest speedup for GCC 7.1.0 on RPi3 
(see "Copy CID to clipboard  for a given optimization solution at this 
\href{http://cknowledge.org/repo/web.php?wcid=8289e0cf24346aa7:79bca2b76876b5c6}{link})
for their own programs using CK:

\begin{lstlisting}[breaklines]
$ ck benchmark program:{new program}
  --shared_solution_cid=27bc42ee449e880e:
  79bca2b76876b5c6-8289e0cf24346aa7-
  f49649288ab0accd
\end{lstlisting}

or

\begin{lstlisting}[breaklines]
$ ck benchmark program:{new program} 
  -O27bc42ee449e880e:79bca2b76876b5c6-
  8289e0cf24346aa7-f49649288ab0accd
\end{lstlisting}


\section{Autotuning and crowd-tuning real workloads}
\label{sec:collaborative}
In this section we would like to show how we can apply universal autotuning 
and collaboratively found optimization solutions to several popular workloads
used by RPi community: \textit{zlib decode, zlib encode, 
7z encode, aubio, ccrypt, gzip decode, gzip encode, minigzip decode, 
minigzip encode, rhash, sha512sum, unrar}.
We added the latest versions of these real programs 
to the CK describing how to compile and run them 
using CK JSON meta data:

\begin{flushleft}
\texttt{\$ ck ls ck-rpi-optimization:program:*}
\end{flushleft}

We can now autotune any of these programs via CK as described in Section~\ref{sec:flag_autotuning}.
For example, the following command will autotune \textit{zlib decode} workload
with 150 random combinations of compiler flags including parametric and architecture 
specific ones, and will record results in a local repository:

\begin{flushleft}
\texttt{\$ ck autotune program:zlib --cmd\_key=decode
  --iterations=150 --repetitions=3 
  --scenario=experiment.tune.compiler.flags.gcc
  --parametric\_flags --cpu\_flags --base\_flags 
  --record\_uoa=tmp-rpi3-zlib-decode-gcc4-150bpc-rnd}
\end{flushleft}

\begin{figure*}[!htbp]
     \centering
      \includegraphics[width=5.2in]
      {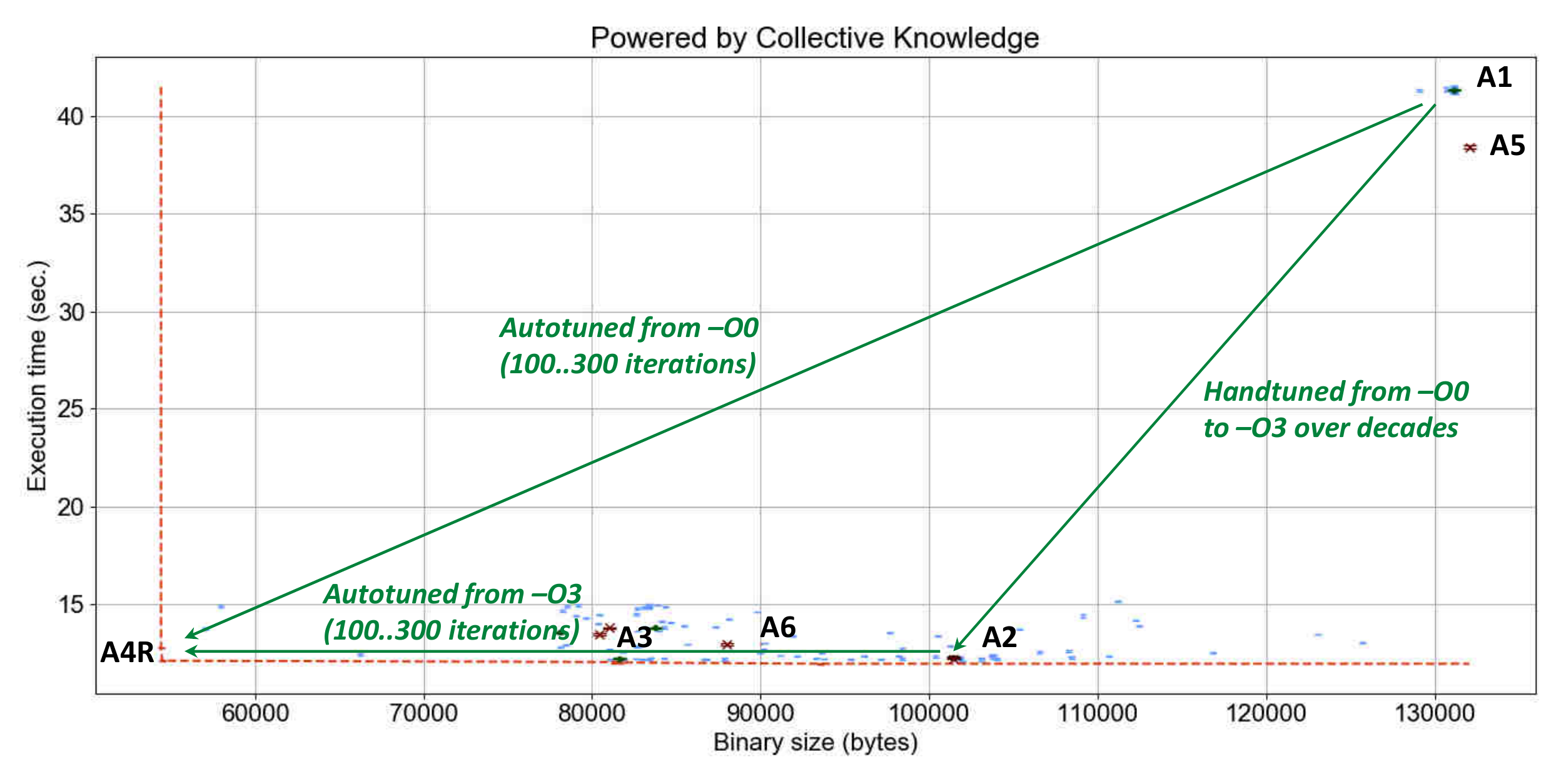} 
      \vspace{0.1in}
          \begin{tabular}{|l|l|l|l|p{3.2in}|}
     \hline
      \textbf{ID} & \textbf{Compiler} & \textbf{Time (sec.)} & \textbf{Size (bytes)} & \textbf{Flags} \\ 
     \hline
      \textbf{ \href{http://cknowledge.org/repo/web.php?wcid=experiment:07969231dd8f9474\&subpoint=cfcb9327f42ddabd}{A1} } &  GCC 4.9.2  &  41.3 $\pm$ 0.0  &  131140  & {\small  }\\
     \hline
      \textbf{ \href{http://cknowledge.org/repo/web.php?wcid=experiment:9b2b24a80c45aa9b\&subpoint=eb28149e9a71762d}{A2} } &  GCC 4.9.2  &  12.2 $\pm$ 0.0  &  101448  & {\small -O3 }\\
     \hline
      \textbf{ \href{http://cknowledge.org/repo/web.php?wcid=experiment:9195c9fa4d5d89af\&subpoint=2432890269556b39}{A3} } &  GCC 4.9.2  &  13.6 $\pm$ 0.0  &  78116  & {\small -Os }\\
     \hline
      \textbf{ \href{http://cknowledge.org/repo/web.php?wcid=experiment:f5489592a3a15bf3\&subpoint=6236b2e4742629aa}{A4R} } &  GCC 4.9.2  &  12.1 $\pm$ 0.1  &  54272  & {\small -O2 -flto -fno-tree-fre }\\
     \hline
      \textbf{ \href{http://cknowledge.org/repo/web.php?wcid=experiment:eca2f4aa2a3ab852\&subpoint=a624bb4c10a1619f}{A5} } &  CLANG 3.8.1  &  38.5 $\pm$ 0.0  &  132080  & {\small  }\\
     \hline
      \textbf{ \href{http://cknowledge.org/repo/web.php?wcid=experiment:777b58443e536152\&subpoint=e13b1e85e3e0c5d6}{A6} } &  CLANG 3.8.1  &  12.9 $\pm$ 0.1  &  90076  & {\small -O3 }\\
     \hline
    \end{tabular}     
      \vspace{0.1in}
     \caption{
      Results of GCC 4.9.2 random compiler flag autotuning of a zlib decode workload on RPi3
      device using CK with a highlighted frontier (trading-off execution time and code size) 
      and the best found combinations of flags on this frontier.
     }
     \label{fig:autotuning-zlib-decode-gcc4}
   \end{figure*}

Figure~\ref{fig:autotuning-zlib-decode-gcc4} 
(\href{http://cknowledge.org/repo/web.php?wcid=graph:3a97d1f6494f9d45&subgraph=rpi3-autotuning-zlib-decode-gcc4-interactive}{link with interactive graph}) 
shows a manually annotated graph with the outcome of such autotuning 
when using GCC 4.9.2 compiler on RPi3 device
in terms of execution time with variation and code size.
Each blue point on this graph is related to one combination of random compiler flags.
The red line highlights the frontier of all autotuning results 
to let users trade off execution time and code size 
during multi-objective optimization.
Similar to graphs in Section~\ref{sec:flag_autotuning}, we also plotted points 
when using several main GCC and Clang optimization levels.

In contrast with \textit{susan corners} workload, autotuning did not improve execution time 
of \textit{zlib decode} over \textit{-O3} level most likely because this algorithm is present
in many benchmarking suits. 
On the other hand, autotuning impressively improved code size over \textit{-O3} 
by nearly 2x without sacrificing execution time, and by ~1.5x with 11\% execution time
improvement over \textit{-Os} (reduced optimization solution \textbf{A4R}), 
showing that code size optimization is still a second class citizen.

Since local autotuning can still be quite costly (150 iterations to achieve above results),
we can now first check 10..20 most efficient combinations of compiler flags 
already found and shared by the community for this compiler and hardware
(Figure~\ref{fig:ck-snapshot-of-results-gcc4}).
Note that programs from this section did not participate in crowd-tuning
to let us have a fair evaluation of the influence of shared optimizations 
on these programs similar to leave-one-out cross-validation in machine learning.

\begin{figure*}[!htbp]
     \centering
      \includegraphics[width=5.2in]
      {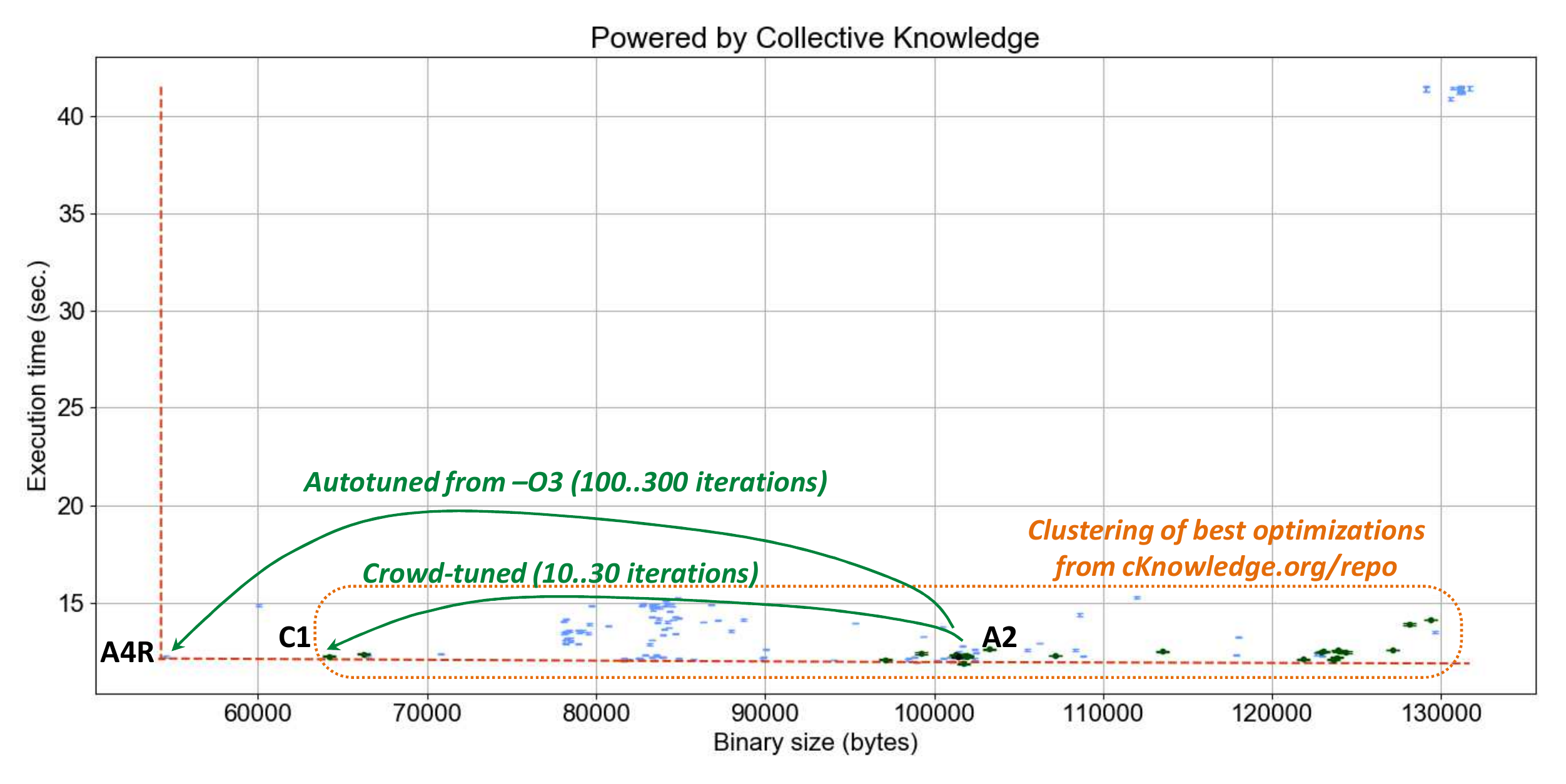} 
      \vspace{0.1in}
          \begin{tabular}{|l|l|l|l|p{3.2in}|}
     \hline
      \textbf{ID} & \textbf{Compiler} & \textbf{Time (sec.)} & \textbf{Size (bytes)} & \textbf{Flags} \\ 
     \hline
      \textbf{ \href{http://cknowledge.org/repo/web.php?wcid=experiment:9b2b24a80c45aa9b\&subpoint=eb28149e9a71762d}{A2} } &  GCC 4.9.2  &  12.2 $\pm$ 0.0  &  101448  & {\small -O3 }\\
     \hline
      \textbf{ \href{http://cknowledge.org/repo/web.php?wcid=experiment:f5489592a3a15bf3\&subpoint=6236b2e4742629aa}{A4R} } &  GCC 4.9.2  &  12.1 $\pm$ 0.1  &  54272  & {\small -O2 -flto -fno-tree-fre }\\
     \hline
      \textbf{ \href{http://cknowledge.org/repo/web.php?wcid=experiment:dfc49b5be33c1813\&subpoint=ec6a2e99da2e3445}{C1} } &  GCC 4.9.2  &  12.2 $\pm$ 0.1  &  64184  & {\small -O3 -fno-inline -flto }\\
     \hline
    \end{tabular}     
      \vspace{0.1in}
     \caption{
      Speeding up GCC 4.9.2 autotuning of a zlib decode workload on RPi3 device using 
      10..20 best performing combinations of compiler flags already found and shared by the community
      during crowd-tuning.
     }
     \label{fig:autotuning-zlib-decode-gcc4-reactions}
   \end{figure*}

Figure~\ref{fig:autotuning-zlib-decode-gcc4-reactions} shows "reactions" 
of \textit{zlib decode} to these optimizations in terms of execution time and code size 
(\href{http://cknowledge.org/repo/web.php?wcid=graph:47f0b282396776c4&subgraph=rpi3-autotuning-zlib-decode-gcc4-reactions-interactive}{the online interactive graph}).
We can see that crowd-tuning solutions indeed cluster in a relatively small area 
close to \textit{-O3} with one collaborative solution (\textbf{C1}) close to the 
best optimization solution found during lengthy autotuning (\textbf{A4R}) 
thus providing a good trade off between autotuning time, execution time and code size.

\begin{figure*}[!htbp]
     \centering
      \includegraphics[width=5.2in]
      {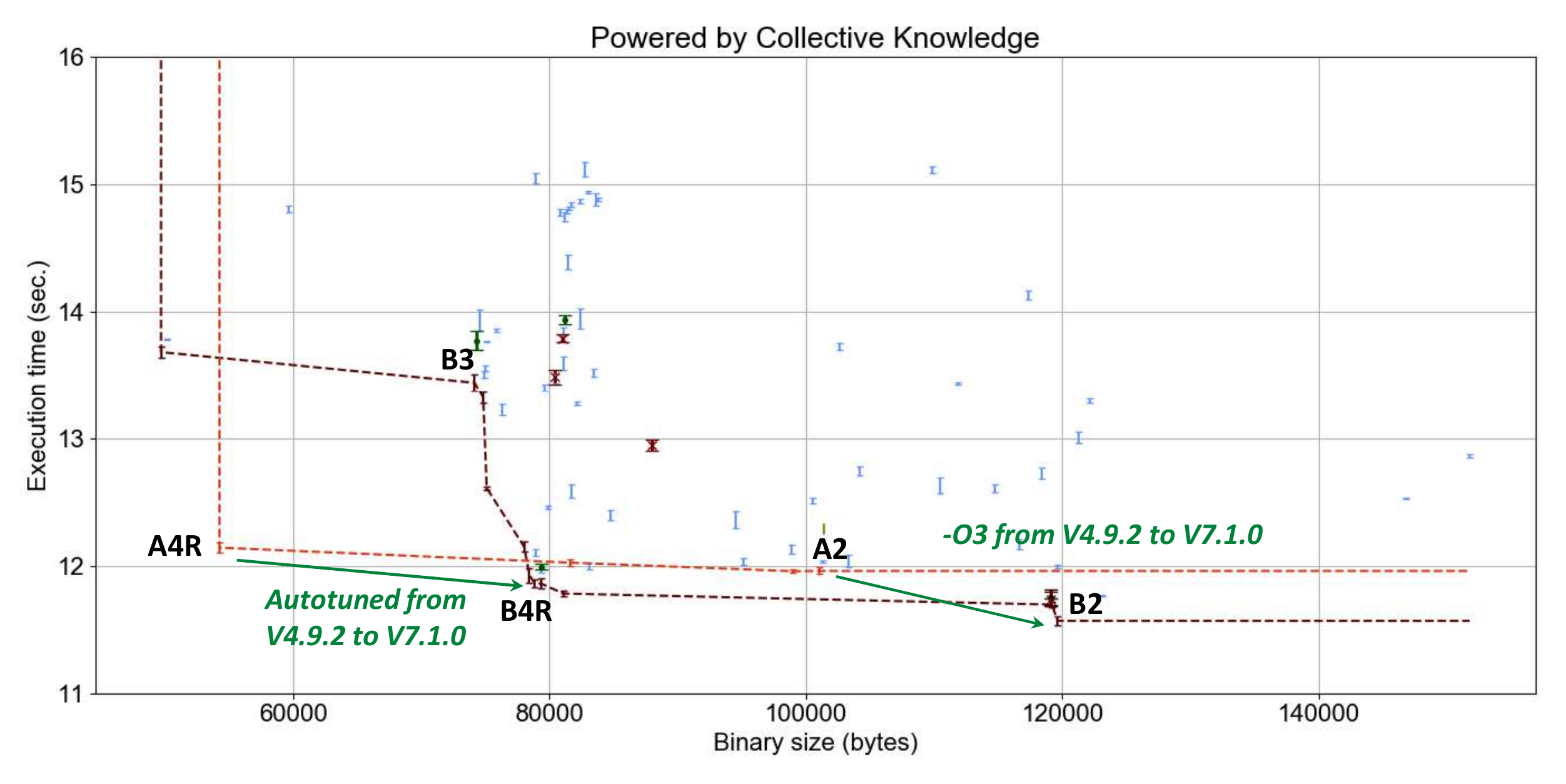} 
      \vspace{0.1in}
          \begin{tabular}{|l|l|l|l|p{3.2in}|}
     \hline
      \textbf{ID} & \textbf{Compiler} & \textbf{Time (sec.)} & \textbf{Size (bytes)} & \textbf{Flags} \\ 
     \hline
      \textbf{ \href{http://cknowledge.org/repo/web.php?wcid=experiment:9b2b24a80c45aa9b\&subpoint=eb28149e9a71762d}{A2} } &  GCC 4.9.2  &  12.2 $\pm$ 0.0  &  101448  & {\small -O3 }\\
     \hline
      \textbf{ \href{http://cknowledge.org/repo/web.php?wcid=experiment:f5489592a3a15bf3\&subpoint=6236b2e4742629aa}{A4R} } &  GCC 4.9.2  &  12.1 $\pm$ 0.1  &  54272  & {\small -O2 -flto -fno-tree-fre }\\
     \hline
      \textbf{ \href{http://cknowledge.org/repo/web.php?wcid=experiment:1d1b423cb2567413\&subpoint=f42caae168d92907}{B1} } &  GCC 7.1.0  &  41.3 $\pm$ 0.0  &  128376  & {\small  }\\
     \hline
      \textbf{ \href{http://cknowledge.org/repo/web.php?wcid=experiment:45b844dc97bc88bb\&subpoint=c88f2a728405d8eb}{B2} } &  GCC 7.1.0  &  11.7 $\pm$ 0.1  &  119084  & {\small -O3 }\\
     \hline
      \textbf{ \href{http://cknowledge.org/repo/web.php?wcid=experiment:a1e26d8e4858adf5\&subpoint=bbf4d22b4e9b34d5}{B3} } &  GCC 7.1.0  &  13.7 $\pm$ 0.1  &  74280  & {\small -Os }\\
     \hline
      \textbf{ \href{http://cknowledge.org/repo/web.php?wcid=experiment:b642fbace509ae5a\&subpoint=cd944d398d208b53}{B4R} } &  GCC 7.1.0  &  11.9 $\pm$ 0.1  &  78700  & {\small -O2 -fno-early-inlining -fno-tree-fre }\\
     \hline
    \end{tabular}     
      \vspace{0.1in}
     \caption{
       Results of GCC 7.1.0 random compiler flag autotuning of zlib decode on RPi3 device 
       with a highlighted frontier (trading-off execution time and code size), 
       the best combinations of flags on this frontier, and comparison with the results from GCC 4.9.2.
     }
     \label{fig:autotuning-zlib-decode-gcc7}
   \end{figure*}

Autotuning \textit{zlib decode} using \textit{GCC 7.1.0} revels even more interesting results
in comparison with \textit{susan corners} as shown in Figure~\ref{fig:autotuning-zlib-decode-gcc7} 
(\href{http://cknowledge.org/repo/web.php?wcid=graph:2bf38fd88a0e3ba1&subgraph=rpi3-autotuning-zlib-decode-gcc7-interactive}{the online interactive graph}).
While there is practically no execution time improvements when switching from \textit{GCC 4.9.2} to \textit{GCC 7.1.0}
on \textit{-O3} and \textit{-Os} optimization levels, \textit{GCC 7.1.0 -O3} considerably degraded code size by nearly 20\%.
Autotuning also shows few opportunities on \textit{GCC 7.1.0} in comparison with \textit{GCC 4.9.2}
where the best found optimization \textbf{B4R} is worse in terms of a code size than \textbf{A4R} also by around 20\%.
These results highlight issues which both end-users and compiler designers face
when searching for efficient combinations of compiler flags or preparing the 
default optimization levels -Ox.

\begin{figure*}[!htbp]
     \centering
      \includegraphics[width=5.2in]
      {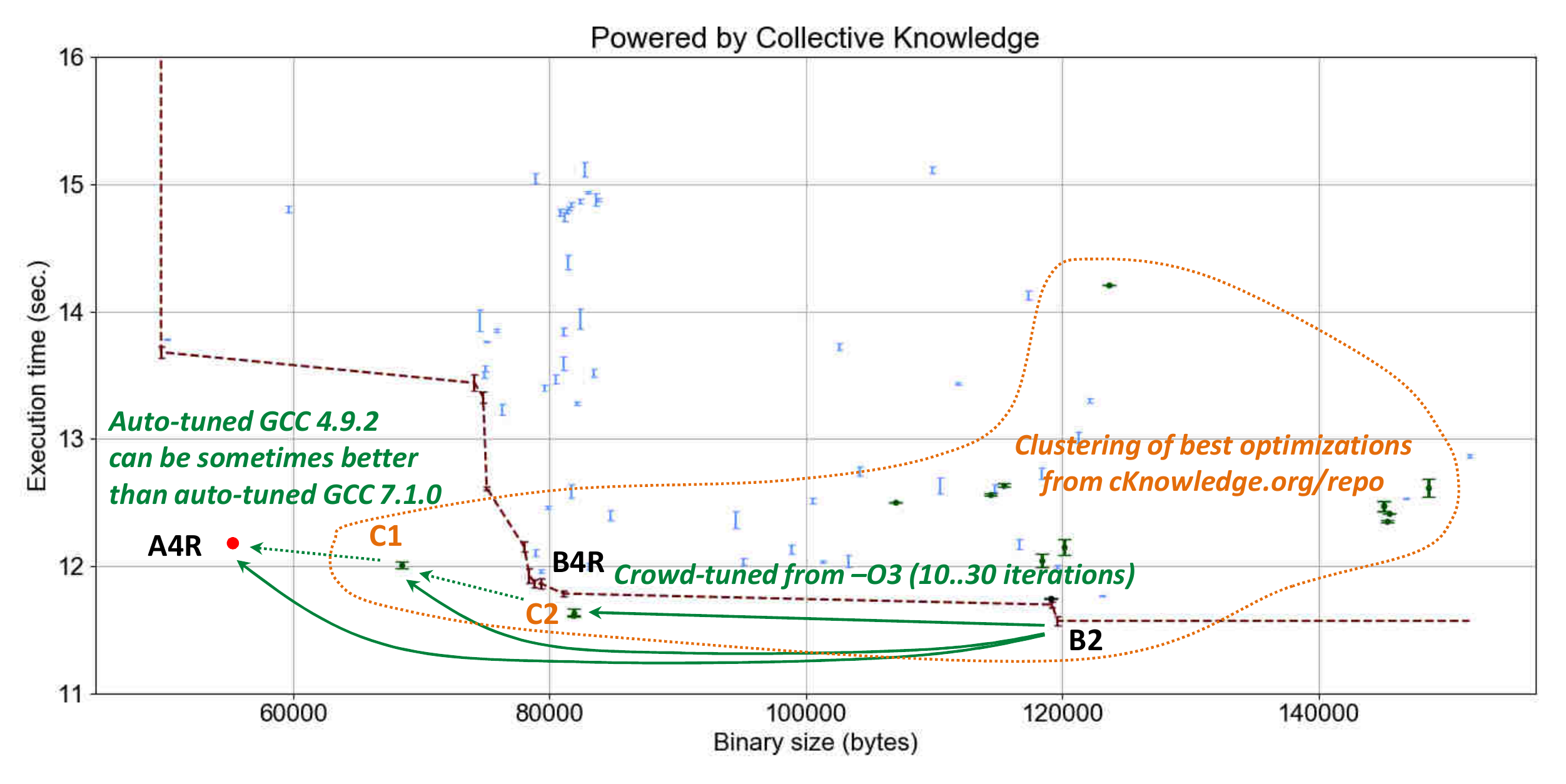} 
      \vspace{0.1in}
          \begin{tabular}{|l|l|l|l|p{3.2in}|}
     \hline
      \textbf{ID} & \textbf{Compiler} & \textbf{Time (sec.)} & \textbf{Size (bytes)} & \textbf{Flags} \\ 
     \hline
      \textbf{ \href{http://cknowledge.org/repo/web.php?wcid=experiment:f5489592a3a15bf3\&subpoint=6236b2e4742629aa}{A4R} } &  GCC 4.9.2  &  12.1 $\pm$ 0.1  &  54272  & {\small -O2 -flto -fno-tree-fre }\\
     \hline
      \textbf{ \href{http://cknowledge.org/repo/web.php?wcid=experiment:45b844dc97bc88bb\&subpoint=c88f2a728405d8eb}{B2} } &  GCC 7.1.0  &  11.7 $\pm$ 0.1  &  119084  & {\small -O3 }\\
     \hline
      \textbf{ \href{http://cknowledge.org/repo/web.php?wcid=experiment:b642fbace509ae5a\&subpoint=cd944d398d208b53}{B4R} } &  GCC 7.1.0  &  11.9 $\pm$ 0.1  &  78700  & {\small -O2 -fno-early-inlining -fno-tree-fre }\\
     \hline
      \textbf{ \href{http://cknowledge.org/repo/web.php?wcid=experiment:f89c04c152682687\&subpoint=9a4548dd347699b7}{C1} } &  GCC 7.1.0  &  12.0 $\pm$ 0.0  &  68464  & {\small -O3 -fno-inline -flto }\\
     \hline
      \textbf{ \href{http://cknowledge.org/repo/web.php?wcid=experiment:f89c04c152682687\&subpoint=b5a6ade146d9b028}{C2} } &  GCC 7.1.0  &  11.6 $\pm$ 0.1  &  81880  & {\small -O3 -flto }\\
     \hline
    \end{tabular}     
      \vspace{0.1in}
     \caption{
      Testing reactions of zlib decode to top most efficient GCC 7.1.0 optimizations shared by the community for RPi3 devices vs GCC 4.9.2.
     }
     \label{fig:autotuning-zlib-decode-gcc7-reactions}
   \end{figure*}

CK crowd-tuning can assist in this case too - Figure~\ref{fig:autotuning-zlib-decode-gcc7-reactions}
shows reactions of \textit{zlib decode} to the most efficient combinations of GCC 7.1.0 compiler flags
shared by the community for RPi3 
(\href{http://cknowledge.org/repo/web.php?wcid=graph:a53089441c68c978&subgraph=rpi3-autotuning-zlib-decode-gcc7-reactions-interactive}{the online interactive graph}).
Shared optimization solution~\textbf{C2} achieved the same results in terms of execution time and code size
as reduced solution \textbf{B4R} found during 150 random autotuning iterations.
Furthermore, another shared optimization solution~\textbf{C1} improved code size by ~15\% in comparison
with GCC 7.1.0 autotuning solution~\textbf{B4R} and is close to the best solution GCC 4.9.2 autotuning solution~\textbf{A4R}.
These results suggest that 150 iterations with random combinations of compiler flags 
may not be enough to find an efficient solution for \textit{zlib decode}.
In turn, crowd-tuning can help considerably accelerate and focus such optimization space exploration.

\begin{figure*}[!htbp]
     \centering
      \includegraphics[width=5.2in]
      {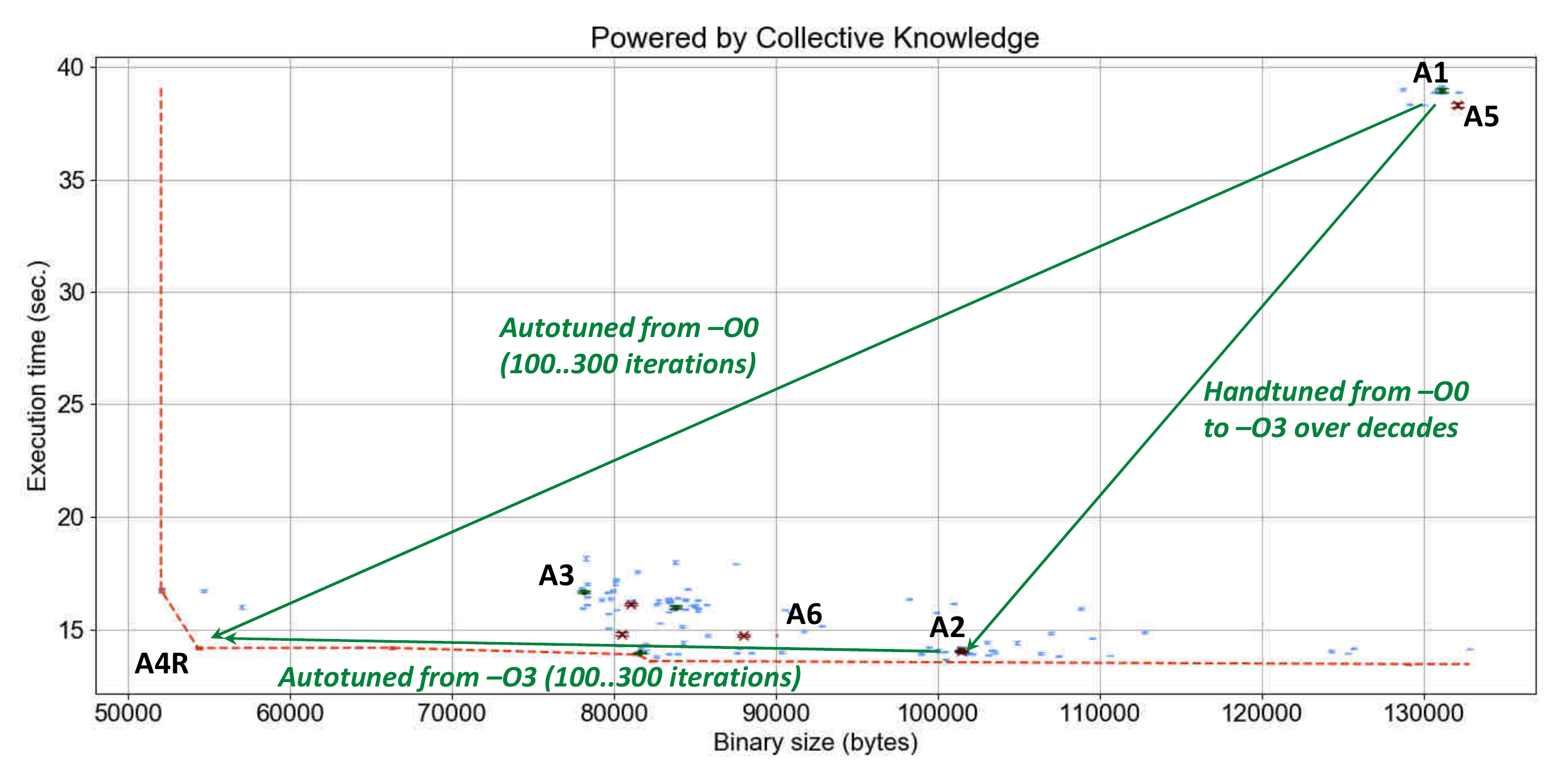} 
      \vspace{0.1in}
          \begin{tabular}{|l|l|l|l|p{3.2in}|}
     \hline
      \textbf{ID} & \textbf{Compiler} & \textbf{Time (sec.)} & \textbf{Size (bytes)} & \textbf{Flags} \\ 
     \hline
      \textbf{ \href{http://cknowledge.org/repo/web.php?wcid=experiment:28ee633f4192a488\&subpoint=463438f9b5e50e55}{A1} } &  GCC 4.9.2  &  39.0 $\pm$ 0.1  &  131140  & {\small  }\\
     \hline
      \textbf{ \href{http://cknowledge.org/repo/web.php?wcid=experiment:21f631290c7846ee\&subpoint=b3d50b1184e6ebed}{A2} } &  GCC 4.9.2  &  14.0 $\pm$ 0.1  &  101448  & {\small -O3 }\\
     \hline
      \textbf{ \href{http://cknowledge.org/repo/web.php?wcid=experiment:2bdb07edb41aabb0\&subpoint=877f5d27c47b8845}{A3} } &  GCC 4.9.2  &  16.7 $\pm$ 0.1  &  78116  & {\small -Os }\\
     \hline
      \textbf{ \href{http://cknowledge.org/repo/web.php?wcid=experiment:85a9d07941d187e4\&subpoint=c59cc63440c795a7}{A4R} } &  GCC 4.9.2  &  14.2 $\pm$ 0.1  &  54284  & {\small -O2 -flto }\\
     \hline
      \textbf{ \href{http://cknowledge.org/repo/web.php?wcid=experiment:731a468f9643496a\&subpoint=9ca6c418285e121a}{A5} } &  CLANG 3.8.1  &  38.2 $\pm$ 0.1  &  132080  & {\small  }\\
     \hline
      \textbf{ \href{http://cknowledge.org/repo/web.php?wcid=experiment:90dc03e42974d27a\&subpoint=47ebb3b545c2e220}{A6} } &  CLANG 3.8.1  &  14.7 $\pm$ 0.1  &  90076  & {\small -O3 }\\
     \hline
    \end{tabular}     
      \vspace{0.1in}
     \caption{
      Results of GCC 4.9.2 random compiler flag autotuning of a zlib encode workload on RPi3
      device using CK with a highlighted frontier (trading-off execution time and code size) 
      and the best found combinations of flags on this frontier.
     }
     \label{fig:autotuning-zlib-encode-gcc4}
   \end{figure*}

\begin{figure*}[!htbp]
     \centering
      \includegraphics[width=5.8in]
      {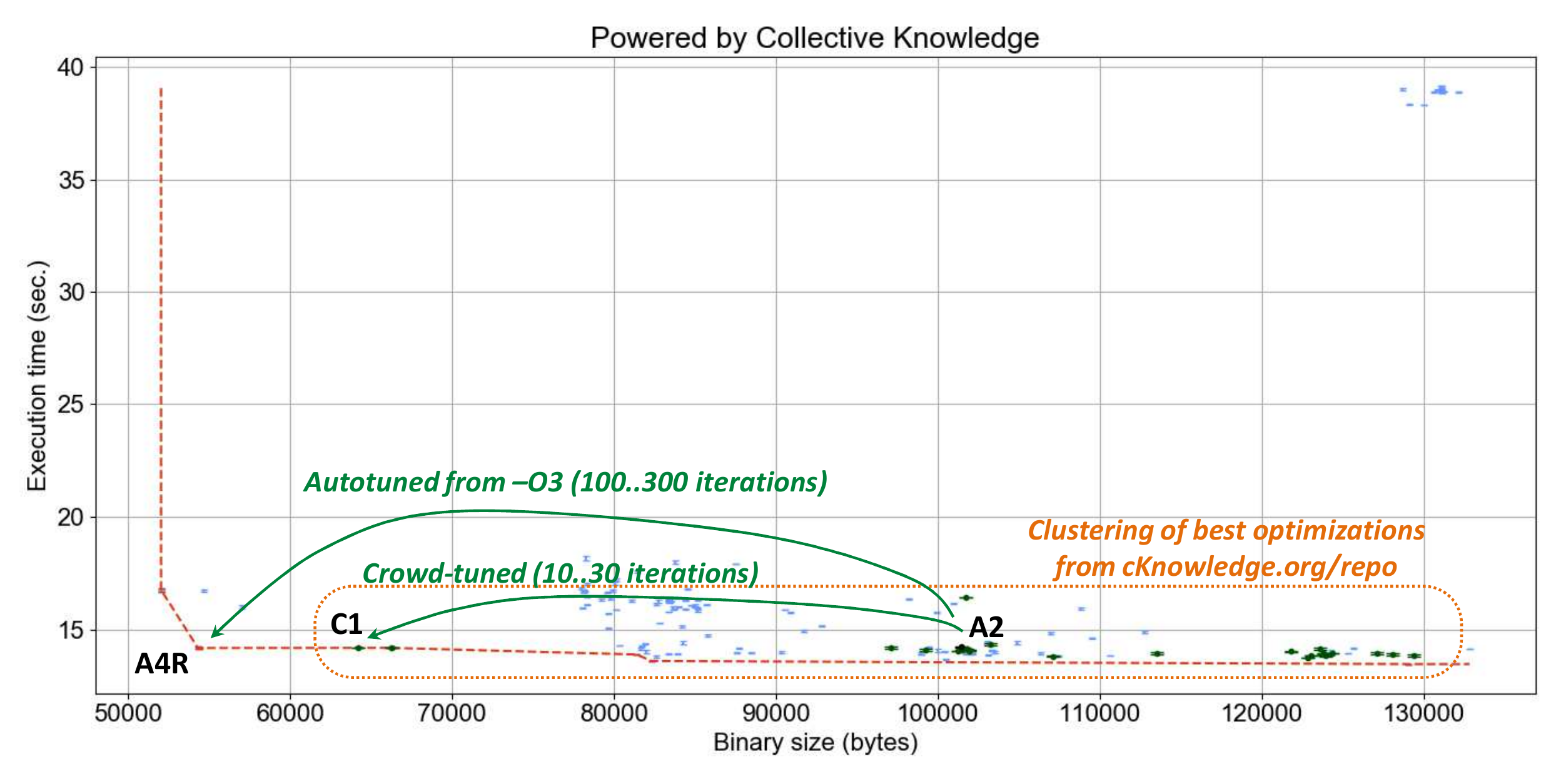} 
      \vspace{0.1in}
          \begin{tabular}{|l|l|l|l|p{3.2in}|}
     \hline
      \textbf{ID} & \textbf{Compiler} & \textbf{Time (sec.)} & \textbf{Size (bytes)} & \textbf{Flags} \\ 
     \hline
      \textbf{ \href{http://cknowledge.org/repo/web.php?wcid=experiment:21f631290c7846ee\&subpoint=b3d50b1184e6ebed}{A2} } &  GCC 4.9.2  &  14.0 $\pm$ 0.1  &  101448  & {\small -O3 }\\
     \hline
      \textbf{ \href{http://cknowledge.org/repo/web.php?wcid=experiment:85a9d07941d187e4\&subpoint=c59cc63440c795a7}{A4R} } &  GCC 4.9.2  &  14.2 $\pm$ 0.1  &  54284  & {\small -O2 -flto }\\
     \hline
      \textbf{ \href{http://cknowledge.org/repo/web.php?wcid=experiment:872541a6bf29037e\&subpoint=f9fa9a3effb5c863}{C1} } &  GCC 4.9.2  &  14.2 $\pm$ 0.0  &  64184  & {\small -O3 -fno-inline -flto }\\
     \hline
    \end{tabular}     
      \vspace{0.1in}
     \caption{
      Accelerating GCC 4.9.2 autotuning of a zlib encode workload on RPi3 device using 
      10..20 best performing combinations of compiler flags already found 
      and shared by the community during collaborative optimization.
     }
     \label{fig:autotuning-zlib-encode-gcc4-reactions}
   \end{figure*}

\begin{figure*}[!htbp]
     \centering
      \includegraphics[width=5.2in]
      {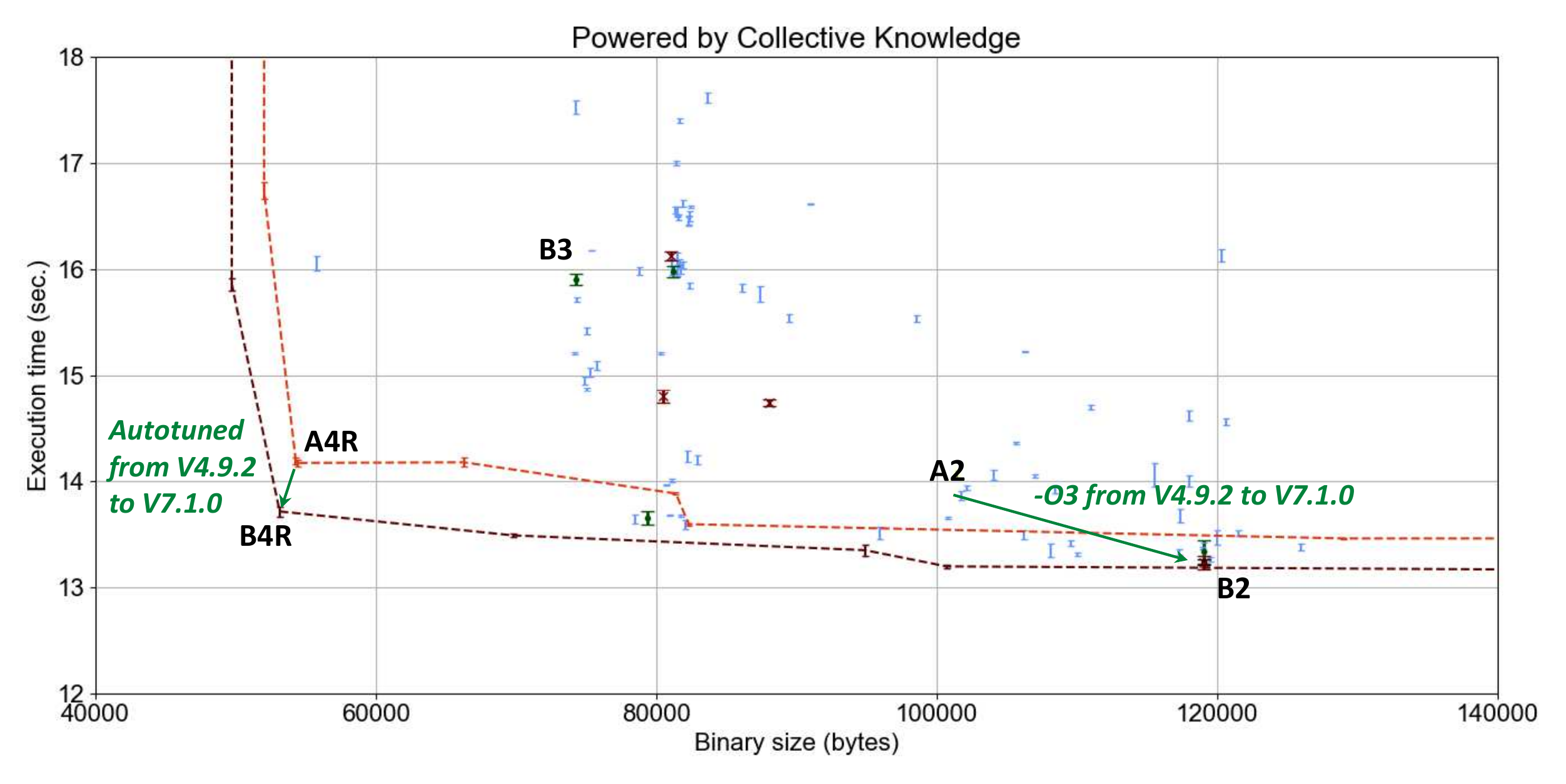} 
      \vspace{0.1in}
          \begin{tabular}{|l|l|l|l|p{3.2in}|}
     \hline
      \textbf{ID} & \textbf{Compiler} & \textbf{Time (sec.)} & \textbf{Size (bytes)} & \textbf{Flags} \\ 
     \hline
      \textbf{ \href{http://cknowledge.org/repo/web.php?wcid=experiment:21f631290c7846ee\&subpoint=b3d50b1184e6ebed}{A2} } &  GCC 4.9.2  &  14.0 $\pm$ 0.1  &  101448  & {\small -O3 }\\
     \hline
      \textbf{ \href{http://cknowledge.org/repo/web.php?wcid=experiment:85a9d07941d187e4\&subpoint=c59cc63440c795a7}{A4R} } &  GCC 4.9.2  &  14.2 $\pm$ 0.1  &  54284  & {\small -O2 -flto }\\
     \hline
      \textbf{ \href{http://cknowledge.org/repo/web.php?wcid=experiment:755bdc4154a3240e\&subpoint=479684c44854800c}{B1} } &  GCC 7.1.0  &  38.8 $\pm$ 0.0  &  128376  & {\small  }\\
     \hline
      \textbf{ \href{http://cknowledge.org/repo/web.php?wcid=experiment:50948cede943469a\&subpoint=381aef856bc24d3d}{B2} } &  GCC 7.1.0  &  13.2 $\pm$ 0.1  &  119084  & {\small -O3 }\\
     \hline
      \textbf{ \href{http://cknowledge.org/repo/web.php?wcid=experiment:4cc78e1a736bc05e\&subpoint=7177c749f6b5004a}{B3} } &  GCC 7.1.0  &  15.9 $\pm$ 0.1  &  74280  & {\small -Os }\\
     \hline
      \textbf{ \href{http://cknowledge.org/repo/web.php?wcid=experiment:3a9a0b4e4740a607\&subpoint=a8e09e075b5b4a38}{B4R} } &  GCC 7.1.0  &  13.7 $\pm$ 0.0  &  52424  & {\small -O2 -fgcse-after-reload -flto -fschedule-fusion -fno-ssa-phiopt -fno-tree-fre }\\
     \hline
    \end{tabular}     
      \vspace{0.1in}
     \caption{
       Results of GCC 7.1.0 random compiler flag autotuning of zlib encode on RPi3 device 
       with a highlighted frontier (trading-off execution time and code size), 
       the best combinations of flags on this frontier, and comparison with the results from GCC 4.9.2.
     }
     \label{fig:autotuning-zlib-encode-gcc7}
   \end{figure*}

\begin{figure*}[!htbp]
     \centering
      \includegraphics[width=5.8in]
      {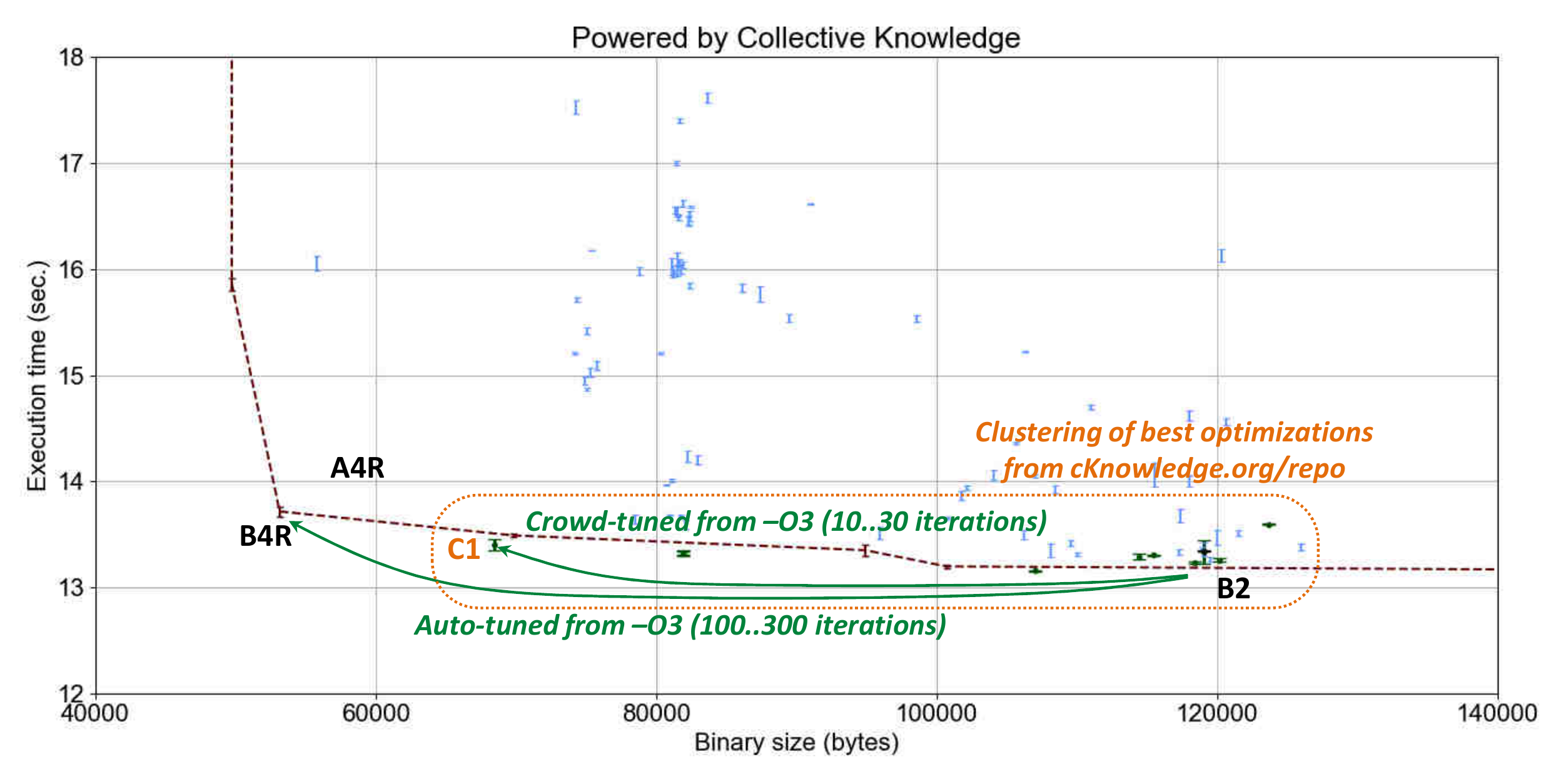} 
      \vspace{0.1in}
          \begin{tabular}{|l|l|l|l|p{3.2in}|}
     \hline
      \textbf{ID} & \textbf{Compiler} & \textbf{Time (sec.)} & \textbf{Size (bytes)} & \textbf{Flags} \\ 
     \hline
      \textbf{ \href{http://cknowledge.org/repo/web.php?wcid=experiment:85a9d07941d187e4\&subpoint=c59cc63440c795a7}{A4R} } &  GCC 4.9.2  &  14.2 $\pm$ 0.1  &  54284  & {\small -O2 -flto }\\
     \hline
      \textbf{ \href{http://cknowledge.org/repo/web.php?wcid=experiment:50948cede943469a\&subpoint=381aef856bc24d3d}{B2} } &  GCC 7.1.0  &  13.2 $\pm$ 0.1  &  119084  & {\small -O3 }\\
     \hline
      \textbf{ \href{http://cknowledge.org/repo/web.php?wcid=experiment:3a9a0b4e4740a607\&subpoint=a8e09e075b5b4a38}{B4R} } &  GCC 7.1.0  &  13.7 $\pm$ 0.0  &  52424  & {\small -O2 -fgcse-after-reload -flto -fschedule-fusion -fno-ssa-phiopt -fno-tree-fre }\\
     \hline
      \textbf{ \href{http://cknowledge.org/repo/web.php?wcid=experiment:047ddbe5ef0ab588\&subpoint=0b17daf9418aaa6d}{C1} } &  GCC 4.9.2  &  13.3 $\pm$ 0.1  &  68464  & {\small -O3 -fno-inline -flto }\\
     \hline
    \end{tabular}     
      \vspace{0.1in}
     \caption{
      Analyzing reactions of zlib encode to top most efficient GCC 7.1.0 optimizations shared by the community for RPi3 devices vs GCC 4.9.2.
     }
     \label{fig:autotuning-zlib-encode-gcc7-reactions}
   \end{figure*}

We performed the same autotuning and crowd-tuning experiments for \textit{zlib encode} workload
with the results shown in Figures~\ref{fig:autotuning-zlib-encode-gcc4},~\ref{fig:autotuning-zlib-encode-gcc4-reactions},~\ref{fig:autotuning-zlib-encode-gcc7},~\ref{fig:autotuning-zlib-encode-gcc7-reactions}.
The results show similar trend that \textit{-O3} optimization level of both \textit{GCC 4.7.2} and \textit{GCC 7.1.0} 
perform well in terms of execution time, while there is the same degradation in the code size when moving to a new compiler
(since we monitor the whole zlib binary size for both decode and encode functions).
Crowd-tuning also helped improve the code size though optimizations~\textbf{A4R},~\textbf{B4R} and \textbf{C1}
are not the same as in case of \textit{zlib decode}.
The reason is that algorithms are different and need different optimizations 
to keep execution time intact while improving code size.
Such result provides an extra motivation for function-level optimizations
already available in GCC.

Besides \textit{zlib}, we applied crowd-tuning with the best found and shared optimizations 
to other RPi programs using \textit{GCC 4.9.2} and \textit{GCC 7.1.0}.
Table~\ref{fig:crowdtuning-all-rpi3-progs} shows reactions of these optimizations
with the best trade-offs for execution time and code size.
One may notice that though \textit{GCC 7.1.0 -O3} level improves execution time
of most of the programs apart from a few exceptions, it also considerably degrades
code size in comparison with \textit{GCC 4.9.2 -O3} level.
These results also confirm that neither \textit{-O3} nor \textit{-Os} on both 
\textit{GCC 4.9.2} and \textit{GCC 7.1.0} achieves the best trade-offs for execution
time and code size thus motivating again our collaborative and continuous optimization
approach.

  \begin{table*}[]
    \centering
         \begin{tabular}{|l|l|p{1.2in}|p{0.9in}|p{1.8in}|}
     \hline
      \textbf{Workload} & \textbf{Compiler} & \textbf{Time improvement over -O3 (-O3 time in brackets)} & \textbf{Binary size improvement over -O3 (-O3 size in brackets)} & \textbf{Flags} \\ 
     \hline
      \textbf{ \href{http://cknowledge.org/repo/web.php?wcid=experiment:860174fc1b709377\&subpoint=ec934e93826572d2}{7z encode} } &  GCC 4.9.2  &  ~ 1.02 (5.5 $\pm$ 0.1)  &  ~ 1.52 (859728)  & {\small -O3 -fno-inline -flto }\\
     \hline
      \textbf{ \href{http://cknowledge.org/repo/web.php?wcid=experiment:03af5407d14468eb\&subpoint=6d4d4e8e1c6daac6}{7z encode} } &  GCC 7.1.0  &  no (6.0 $\pm$ 1.0)  &  no (887464)  & {\small -O3 }\\
     \hline
      \textbf{ \href{http://cknowledge.org/repo/web.php?wcid=experiment:89fd652535db438e\&subpoint=24844209902c136a}{ccrypt encrypt} } &  GCC 4.9.2  &  no (7.0 $\pm$ 2.0)  &  no (61772)  & {\small -O3 }\\
     \hline
      \textbf{ \href{http://cknowledge.org/repo/web.php?wcid=experiment:264c8a9a86776fcb\&subpoint=f8130c563953d484}{ccrypt encrypt} } &  GCC 7.1.0  &  ~ 1.16 (7.6 $\pm$ 0.1)  &  ~ 1.00 (59996)  & {\small -O3 -fno-auto-inc-dec -fguess-branch-probability -fipa-pure-const -freorder-blocks -fselective-scheduling2 -ftree-ccp -fno-tree-pre -ftree-tail-merge }\\
     \hline
      \textbf{ \href{http://cknowledge.org/repo/web.php?wcid=experiment:59dbd50c4fa98b6c\&subpoint=f4f3baa9717f1cee}{gzip decode} } &  GCC 4.9.2  &  ~ 1.04 (4.2 $\pm$ 0.0)  &  ~ 1.12 (85956)  & {\small -O3 -fno-inline -flto }\\
     \hline
      \textbf{ \href{http://cknowledge.org/repo/web.php?wcid=experiment:d2044082a58f5f78\&subpoint=64f3736c4460e87f}{gzip decode} } &  GCC 7.1.0  &  ~ 1.04 (4.2 $\pm$ 0.0)  &  ~ 1.18 (90568)  & {\small -O3 -fno-inline -flto }\\
     \hline
      \textbf{ \href{http://cknowledge.org/repo/web.php?wcid=experiment:d2044082a58f5f78\&subpoint=d6ae97514d86a01a}{gzip decode} } &  GCC 7.1.0  &  ~ 1.08 (4.2 $\pm$ 0.0)  &  ~ 0.81 (90568)  & {\small -O3 -fno-cprop-registers -flto -funroll-all-loops }\\
     \hline
      \textbf{ \href{http://cknowledge.org/repo/web.php?wcid=experiment:819a33f3e0a4efc3\&subpoint=d8685bf484cb1d9c}{gzip encode} } &  GCC 4.9.2  &  ~ 0.98 (12.3 $\pm$ 0.1)  &  ~ 1.10 (85956)  & {\small -O3 -fno-omit-frame-pointer -fno-tree-loop-optimize }\\
     \hline
      \textbf{ \href{http://cknowledge.org/repo/web.php?wcid=experiment:7bd35343c393218c\&subpoint=ea66fb473741f8ba}{gzip encode} } &  GCC 7.1.0  &  ~ 1.01 (12.3 $\pm$ 0.8)  &  ~ 1.18 (90568)  & {\small -O3 -fno-inline -flto }\\
     \hline
      \textbf{ \href{http://cknowledge.org/repo/web.php?wcid=experiment:3308b411e99c6c19\&subpoint=1970017ea1e6fd3f}{minigzip decode} } &  GCC 4.9.2  &  ~ 1.24 (10.0 $\pm$ 4.0)  &  ~ 1.60 (101432)  & {\small -O3 -fno-inline -flto }\\
     \hline
      \textbf{ \href{http://cknowledge.org/repo/web.php?wcid=experiment:3308b411e99c6c19\&subpoint=704e0a8e94359933}{minigzip decode} } &  GCC 4.9.2  &  ~ 1.32 (10.0 $\pm$ 4.0)  &  ~ 1.00 (101432)  & {\small -O3 -fselective-scheduling2 -fno-tree-pre }\\
     \hline
      \textbf{ \href{http://cknowledge.org/repo/web.php?wcid=experiment:b9728f4c7b36ac8a\&subpoint=f72e2b4f5e47b98c}{minigzip decode} } &  GCC 7.1.0  &  ~ 1.14 (8.0 $\pm$ 3.0)  &  ~ 1.76 (119088)  & {\small -O3 -fno-inline -flto }\\
     \hline
      \textbf{ \href{http://cknowledge.org/repo/web.php?wcid=experiment:8c4b2b03f4aa21a7\&subpoint=243c5a4fc20ee826}{minigzip encode} } &  GCC 4.9.2  &  ~ 0.89 (9.9 $\pm$ 0.0)  &  ~ 1.60 (101432)  & {\small -O3 -fno-inline -flto }\\
     \hline
      \textbf{ \href{http://cknowledge.org/repo/web.php?wcid=experiment:18828143558b3c2f\&subpoint=0e1438a8a38e6fd6}{minigzip encode} } &  GCC 7.1.0  &  ~ 1.00 (9.6 $\pm$ 0.0)  &  ~ 1.76 (119088)  & {\small -O3 -fno-inline -flto }\\
     \hline
      \textbf{ \href{http://cknowledge.org/repo/web.php?wcid=experiment:7ca1118331485cb4\&subpoint=4b74181efc011cbd}{rhash sha3} } &  GCC 4.9.2  &  ~ 1.00 (4.8 $\pm$ 0.0)  &  ~ 1.12 (14848)  & {\small -O3 -flto }\\
     \hline
      \textbf{ \href{http://cknowledge.org/repo/web.php?wcid=experiment:94a5792a3a4b925f\&subpoint=fc8e3a495ab4298c}{rhash sha3} } &  GCC 7.1.0  &  ~ 1.35 (5.2 $\pm$ 0.0)  &  ~ 1.30 (16396)  & {\small -O3 -fno-inline -flto }\\
     \hline
      \textbf{ \href{http://cknowledge.org/repo/web.php?wcid=experiment:94a5792a3a4b925f\&subpoint=dab8b6c57a392ca5}{rhash sha3} } &  GCC 7.1.0  &  ~ 1.48 (5.2 $\pm$ 0.0)  &  ~ 1.07 (16396)  & {\small -O3 -fno-schedule-insns -ftracer }\\
     \hline
      \textbf{ \href{http://cknowledge.org/repo/web.php?wcid=experiment:b8ab612f1524fa3b\&subpoint=042129a7110f3e31}{sha512sum sha512} } &  GCC 4.9.2  &  ~ 1.12 (7.8 $\pm$ 0.0)  &  ~ 1.06 (125372)  & {\small -O3 -fno-schedule-insns -fselective-scheduling2 }\\
     \hline
      \textbf{ \href{http://cknowledge.org/repo/web.php?wcid=experiment:20995294bf184644\&subpoint=fa8f9e4af29f3a8c}{sha512sum sha512} } &  GCC 7.1.0  &  ~ 1.22 (7.3 $\pm$ 0.0)  &  ~ 1.07 (121180)  & {\small -O3 -fno-predictive-commoning -fno-schedule-insns -funroll-loops }\\
     \hline
      \textbf{ \href{http://cknowledge.org/repo/web.php?wcid=experiment:1a89c378d17ce4d0\&subpoint=981b7b9de6f50286}{unrar} } &  GCC 4.9.2  &  ~ 0.97 (18.0 $\pm$ 4.0)  &  ~ 1.38 (326572)  & {\small -O3 -fno-inline -flto }\\
     \hline
      \textbf{ \href{http://cknowledge.org/repo/web.php?wcid=experiment:1a89c378d17ce4d0\&subpoint=94451a72a5aadda2}{unrar} } &  GCC 4.9.2  &  ~ 1.13 (18.0 $\pm$ 4.0)  &  ~ 0.80 (326572)  & {\small -O3 -fno-section-anchors -fselective-scheduling2 -fno-tree-forwprop -funroll-all-loops }\\
     \hline
      \textbf{ \href{http://cknowledge.org/repo/web.php?wcid=experiment:cbeb48dd9eaec742\&subpoint=82db107f34066a74}{unrar} } &  GCC 7.1.0  &  ~ 0.96 (18.0 $\pm$ 6.0)  &  ~ 1.38 (326572)  & {\small -O3 -fno-inline -flto }\\
     \hline
      \textbf{ \href{http://cknowledge.org/repo/web.php?wcid=experiment:cbeb48dd9eaec742\&subpoint=8a5a431da242467a}{unrar} } &  GCC 7.1.0  &  ~ 1.07 (18.0 $\pm$ 6.0)  &  ~ 0.78 (326572)  & {\small -O3 -fno-tree-ter -funroll-all-loops }\\
     \hline
    \end{tabular}     
    \caption{
      The highest found improvements (degradations) in execution time and binary size 
      for several important RPi3 programs 
      as reactions to top most efficient shared optimizations 
      for GCC 4.9.2 and GCC 7.1.0.
    }
    \label{fig:crowdtuning-all-rpi3-progs}
  \end{table*}

Indeed, a dozen of shared most efficient optimizations at \href{http://cKnowledge.org/repo}{cKnowledge.org/repo} 
is enough to either improve execution time of above programs by up to 1.5x or code size by up to 1.8x or even
improve both size and speed at the same time.
It also helps end-users find the most efficient optimization no matter which compiler, environment and hardware are used.

We can also notice that 11 workloads (computational species) 
share \textit{-O3 -fno-inline -flto} combination of flags 
to achieve the best trade-off between execution time and code size.
This result supports our original research to use workload features, 
hardware properties, crowd-tuning and  machine learning
to predict such optimizations~\cite{Fur2009,29db2248aba45e59:a31e374796869125,cm:29db2248aba45e59:cd11e3a188574d80}.
However, in contrast with the past work, we are now able
to gradually collect a large, realistic (i.e. not randomly
synthesized) set of diverse workloads with the help of the
community to make machine learning statistically meaningful.

All scripts to reproduce experiments from this section are available in the following CK entries:

\begin{flushleft}
\texttt{\$ ck find script:rpi3-zlib-decode*\newline
\$ ck find script:rpi3-zlib-encode*\newline
\$ ck find script:rpi3-all-autotune}
\end{flushleft}


\section{Crowd-fuzzing compilers}
\label{sec:crowdfuzzing}
When distributing compiler autotuning and learning across diverse environments, 
compilers and devices~\cite{Fur2009,cm:29db2248aba45e59:cd11e3a188574d80} 
we noticed that about 10..15\% of randomly generated combinations 
of flags can crash a compiler or produce wrong code with segmentation faults
or incorrect output.
Indeed our approach stresses various unexpected combinations 
of compiler optimizations across diverse and possibly untested platforms and workloads
thus helping automatically detect software and hardware bugs.
It complements well-known fuzzing techniques for automatic software 
testing~\cite{Duran:1981:RRT:800078.802530,Takanen:2008:FSS:1404500,Yang:2011:FUB:1993498.1993532}.

Our CK-based customizable autotuning workflow can assist in creating, 
learning and improving such collaborative fuzzers 
which can distribute testing across diverse platforms and workloads 
provided by volunteers while sharing and reproducing bugs.
We just need to retarget our autotuning workflow to search for
bugs instead of or together with improvements in performance, 
energy, size and other characteristics.

We prepared an example scenario~\emph{experiment.tune.compiler.flags.gcc.fuzz}
to randomly generate compiler flags for any GCC and record only cases
with failed program pipeline.
One can use it in a same way as any CK autotuning while selecting 
above scenario as following:

\begin{flushleft}
\texttt{\$ ck autotune program:cbench-automotive-susan --iterations=150 --repetitions=3 
  --scenario=experiment.tune.compiler.flags.gcc.fuzz
  --cmd\_key=corners --record\_uoa=tmp-susan-corners-gcc7-150-rnd-fuzz}
\end{flushleft}

It is then possible to view all results with unexpected behavior 
in a web browser and reproduce individual cases 
on a local or different machine as following: 

\begin{flushleft}
\texttt{\$ ck browser experiment:tmp-susan-corners-gcc7-150-rnd-fuzz}
\texttt{\$ ck replay experiment:tmp-susan-corners-gcc7-150-rnd-fuzz}
\end{flushleft}

We performed the same auto-fuzzing experiments for \textit{susan corners} program 
with both \textit{GCC 4.9.2} and \textit{GCC 7.1.0} as in Section~\ref{sec:flag_autotuning}.
These results are available in the following CK entries:
\begin{flushleft}
\texttt{\$ ck search experiment:rpi3-*fuzz*}
\end{flushleft}
It is also possible to browse them~\href{http://cknowledge.org/repo/web.php?wcid=experiment:rpi3-*fuzz*}{online}.

   \begin{figure*}[htbp]
     \centering
      \includegraphics[width=5.2in]
      {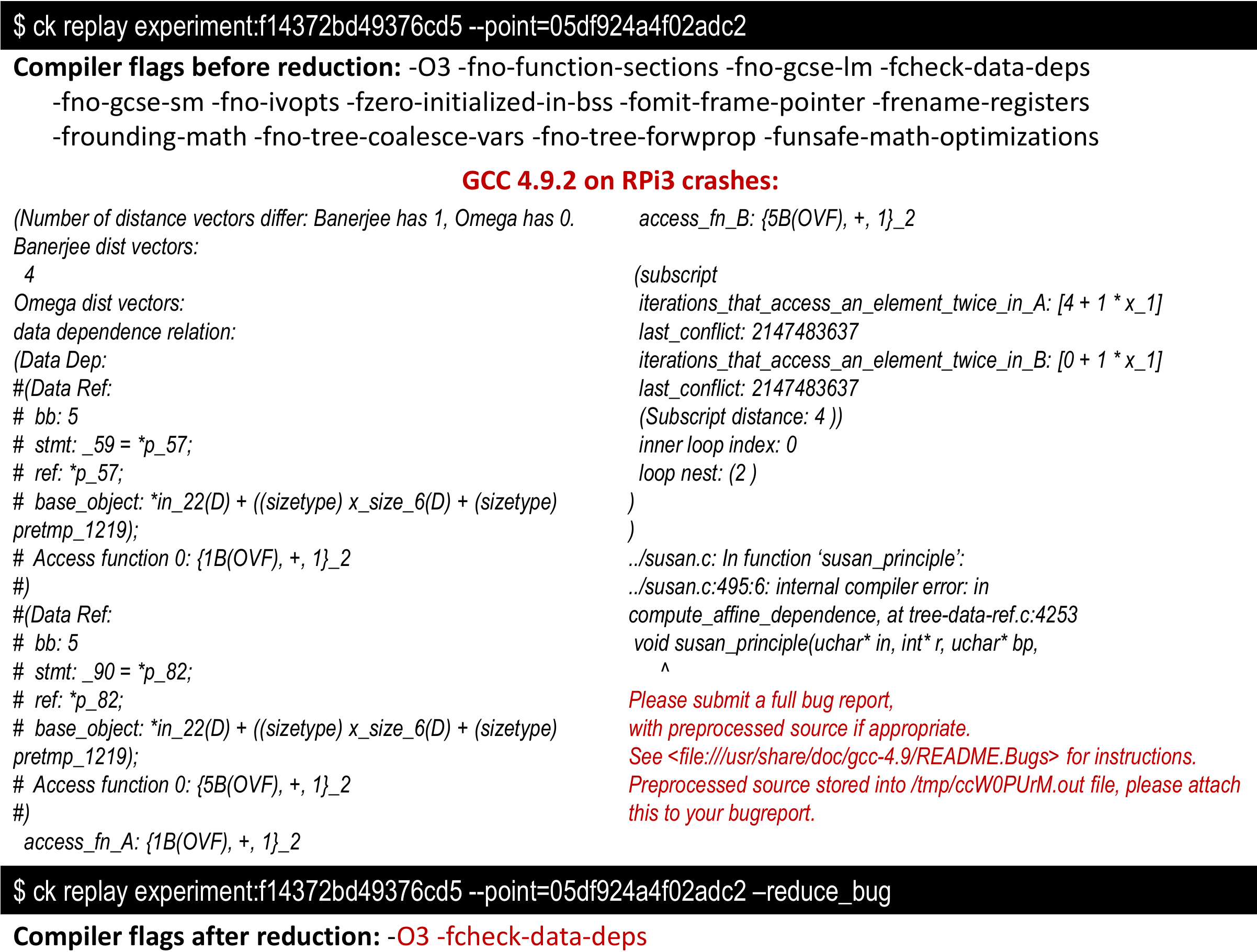} 
     \caption{
       Basic example of reproducing and reducing GCC bugs after random compiler flag autotuning.
     }
     \label{fig:ck-trivial-fuzzing-example}
   \end{figure*}

Figure~\ref{fig:ck-trivial-fuzzing-example} shows a simple example of reproducing 
a GCC bug using CK together with the original random combination of flags 
and the reduced one.
GCC flag \emph{-fcheck-data-deps} compares several passes for dependency analysis
and report a bug in case of discrepancy.
Such discrepancy was automatically found when autotuning \emph{susan corners} 
using \emph{GCC 4.9.2} on RPi3.

Since CK automatically adapts to a user environment, it is also possible
to reproduce the same bug using a different compiler version.
Compiling the same program with the same combination of flags on the same platform
using \emph{GCC 7.1.0} showed that this bug has been fixed in the latest compiler.

We hope that our extensible and portable benchmarking workflow will help students and engineers 
prototype and crowdsource different types of fuzzers.
It may also assist even existing projects~\cite{microsoft-fuzz,oss-fuzz}
to crowdsource fuzzing across diverse platforms and workloads.
For example, we collaborate with colleagues from Imperial College London
to develop CK-based, continuous and collaborative OpenGL and OpenCL compiler 
fuzzers~\cite{Lidbury:2015:MCF:2737924.2737986,DBLP:journals/corr/LascuD15,ck-clsmith}
while aggregating results from users in public or private repositories 
(~\href{http://cknowledge.org/repo/web.php?template=cknowledge&wcid=bc0409fb61f0aa82:1b437e72c74fe782&table_sort=2}{link to public OpenCL fuzzing results across diverse desktop and mobile platforms}).

\textit{All scripts to reproduce experiments from this section are available in the following CK entry:}

\begin{flushleft}
\texttt{\$ ck find script:rpi3-susan-fuzz-bugs}
\end{flushleft}


\section{Unifying and crowdsourcing machine learning}
\label{sec:crowdmodeling}
Having all optimization statistics continuously aggregated in a repository 
in a common format with JSON meta description 
makes it relatively straightforward to apply various machine learning 
and predictive analytics techniques including 
decision trees, nearest neighbor classifiers, support vector machines (SVM)
and deep learning~\cite{citeulike:873540,sammutencyclopedia}.
These techniques can help automate detection of regularities and consistent patterns in program behavior,
build models, and predict efficient optimizations rather than
continuously re-optimizing each new program as we previously demonstrated 
in the MILEPOST project~\cite{29db2248aba45e59:a31e374796869125,CFAP2007}.
Furthermore, we can now teach students how to collaboratively model 
the behavior of all computer systems, speed up optimization space exploration, 
and improve predictions of the most efficient software and hardware optimizations
based on various program, data set, platform and run-time 
features~\cite{fursin:hal-01054763,cm:29db2248aba45e59:cd11e3a188574d80}.

   \begin{figure*}[!htbp]
     \centering
      \includegraphics[width=6.6in]
      {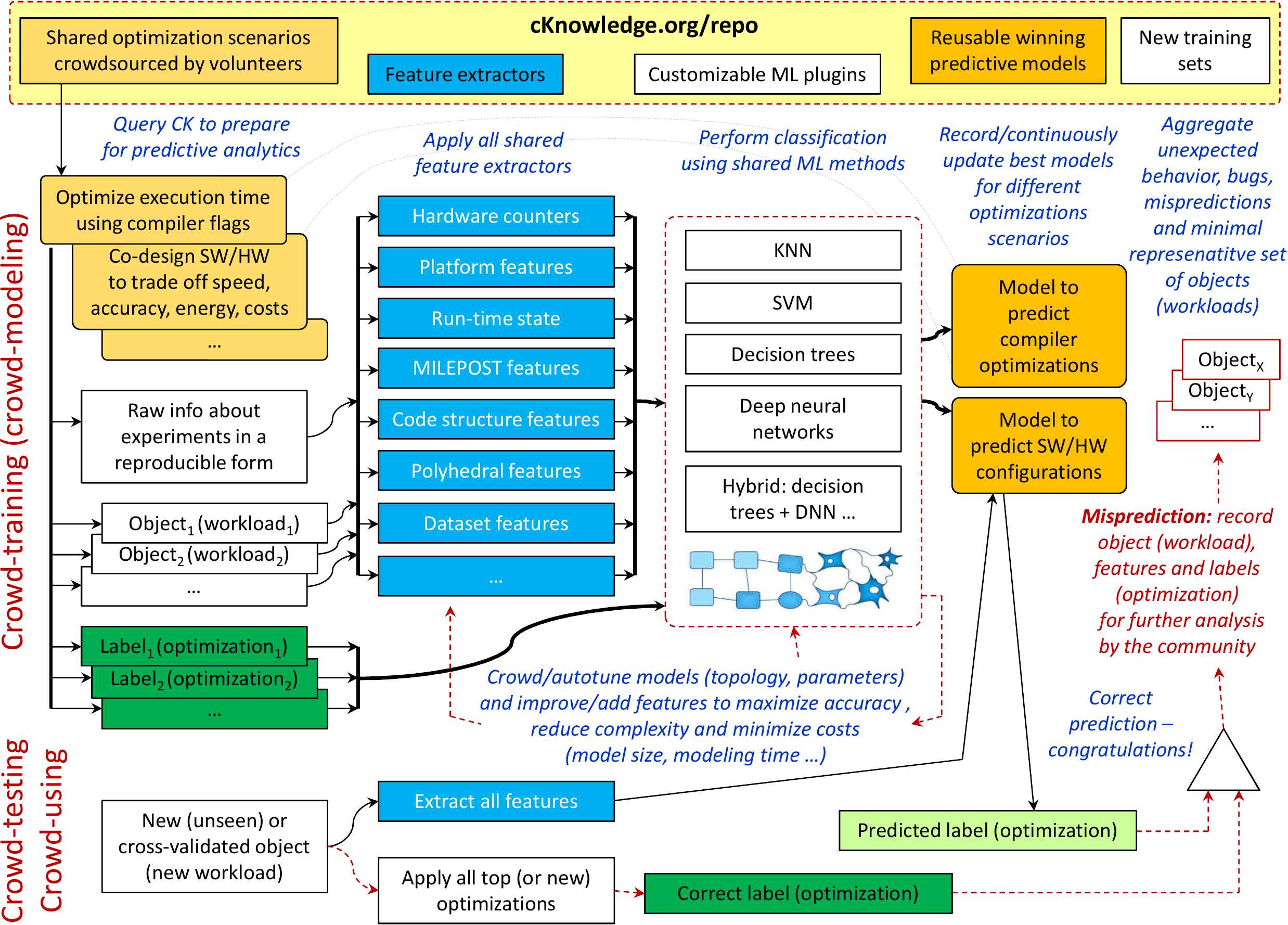} 
     \caption{
       Universal and high-level Collective Knowledge workflow to connect various communities 
       for collaborative, continuous and semi-automatic learning of multi-objective optimizations
       using shared machine learning modules (plugins) with the unified CK API.
     }
     \label{fig:ck-crowdmodeling}
   \end{figure*}

  \begin{table*}[!htbp]
    \centering
        \begin{tabular}{|l|p{1.2in}|p{0.9in}|p{0.9in}|}
     \hline
      \textbf{Model} & \textbf{Features} & \textbf{Accuracy (GCC 4.9.2)} & \textbf{Accuracy (GCC 7.1.0)} \\ 
     \hline
      \textbf{ milepost nn } &  ft1 .. ft56  &  0.37  &  0.30 \\
     \hline
    \end{tabular}     
    \caption{
     Accuracy of the nearest neighbor classifier with MILEPOST features 
     to predict the most efficient combinations of compiler flags 
     for GCC 4.9.2 and GCC 7.1.0 flags on RPi3 device.
    }
    \label{fig:crowdmodeling-milepost-all-rpi3-progs}
  \end{table*}

To demonstrate our approach, we converted all our past research artifacts 
on machine learning based optimization and SW/HW co-design
to CK modules.
We then assembled them to a universal Collective Knowledge workflow 
shown in Figure~\ref{fig:ck-crowdmodeling}.
If you do not know about machine learning based compiler optimizations, 
we suggest that you start from our MILEPOST GCC paper~\cite{29db2248aba45e59:a31e374796869125}
to make yourself familiar with terminology 
and methodology for machine learning training 
and prediction used further.
Next, we will briefly demonstrate the use of this customizable workflow 
to continuously classify shared workloads presented in this report 
in terms of the most efficient compiler optimizations
while using MILEPOST models and features.

   \begin{figure*}[!htbp]
     \centering
      \includegraphics[width=6.6in]
      {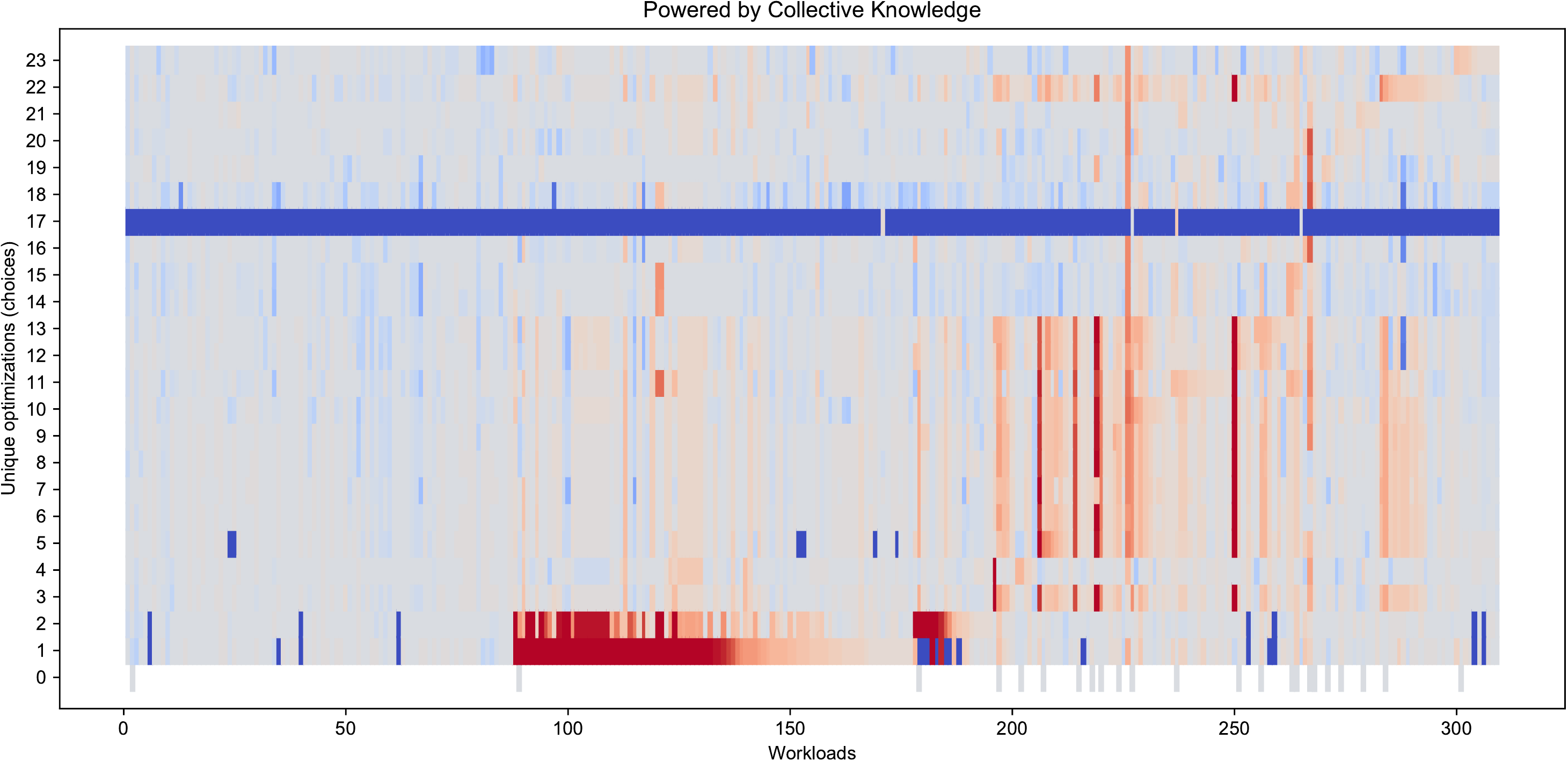} 
      \includegraphics[width=6.6in]
      {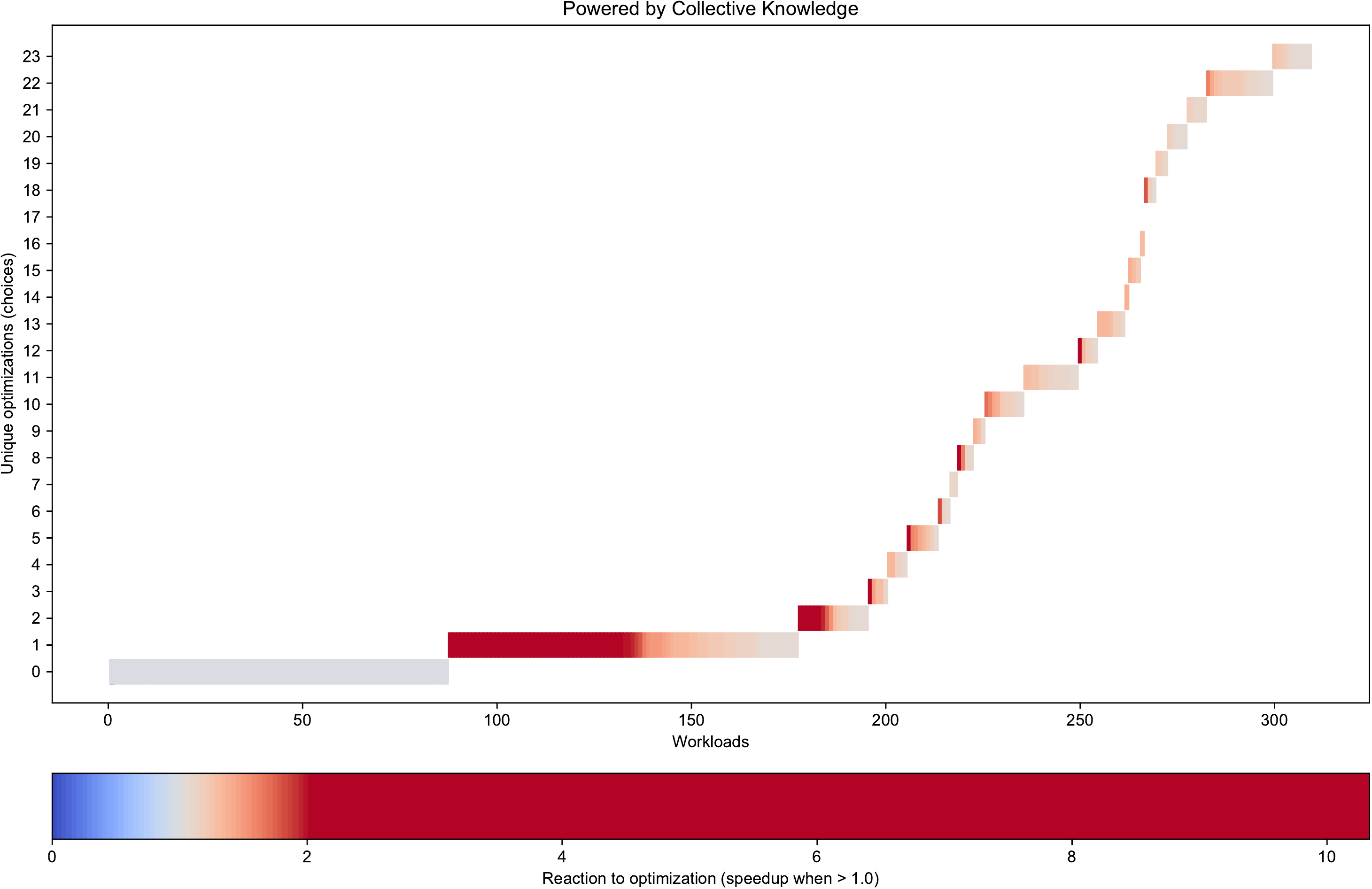} 
     \caption{
       Top graph: reactions of all workloads 
          to all top performing combinations of optimizations
          for GCC 4.9.2 on RPi3 device (speedups if value is more than 1.0).
       Bottom graph: groups of workloads achieving the highest speedup
          for a given unique combination of optimizations.
     }
     \label{fig:ck-reactions-gcc4}
   \end{figure*}

   \begin{figure*}[!htbp]
     \centering
      \includegraphics[width=6.6in]
      {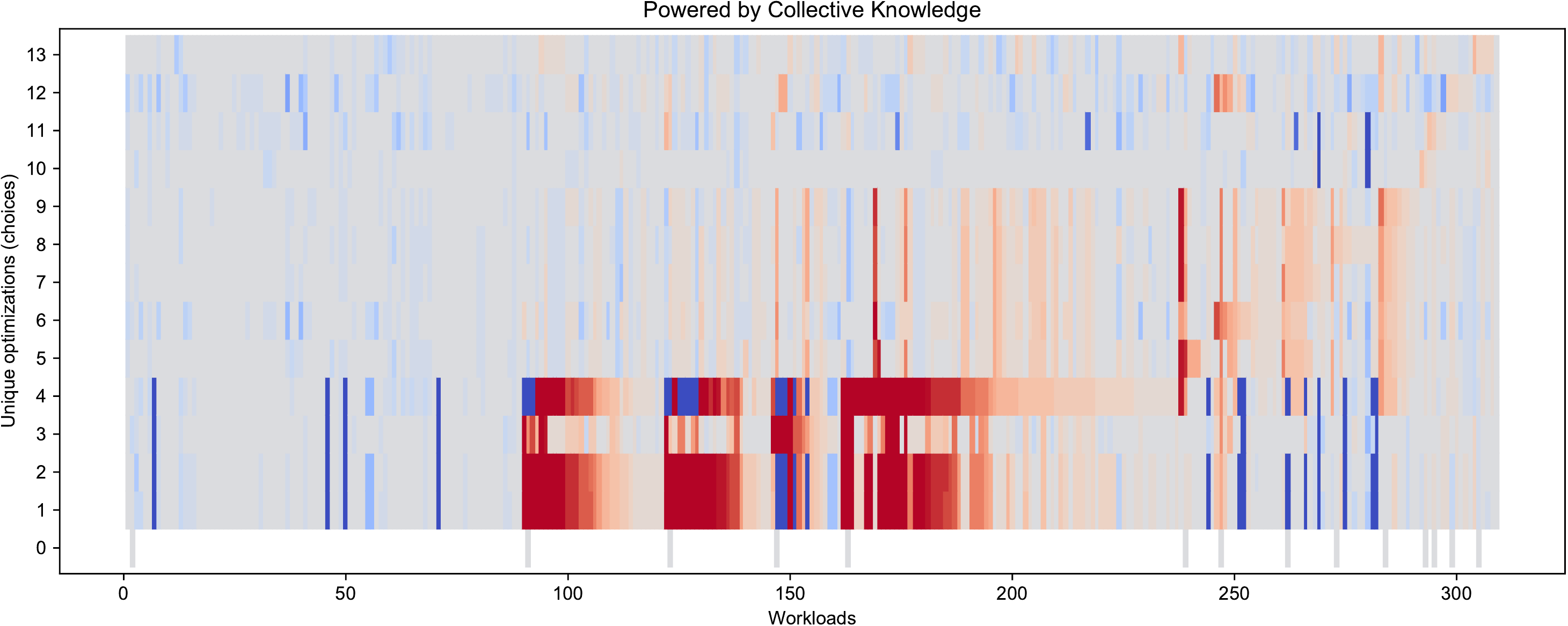} 
      \vspace{0.1in}
      \includegraphics[width=6.6in]
      {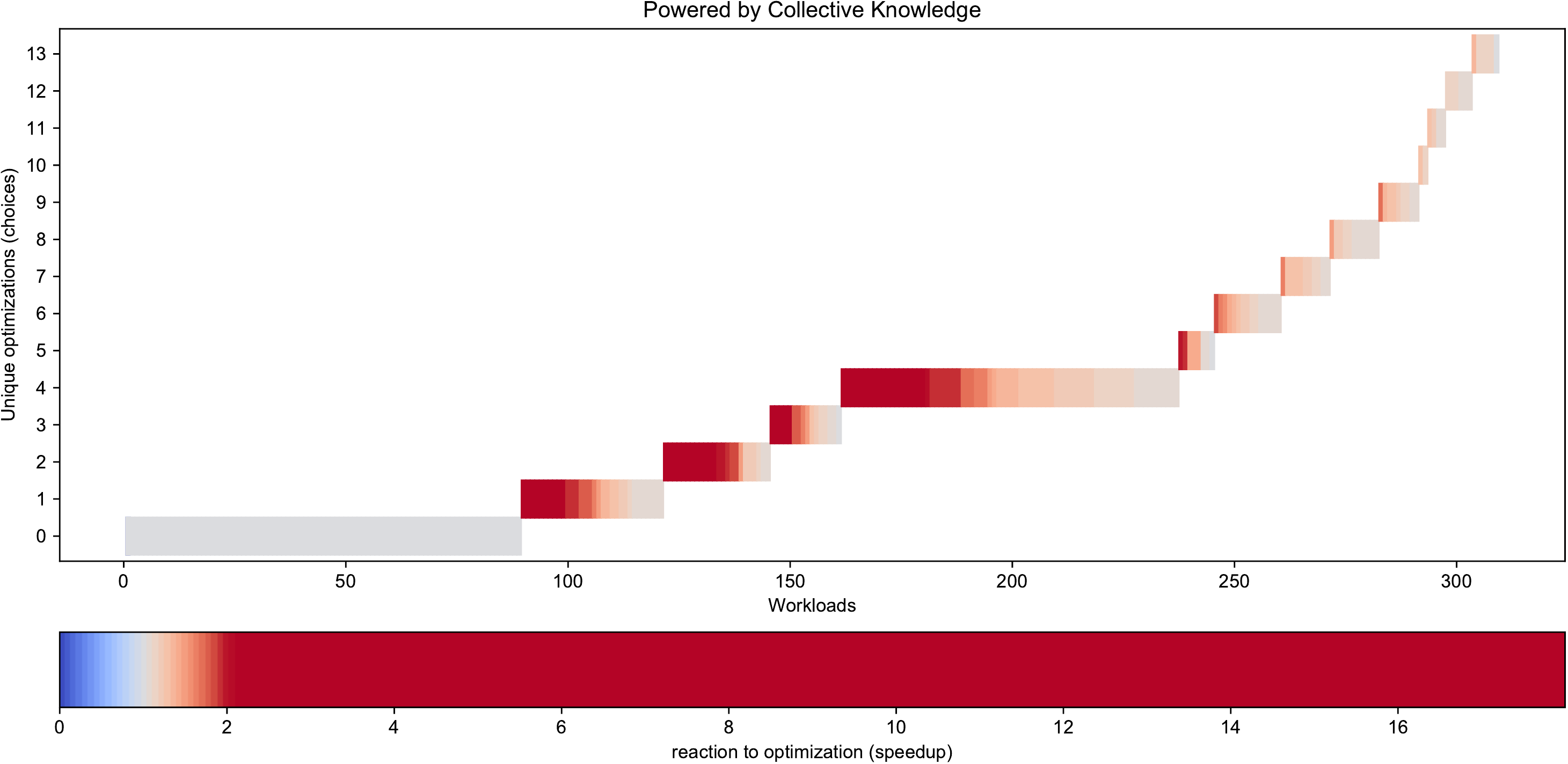} 
     \caption{
       Top graph: reactions of all workloads 
          to all top performing combinations of optimizations
          for GCC 7.1.0 on RPi3 device (speedup if value is more than 1.0).
       Bottom graph: groups of workloads achieving the highest speedup
          for a given unique combination of optimizations.
     }
     \label{fig:ck-reactions-gcc7}
   \end{figure*}

First, we query the public CK repository~\cite{live-ck-repo}
to collect all optimization statistics together with all associated objects 
(workloads, data sets, platforms) for a given optimization scenario. 
In our compiler flag optimization scenario, we retrieve all
most efficient compiler flags combinations found and shared
by the community when crowd-tuning GCC 4.9.2 on RPi3 device
(Figure~\ref{fig:ck-snapshot-of-results-gcc4}).

Note that our CK crowd-tuning workflow also continuously applies
such optimization to all shared workloads.
This allows us to analyze "reaction" of any given workload 
to all most efficient optimizations.
We can then group together those workloads which exhibit similar reactions.

The top graph in Figure~\ref{fig:ck-reactions-gcc4} shows reactions of all workloads 
to the most efficient optimizations as a ratio of the default execution time (-O3) 
to the execution time of applied optimization.
It confirms yet again (\cite{cm:29db2248aba45e59:cd11e3a188574d80}) that there is no single "winning" 
combination of optimizations and they can either considerably improve or degrade execution time 
on different workloads.
It also confirms that it is indeed possible to group together multiple workloads 
which share the most efficient combination of compiler flags, i.e. which achieve 
the highest speedup for a common optimization as shown in the bottom graph 
in Figure~\ref{fig:ck-reactions-gcc4}.
Figure~\ref{fig:ck-reactions-gcc7} shows similar trends for GCC 7.1.0 on the same RPi3 device
even though the overall number of the most efficient combinations of compiler flags is smaller 
than for GCC 4.9.2 likely due to considerably improved internal optimization heuristics over the past years
(see Figure~\ref{fig:ck-snapshot-of-results-gcc7}).

Having such groups of labeled objects (where labels are the most efficient optimizations
and objects are workloads) allows us to use standard machine learning classification methodology.
One must find such a set of objects' features and a model which maximizes 
correct labeling of previously unseen objects, or in our cases can correctly predict 
the most efficient software optimization and hardware design for a given workload.
As example, we extracted 56 so-called MILEPOST features described in~\cite{29db2248aba45e59:a31e374796869125} 
(static program properties extracted from GCC's intermediate representation) 
from all shared programs, stored them in \emph{program.static.features},
and applied simple nearest neighbor classifier to above data.
We then evaluated the quality of such model (ability to predict) using prediction accuracy
during standard leave-one-out cross-validation technique: for each workload we remove it
from the training set, build a model, validate predictions, sum up all correct predictions 
and divide by the total number of workloads.

Table~\ref{fig:crowdmodeling-milepost-all-rpi3-progs} shows this prediction accuracy
of our MILEPOST model for compiler flags from GCC 4.9.2 and GCC 7.1.0 
across all shared workloads on RPi3 device.
One may notice that it is nearly twice lower than in the
original MILEPOST
paper~\cite{29db2248aba45e59:a31e374796869125}.
As we explain in~\cite{cm:29db2248aba45e59:cd11e3a188574d80},
in the MILEPOST project we could only use a dozen of similar
workloads and just a few most efficient optimizations to be
able to perform all necessary experiments within a reasonable
amount of time (6 months).
After brining the community on hoard, we could now use a much larger 
collective training set with more than 300 shared, diverse 
and non-synthesized workloads while analyzing much more optimizations 
by crowdsourcing autotuning.
This helps obtain a more realistic limit of the MILEPOST predictor.

Though relatively low, this number can now become a 
reference point to be further improved by the community.
It is similar in spirit to the ImageNet Large Scale Visual Recognition Competition
(ILSVRC)~\cite{DBLP:journals/corr/RussakovskyDSKSMHKKBBF14}
which reduced image classification error rate from 25\%
in 2011 to just a few percent with the help of the community.
Furthermore, we can also keep just a few representative
workloads for each representative group as well as misclassified ones in
a public repository thus producing a minimized, realistic and
representative training set for systems researchers.

\textit{We shared all demo scripts which we used to generate data and
graphs in this section in the following CK entry (however they are not yet
user-friendly and we will continue improving documentation and
standardizing APIs of reusable CK modules with the help of the community):}

\begin{flushleft}
\texttt{\$ ck find script:rpi3-crowdmodel}
\end{flushleft}


\section{Improving and autotuning models and features}
\label{sec:features}
There are many publications demonstrating interesting machine learning algorithms,
features and models to predict efficient program optimizations and hardware
designs~\cite{Monsifrot,SAMP2003,
Marin:2004:CPP:1012888.1005691, SA2005, soffa2005, ABCP06,
CFAP2007, DJBP2009, JGVP2009,
DBLP:conf/cf/ShenVSAS13,DBLP:journals/ijpp/ParkCPBCS13,
DBLP:journals/taco/LeatherBO14,
DBLP:conf/IEEEpact/CumminsP0L17,
Ashouri:2017:MMC:3132652.3124452}.
Though all these techniques can be potentially useful, the
lack of common interfaces and meta information for artifacts
and experimental workflows makes it extremely challenging 
to compare, reuse and build upon them particularly 
in industrial projects with tough deadlines.

Even artifact evaluation which we introduced at systems
conferences~\cite{ae} to partially solve these issues is not
yet enough because our community does not have a common, 
portable and customizable workflow framework.
Bridging this gap between machine learning and systems research
served as an additional motivation to develop Collective Knowledge
workflow framework.
Our idea is to help colleagues and students share various workloads, 
data sets, machine learning algorithms, models and feature extractors 
as plugins (CK modules) with a common API and meta description.
Plugged to a common machine learning workflow such modules
can then be applied in parallel to continuously compete 
for the most accurate predictions for a given optimization scenario.
Furthermore, the community can continue improving and autotuning models,
analyzing various combination of features, experimenting with hierarchical models, 
and pruning models to reduce their complexity across shared data sets 
to trade off prediction accuracy, speed, size and the ease of interpretation.

   \begin{figure}[!htbp]
     \centering
      \includegraphics[width=3.2in]
      {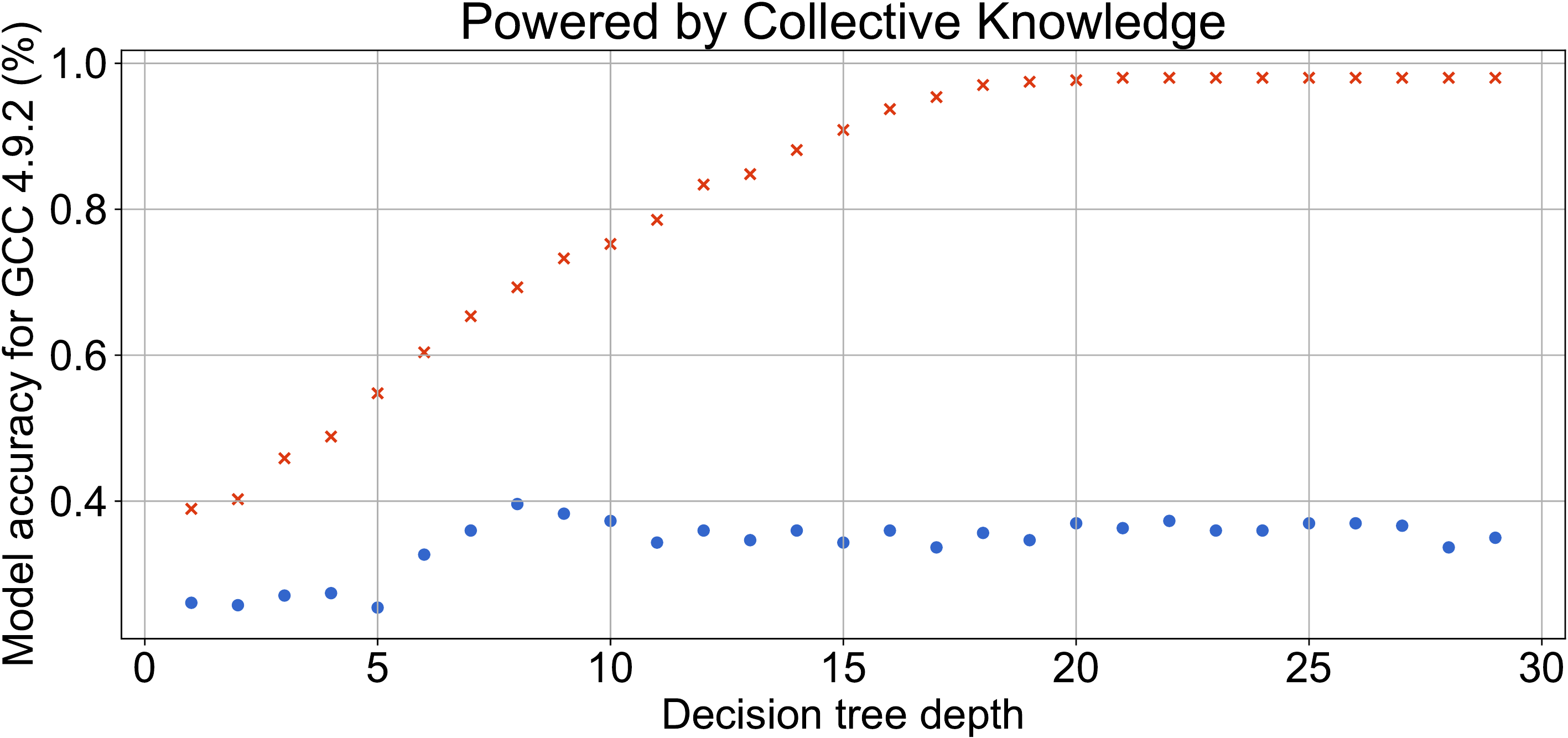} 
      \includegraphics[width=3.2in]
      {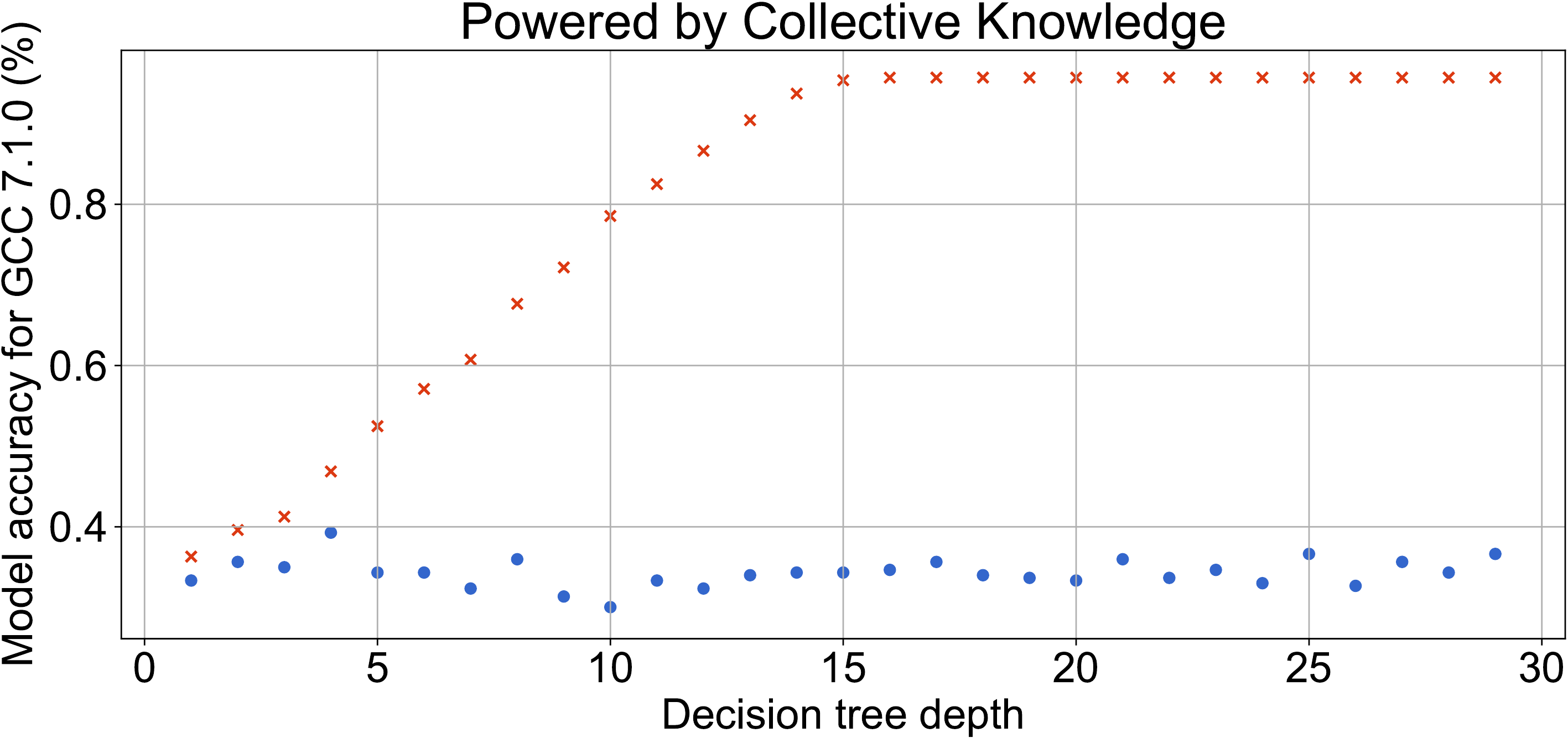} 
     \caption{
      Accuracy of an automatically generated decision tree to predict compiler flags (GCC 4.9.2 on the top graph and GCC 7.1.0 on the bottom graph) 
      on RPi3 when autotuning the tree depth.
      Round blue dots show prediction accuracy with cross-validation while blue crosses show
      prediction accuracy without cross-validation.
     }
     \label{fig:ck-model-crowdtuning-gcc}
   \end{figure}

For a proof-of-concept of such collaborative learning approach, 
we shared a number of customizable CK modules (see~\emph{ck search module:*model*})
for several popular classifiers including the nearest neighbor,
decision trees and deep learning.
These modules serve as wrappers with a common CK API for
TensorFlow, scikit-learn, R and other machine learning frameworks.
We also shared several feature extractors (see \emph{ck search module:*features*}) 
assembling the following groups of program features
which may influence predictions:

\begin{itemize}
  \item {\bf ft1 .. ft56} - original MILEPOST features (see~\cite{29db2248aba45e59:a31e374796869125});
  \item {\bf ft57 .. ft65} - additional features designed and shared by our colleague, Dr. Jeremy Singer~\cite{all-milepost-features};
  \item {\bf ft66 .. ft121} - original MILEPOST features normalized by the total number of instructions (ft24);
\end{itemize}

We then attempted to autotune various parameters
of machine learning algorithms exposed via CK API.
Figure~\ref{fig:ck-model-crowdtuning-gcc}
shows an example of autotuning the depth of a decision tree
(available as customizable CK plugin) with all shared groups of features
and its impact on prediction accuracy of compiler flags using MILEPOST
features from the previous section for GCC 4.9.2 and GCC 7.1.0
on RPi3.
Blue round dots obtained using leave-one-out validation suggest 
that decision trees of depth 8 and 4 are enough 
to achieve maximum prediction accuracy of 0.4\% for GCC 4.9.2 
and GCC 7.1.0 respectively.
Model autotuning thus helped improve prediction accuracy in comparison 
with the original nearest neighbor classifier from the MILEPOST project. 

   \begin{figure*}[!htbp]
     \centering
      \includegraphics[width=2in]
      {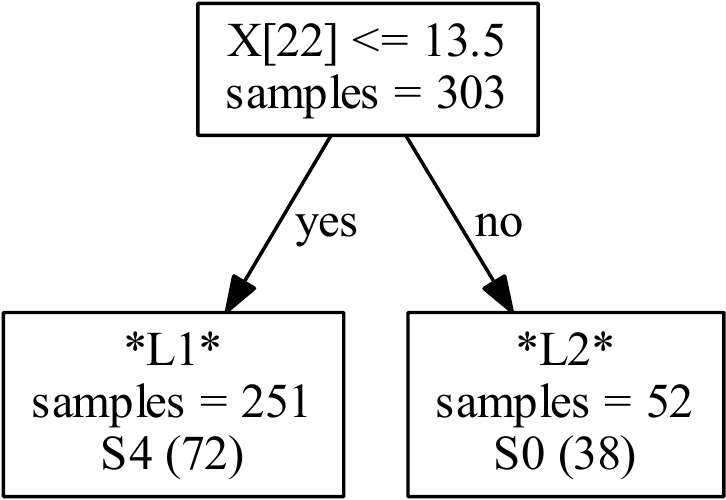} 
      \includegraphics[width=3.6in]
      {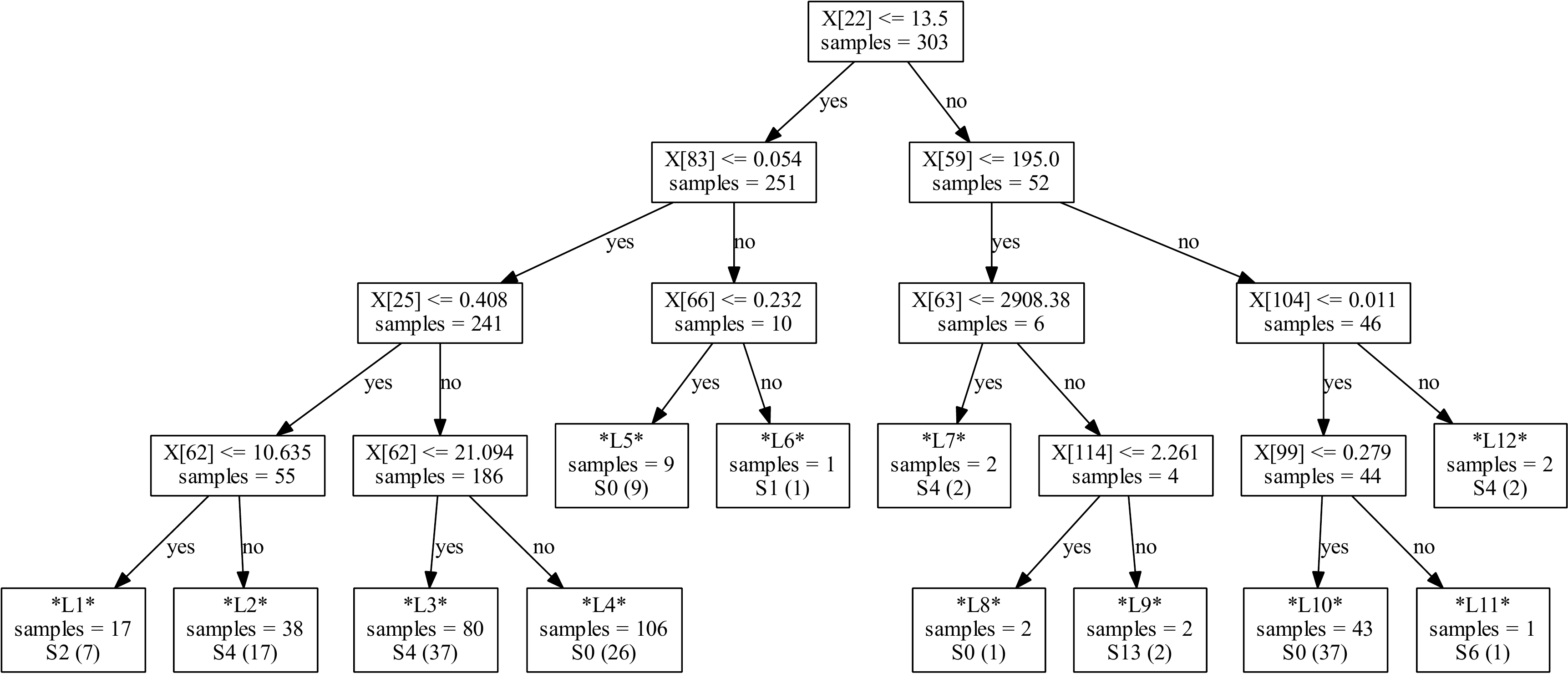} 
      \includegraphics[width=6.6in]
      {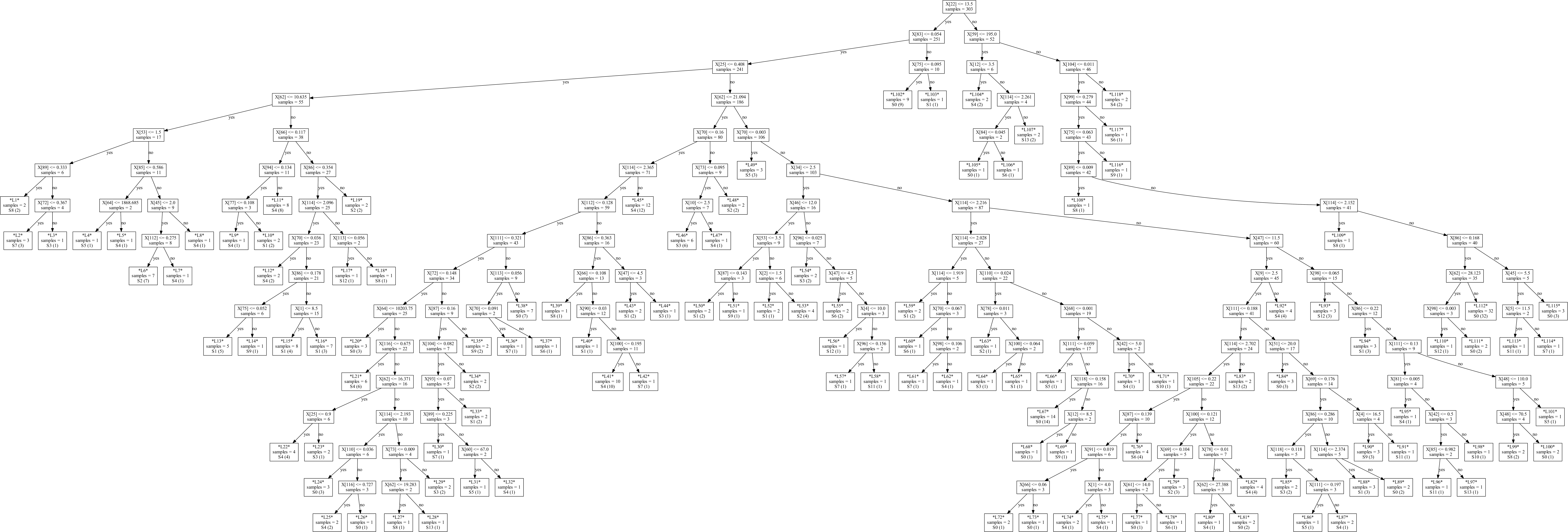} 
     \caption{
      Example of automatically generated decision trees of depth 1 and 4 with leave-one-out cross-validation, 
      and 15 without cross-validation to predict GCC 7.1.0 compiler optimizations using CK modules.
     }
     \label{fig:ck-model-crowdtuning-gcc7-dt}
   \end{figure*}

Figure~\ref{fig:ck-model-crowdtuning-gcc7-dt} shows a few examples of such automatically
generated decision trees with different depths for GCC 7.1.0 using CK.
Such trees are easy to interpret and can therefore help compiler and hardware 
developers quickly understand the most influential features and analyze
relationships between different features and the most efficient
optimizations.
For example, the above results suggest that the number of binary integer operations (ft22) 
and the number of distinct operators (ft59) can help predict optimizations 
which can considerably improve execution time of a given method over -O3.

Turning off cross-validation can also help developers understand 
how well models can perform on all available workloads (in-sample data)
(red dots on Figure~\ref{fig:ck-model-crowdtuning-gcc}).
In our case of GCC 7.1.0, the decision tree of depth 15 shown in Figure~\ref{fig:ck-model-crowdtuning-gcc7-dt})
is enough to capture all compiler optimizations for ~300 available workloads.

  \begin{table*}[!htbp]
    \centering
    {\small
         \begin{tabular}{|l|p{1.2in}|p{0.9in}|p{0.9in}|}
     \hline
      \textbf{Model} & \textbf{Features} & \textbf{Accuracy (GCC 4.9.2)} & \textbf{Accuracy (GCC 7.1.0)} \\ 
     \hline
      \textbf{ decision trees with cross validation; depth 1 } &  ft1 .. ft65  &  0.26  &  0.33 \\
     \hline
      \textbf{ decision trees with cross validation; depth 2 } &  ft1 .. ft65  &  0.26  &  0.36 \\
     \hline
      \textbf{ decision trees with cross validation; depth 4 } &  ft1 .. ft65  &  0.27  &  0.39 \\
     \hline
      \textbf{ decision trees with cross validation; depth 8 } &  ft1 .. ft65  &  0.40  &  0.36 \\
     \hline
      \textbf{ decision trees with cross validation; depth 16 } &  ft1 .. ft65  &  0.36  &  0.35 \\
     \hline
      \textbf{ decision trees with cross validation; depth 20 } &  ft1 .. ft65  &  0.37  &  0.33 \\
     \hline
      \textbf{ decision trees with cross validation; depth 25 } &  ft1 .. ft65  &  0.37  &  0.37 \\
     \hline
      \textbf{ decision trees with cross validation; depth 29 } &  ft1 .. ft65  &  0.35  &  0.37 \\
     \hline
      \textbf{ decision trees without cross validation; depth 1 } &  ft1 .. ft65  &  0.39  &  0.36 \\
     \hline
      \textbf{ decision trees without cross validation; depth 2 } &  ft1 .. ft65  &  0.40  &  0.40 \\
     \hline
      \textbf{ decision trees without cross validation; depth 4 } &  ft1 .. ft65  &  0.49  &  0.47 \\
     \hline
      \textbf{ decision trees without cross validation; depth 8 } &  ft1 .. ft65  &  0.69  &  0.68 \\
     \hline
      \textbf{ decision trees without cross validation; depth 16 } &  ft1 .. ft65  &  0.94  &  0.96 \\
     \hline
      \textbf{ decision trees without cross validation; depth 20 } &  ft1 .. ft65  &  0.98  &  0.96 \\
     \hline
      \textbf{ decision trees without cross validation; depth 25 } &  ft1 .. ft65  &  0.98  &  0.96 \\
     \hline
      \textbf{ decision trees without cross validation; depth 29 } &  ft1 .. ft65  &  0.98  &  0.96 \\
     \hline
      \textbf{ dnn tf with cross validation; iteration 1 } &  ft1 .. ft65  &  0.68  &  0.30 \\
     \hline
      \textbf{ dnn tf with cross validation; iteration 2 } &  ft1 .. ft65  &  0.64  &  0.33 \\
     \hline
      \textbf{ dnn tf with cross validation; iteration 3 } &  ft1 .. ft65  &  0.61  &  0.45 \\
     \hline
      \textbf{ dnn tf with cross validation; iteration 4 } &  ft1 .. ft65  &  0.64  &  0.44 \\
     \hline
      \textbf{ dnn tf without cross validation; iteration 1 } &  ft1 .. ft65  &  0.72  &  0.29 \\
     \hline
      \textbf{ dnn tf without cross validation; iteration 2 } &  ft1 .. ft65  &  0.72  &  0.47 \\
     \hline
      \textbf{ dnn tf without cross validation; iteration 3 } &  ft1 .. ft65  &  0.72  &  0.48 \\
     \hline
      \textbf{ dnn tf without cross validation; iteration 4 } &  ft1 .. ft65  &  0.68  &  0.62 \\
     \hline
      \textbf{ milepost nn } &  ft1 .. ft121  &  0.30  &  0.30 \\
     \hline
      \textbf{ milepost nn } &  ft1 .. ft56  &  0.37  &  0.30 \\
     \hline
      \textbf{ milepost nn } &  ft1 .. ft65  &  0.30  &  0.30 \\
     \hline
      \textbf{ milepost nn } &  ft57 .. ft121  &  0.30  &  0.30 \\
     \hline
      \textbf{ milepost nn } &  ft57 .. ft65  &  0.30  &  0.30 \\
     \hline
      \textbf{ milepost nn } &  ft66 .. ft121  &  0.36  &  0.32 \\
     \hline
      \textbf{ milepost nn } &  ft1 .. ft121\newline(normalized)  &  0.37  &  0.37 \\
     \hline
      \textbf{ milepost nn } &  ft1 .. ft56\newline(normalized)  &  0.37  &  0.33 \\
     \hline
      \textbf{ milepost nn } &  ft1 .. ft65\newline(normalized)  &  0.39  &  0.32 \\
     \hline
      \textbf{ milepost nn } &  ft57 .. ft121\newline(normalized)  &  0.37  &  0.39 \\
     \hline
      \textbf{ milepost nn } &  ft57 .. ft65\newline(normalized)  &  0.37  &  0.35 \\
     \hline
      \textbf{ milepost nn } &  ft66 .. ft121\newline(normalized)  &  0.38  &  0.38 \\
     \hline
      \textbf{ milepost nn (reduce complexity1) } &  ft1 .. ft121\newline(normalized)  &  0.45  &  0.44 \\
     \hline
      \textbf{ milepost nn (reduce complexity2) } &  ft1 .. ft121\newline(normalized)  &  0.45  &  0.40 \\
     \hline
    \end{tabular}     
    }
    \caption{
     Prediction accuracy when autotuning or reducing complexity of decision tree, 
     nearest neighbor and deep learning classifiers
     across different groups of program features.
    }
    \label{fig:crowdmodeling-all-rpi3-progs}
  \end{table*}

To complete our demonstration of CK concepts for collaborative machine learning and optimization,
we also evaluated a deep learning based classifier from TensorFlow~\cite{DBLP:journals/corr/AbadiABBCCCDDDG16}
(see \emph{ck help module:model.tf})
with 4 random configurations of hidden layers ([10,20,10], [21,13,21], [11,30,18,20,13], [17]) 
and training steps (300..3000).
We also evaluated the nearest neighbor classifier used in the MILEPOST project but with different groups of features 
and aggregated all results in Table~\ref{fig:crowdmodeling-all-rpi3-progs}. 
Finally, we automatically reduced the complexity of the nearest neighbor classifier (1) by iteratively removing those features one by one
which do not degrade prediction accuracy and (2) by iteratively adding features one by one to maximize prediction accuracy.
It is interesting to note that our nearest neighbor classifier achieves
a slightly better prediction accuracy with a reduced feature set than with 
a full set of features showing inequality of MILEPOST features
and overfitting.

As expected, deep learning classification achieves a better prediction accuracy of 0.68\%
and 0.45\% for GCC 4.9.2 and GCC 7.1.0 respectively for RPi3 among currently shared models, 
features, workloads and optimizations.
However, since deep learning models are so much more computationally intensive, resource hungry
and difficult to interpret than decision trees, one must carefully balance accuracy vs speed vs size.
That is why we suggest to use hierarchical models where
high-level and coarse-grain program behavior is quickly captured 
using decision trees, while all fine-grain behavior is captured 
by deep learning and similar techniques.
Another possible use of deep learning can be in automatically capturing
influential features from the source code, data sets and hardware.

\textit{All scripts to generate above experiments (require further documentation)
are available in the following CK entry:}

\begin{flushleft}
\texttt{\$ ck find script:rpi3-crowdmodel}
\end{flushleft}


\section{Enabling input-aware optimization}
\label{sec:datasets}
Current prediction accuracy which we achieved for the most efficient compiler flags 
is still disappointing: around 0.45\% for GCC 7.1.0.
We explained this in more detail in~\cite{cm:29db2248aba45e59:cd11e3a188574d80,fursin:hal-01054763}
by missing features particularly available at run-time from data sets and hardware.
Having a customizable experimental workflow with pluggable artifacts 
makes it relatively straightforward to analyze reactions of a given program
to the most efficient optimization across multiple data sets
and search for missing features.

First, we converted 474 different data sets from the MiDataSet suite~\cite{FCOP2007}
as pluggable CK artifacts and shared them as a zip archive (~800MB).
It is possible to download it from the Google Drive
from~\url{https://drive.google.com/open?id=0B-wXENVfIO82OUpZdWIzckhlRk0}
(we plan to move it to a permanent repository in the future)
and then install via CK as following:

\begin{flushleft}
\texttt{\$ ck add repo --zip=ckr-ctuning-datasets.zip --quiet}\newline
\texttt{\$ ck ls dataset --all}\newline
\texttt{\$ ck search dataset --tags=image,jpeg}\newline
\end{flushleft}

All these data sets will be immediately visible to all related programs
via the CK autotuning workflow.
For example, if we now run \textit{susan corners} program, CK will prompt user
a choice of 20 related images from the above data sets:

\begin{flushleft}
\texttt{\$ ck compile program:cbench-automotive-susan --speed}\newline
\texttt{\$ ck run program:cbench-automotive-susan}\newline
\end{flushleft}

   \begin{figure*}[!htbp]
     \centering
      \includegraphics[width=5.8in]
      {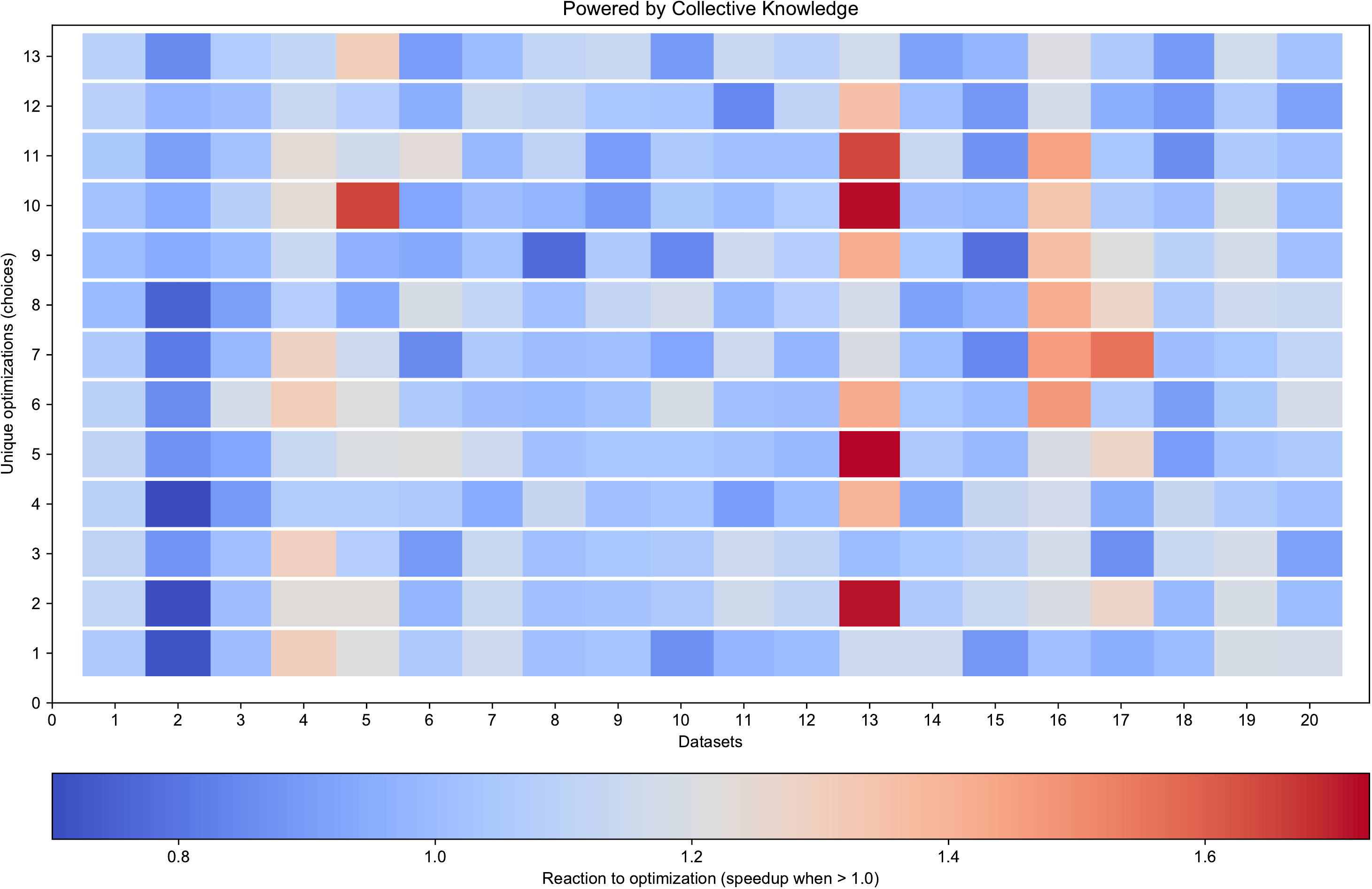} 
     \caption{
       Reactions of a jpeg decoder across 20 distinct data sets (jpeg images)
       to all top performing combinations of compiler optimizations
       for GCC 7.1.0 on RPi3 device (speedups if value is more than 1.0).
     }                                         
     \label{fig:ck-datasets-jpeg-d-reactions}
   \end{figure*}

Next, we can apply all most efficient compiler optimizations 
to a given program with all data sets.
Figure~\ref{fig:ck-datasets-jpeg-d-reactions} shows such reactions 
(ratio of an execution time with a given optimization to an execution time 
with the default -O3 compiler optimization) of a jpeg decoder across 
20 different jpeg images from the above MiDataSet on RPi3.

One can observe that the same combination of compiler flags can both
considerably improve or degrade execution time for the same program
but across different data sets.
For example, data sets 4,5,13,16 and 17 can benefit from the most
efficient combination of compiler flags found by the community
with speedups ranging from 1.2 to 1.7.
On the other hand, it's better to run all other data sets with
the default -O3 optimization level.

Unfortunately, finding data set and other features which could easily differentiate
above optimizations is often very challenging.
Even deep learning may not help if a feature is not yet exposed.
We explain this issue in~\cite{cm:29db2248aba45e59:cd11e3a188574d80}
when optimizing real B\&W filter kernel - we managed to improve
predictions by exposing a "time of the day" feature
only via human intervention.
However, yet again, the CK concept is to bring the interdisciplinary
community on board to share such cases in a reproducible way 
and then collaboratively find various features to improve predictions.

Another aspect which can influence the quality of predictive models,
is that the same combinations of compiler flags are too coarse-grain
and can make different internal optimization decisions 
for different programs.
Therefore, we need to have an access to fine-grain optimizations
(inlining, tiling, unrolling, vectorization, prefetching, etc)
and related features to continue improving our models.
However, this follows our top-down optimization and modeling methodology 
which we implemented in the Collective Knowledge framework.
We want first to analyze, optimize and model coarse-grain behavior of shared workloads
together with the community and students while gradually adding more workloads, 
data sets, models and platforms.
Only when we reached the limit of prediction accuracy, 
we start gradually exposing finer-grain optimizations 
and features via extensible CK JSON interface 
while avoiding explosion in design and optimization spaces
(see details in~\cite{fursin:hal-01054763} for our previous
version of the workflow framework, Collective Mind).
This is much in spirit of how physicists moved from Newton's 
three coarse-grain laws of motion to fine-grain quantum mechanics.

   \begin{figure*}[!htbp]
     \centering
      \includegraphics[width=5.5in]
      {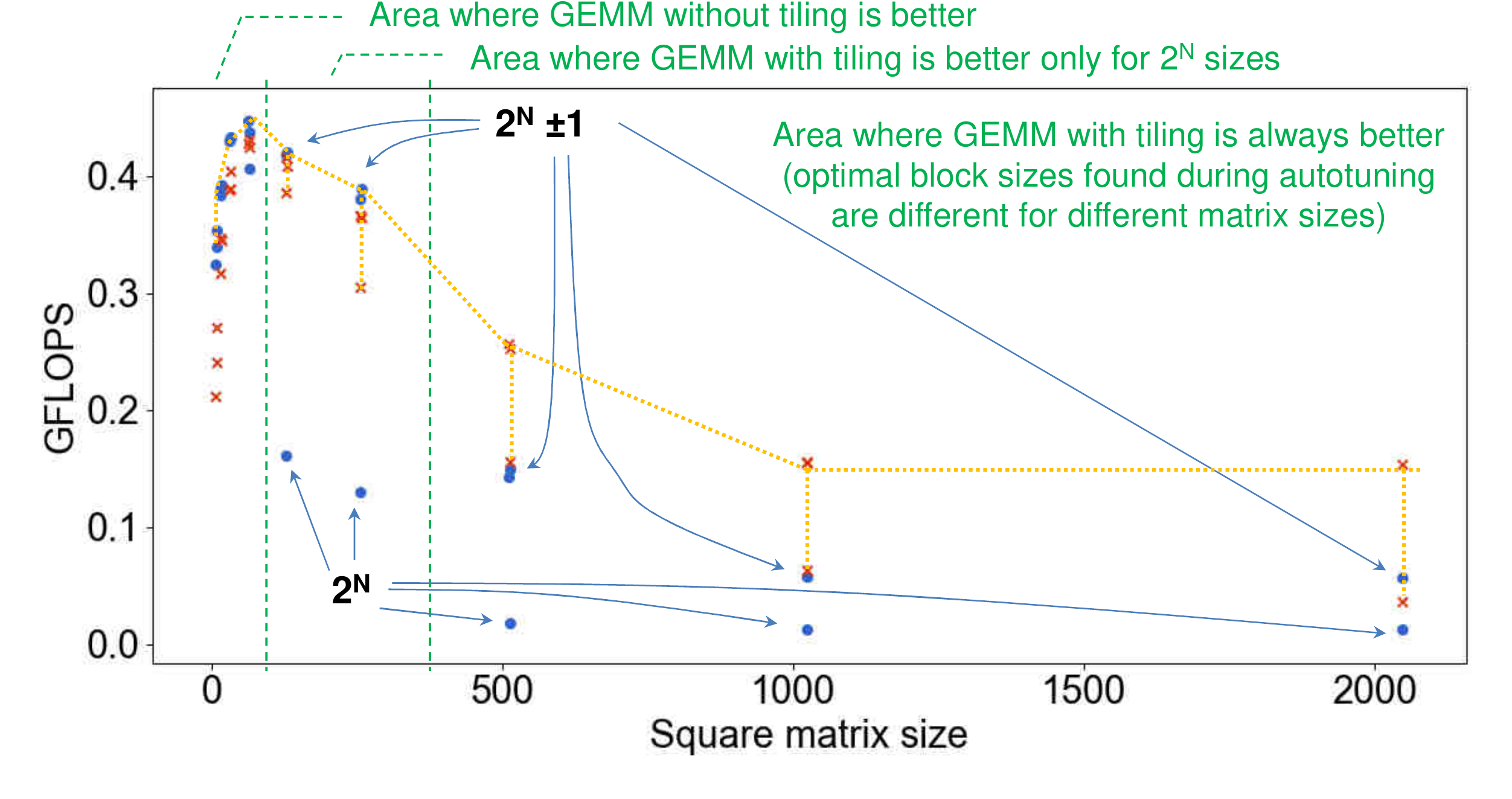} 
     \caption{
      Performance of a tiled matrix multiply in GFLOPS for different square matrix sizes. 
      Blue circles show performance of original (non-blocked) matrix multiply
      while red crosses show best performance found during autotuning on RPi3 device.
     }                                         
     \label{fig:ck-datasets-input-aware-autotuning}
   \end{figure*}

To demonstrate this approach, we shared a simple skeletonized 
matrix multiply kernel from~\cite{Fur2004} in the CK format 
with blocking (tiling) parameter and data set feature 
(square matrix size) exposed via CK API:

\begin{flushleft}
\texttt{\$ ck compile program:shared-matmul-c2 --flags="-DUSE\_BLOCKED\_MATMUL=YES}\newline
\texttt{\$ ck run program:shared-matmul-c2 --env.CT\_MATRIX\_DIMENSION=128 --env.CT\_BLOCK\_SIZE=16}\newline
\end{flushleft}

We can then reuse universal autotuning (exploration) strategies
available as CK modules or implement specialized ones to explore 
exposed fine-grain optimizations versus different data sets.
Figure~\ref{fig:ck-datasets-input-aware-autotuning} shows matmul performance
in GFLOPS during random exploration of a blocking parameter for different square 
matrix sizes on RPi3.
These results are in line with multiple past studies showing that
unblocked matmul is more efficient for small matrix sizes (less than 32
on RPi3) since all data fits cache, or between 32 and 512 (on RPi3) 
if they are not power of 2.
In contrast, the tiled matmul is better on RPi3 for matrix sizes of power of 2 between 32 and 512,
since it can help reduce cache conflict misses, and for all matrix sizes more than 512
where tiling can help optimize access to slow main memory.

   \begin{figure*}[!htbp]
     \centering
      \includegraphics[width=4in]
      {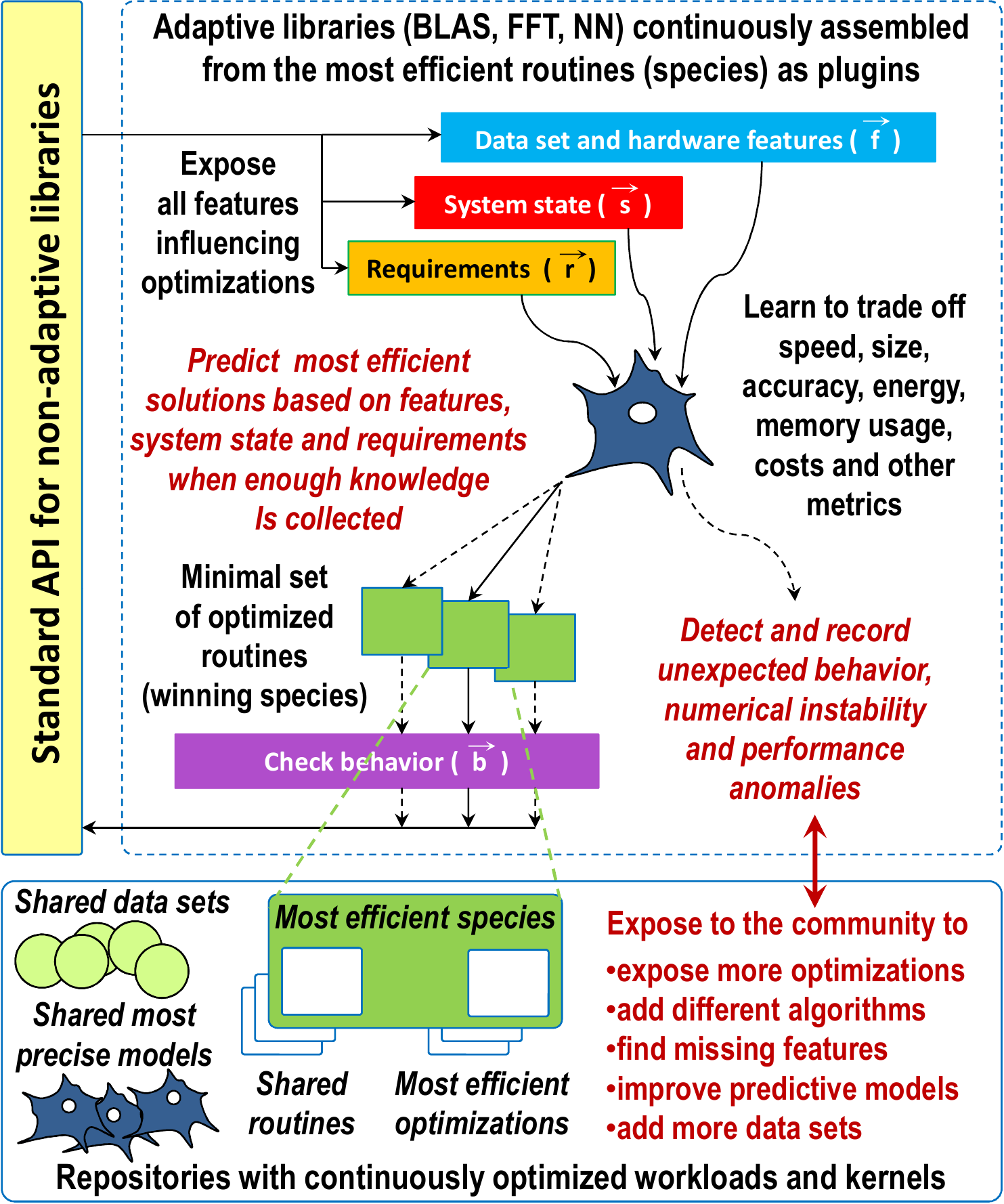} 
     \caption{
      Enabling adaptive and self-optimizing libraries assembled from the most efficient routines
      continuously optimized by the community across different platforms and data sets. 
      These routines are selected automatically at run-time based on platform, data set and other features.
     }
     \label{fig:ck-adaptive-systems}
   \end{figure*}

Our customizable workflow can help teach students how to build efficient,
adaptive and self-optimizing libraries including BLAS, neural networks and FFT.
Such libraries are assembled from the most efficient routines
found during continuous crowd-tuning across numerous data sets and platforms,
and combined with fast and automatically generated decision trees  
or other more precise classifiers~\cite{DBLP:journals/ijhpca/PuschelMSXJPVJ04,5160988,LCWP2009,cm:29db2248aba45e59:cd11e3a188574d80}.
The most efficient routines are then selected at run-time 
depending on data set, hardware and other features as conceptually shown
in Figure~\ref{fig:ck-adaptive-systems}..

\textit{All demo scripts to generate data and graphs in this section are available in the following CK entries:}
\begin{flushleft}
\texttt{\$ ck find script:rpi3-all-autotune-multiple-datasets}\newline
\texttt{\$ ck find script:rpi3-input-aware-autotune-blas}\newline
\end{flushleft}


\section{Reinventing computer engineering via reproducible competitions}
\label{sec:competitions}
Having a common and customizable workflow framework with "plug\&play" artifacts 
opens up another interesting opportunity for computer engineering.
Researchers can use it to compare and improve their techniques
(optimizations, models, algorithms, architectures) 
against each other via open and reproducible competitions
while being on the same page.

This is in spirit with existing machine learning competitions
such as Kaggle and ImageNet challenge~\cite{kaggle,imagenet-challenge} 
to improve prediction accuracy of various models.
The main difference is that we want to focus on 
optimizing the whole software/hardware/model stack 
while trading off multiple metrics including speed, accuracy,
and costs~\cite{request,cm:29db2248aba45e59:cd11e3a188574d80,lpirc}.

Experimental results from such competitions can be continuously aggregated 
and presented in the live Collective Knowledge scoreboard~\cite{live-ck-repo}.
Other academic and industrial researchers can then pay 
a specific attention to the "winning" techniques close 
to a Pareto frontier in a multi-dimensional 
space of accuracy, execution time, power/energy consumption, 
hardware/code/model footprint, monetary costs etc
thus speeding up technology transfer.
Furthermore, "winning" artifacts and workflows can now be recompiled,
reused and extended on the newer platforms with the latest
environment thus improving overall research sustainability.

   \begin{figure*}[!htbp]
     \centering
      \includegraphics[width=5.8in]
      {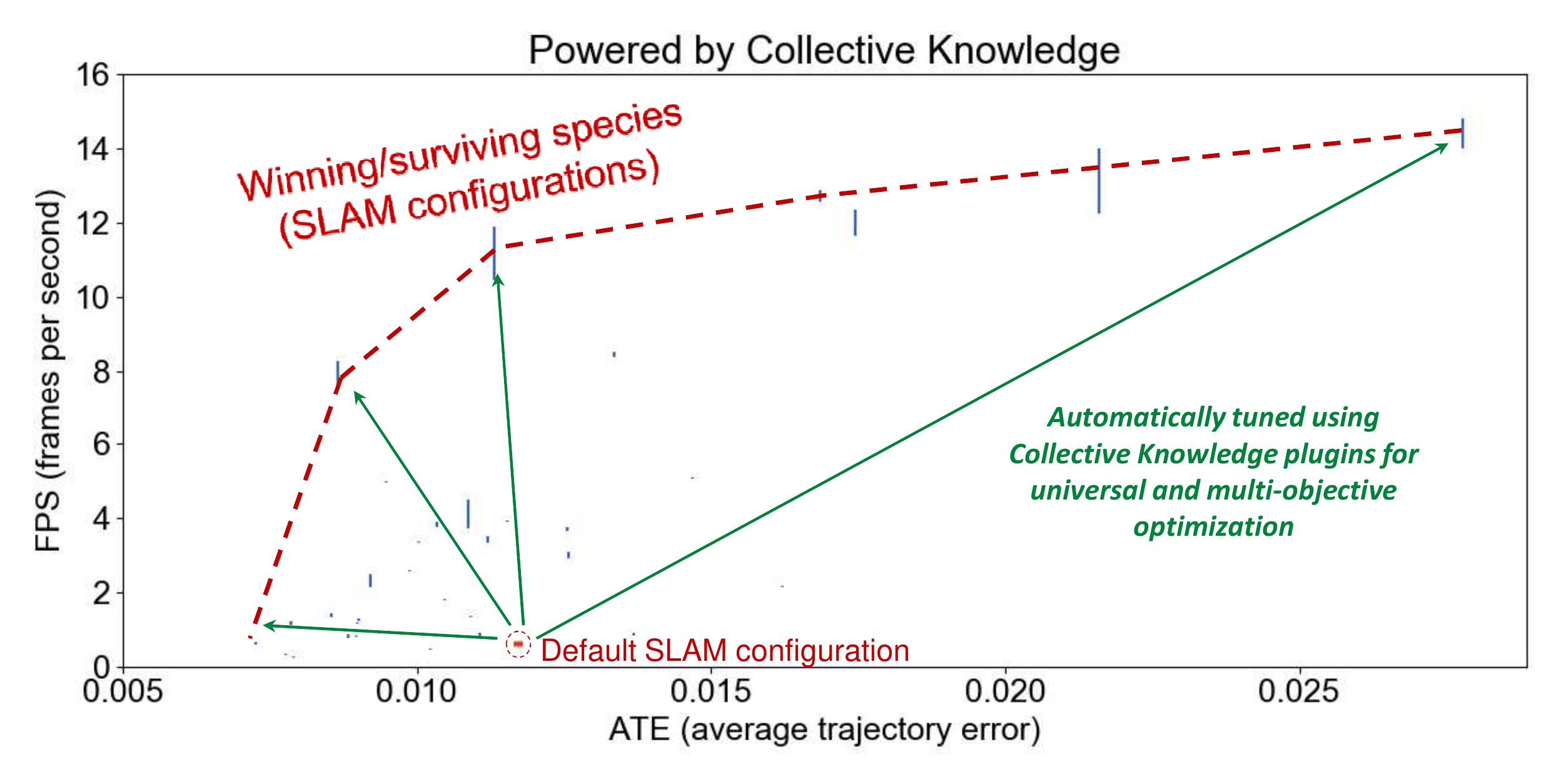} 
     \caption{
      Random exploration of various SLAM algorithms and their parameters (Simultaneous localization and mapping)
      in terms of accuracy (average trajectory error or ATE) versus speed 
      (frames per second) on RPi3 using CK.
     }                                         
     \label{fig:converting-ad-hoc-slambench-to-ck}
   \end{figure*}

   \begin{figure*}[!htbp]
     \centering
      \includegraphics[width=5in]
      {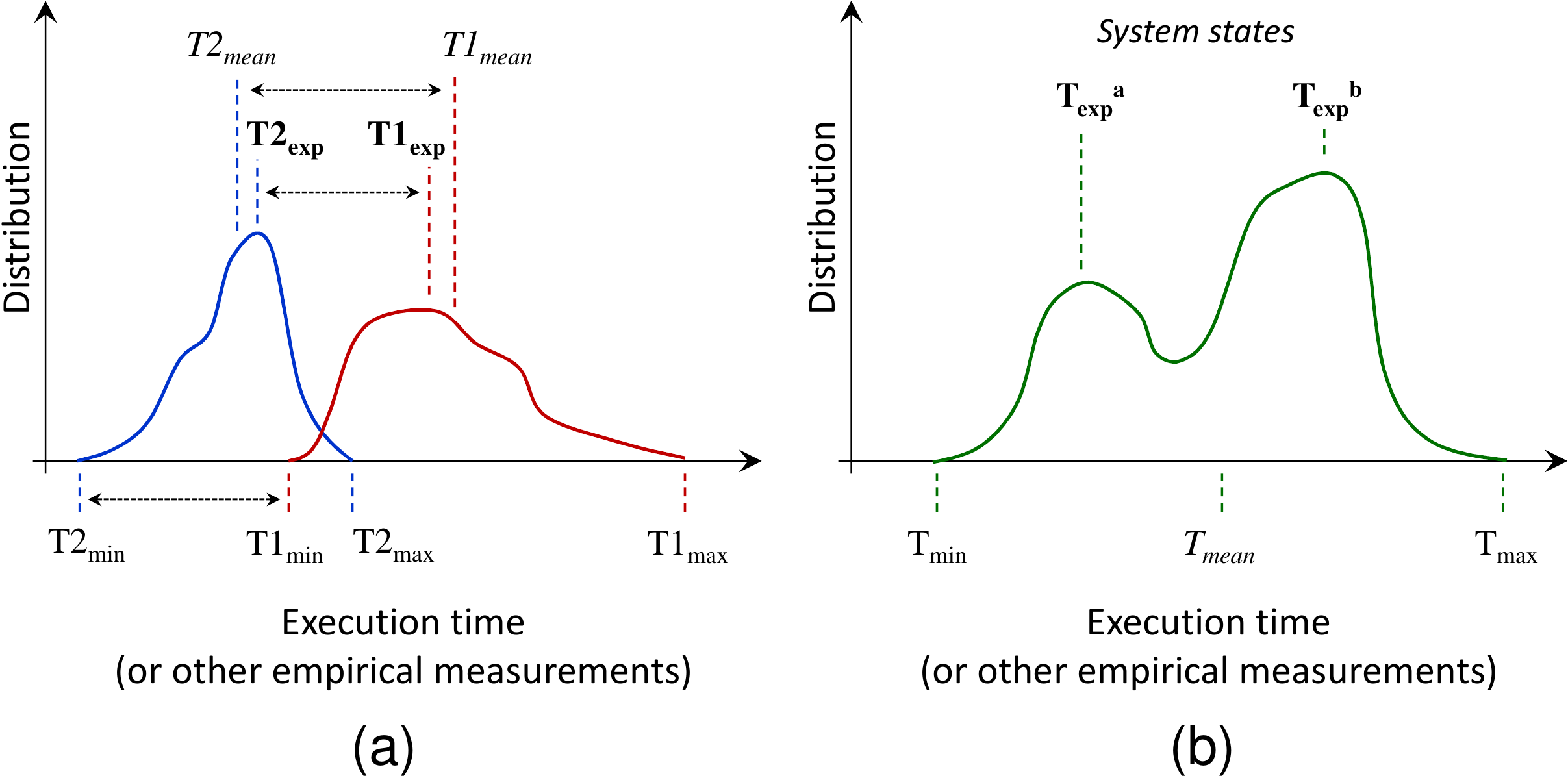} 
     \caption{
       Developing common evaluation methodology for empirical results in systems research:
       (a) Calculating speedups between two optimizations T1 and T2 using min, mean and expected values, and reporting max difference.
       (b) Reporting a problem when several system states are detected
     }
     \label{fig:reproducibility}
   \end{figure*}

For a proof-of-concept, we started helping some authors convert their 
artifacts and experimental workflows to the CK format during 
Artifact Evaluation~\cite{ck-shared-repos,Ainsworth:2017:SPI:3049832.3049865,ctuning-ae1}.
Association for Computing Machinery (ACM)~\cite{acm} 
also recently joined this effort funded by the Alfred P. Sloan Foundation 
to convert already published experimental workflows 
and artifacs from the ACM Digital Library 
to the CK format~\cite{Flick:2015:PCA:2807591.2807619}.

We can then reuse CK functionality to crowdsource benchmarking and multi-objective
autotuning of shared workloads across diverse data sets, models and platforms.
For example, Figure~\ref{fig:converting-ad-hoc-slambench-to-ck} shows 
results from random exploration of various SLAM algorithms (Simultaneous localization and mapping)
and their parameters from~\cite{slambench_paper} in terms of accuracy (average trajectory error or ATE)
versus speed (frames per second) on RPi3 using CK~\cite{ck-slambench-repo}. 
Researchers may easily spend 50\% of their time developing
experimental, benchmarking and autotuning infrastructure 
in such complex projects, and then continuously
updating it to adapt to ever changing software and hardware
instead of innovating.
Worse, such ad-hoc infrastructure may not even survive 
the end of the project or if leading developers leave project.

Using common and portable workflow framework can relieve researchers
from this burden and let them reuse already existing artifacts and
focus on innovation rather than re-developing ad-hoc software from scratch.
Other researchers can also pick up the winning designs on a Pareto frontier,
reproduce results via CK, try them on different platforms
and with different data sets, build upon them, 
and eventually try to develop more efficient algorithms.
Finally, researchers can implement a common experimental methodology 
to evaluate empirical results in systems research similar to physics 
within a common workflow framework rather than writing their own
ad-hoc scripts.
Figure~\ref{fig:reproducibility} shows statistical analysis 
of experimental results implemented in the CK to compare different
optimizations depending on research scenarios.
For example, we report minimal execution time from multiple experiments
to understand the limits of a given architecture, expected value to see
how a given workload performs on average, and max time to detect
abnormal behavior.
If more than one expected value is detected, it usually means
that system was in several different run-time states during experiments
(often related to adaptive changes in CPU and GPU frequency due to DVFS)
and extra analysis is required.

\begin{figure*}[!htbp]
  \centering
   \includegraphics[width=4.5in]
   {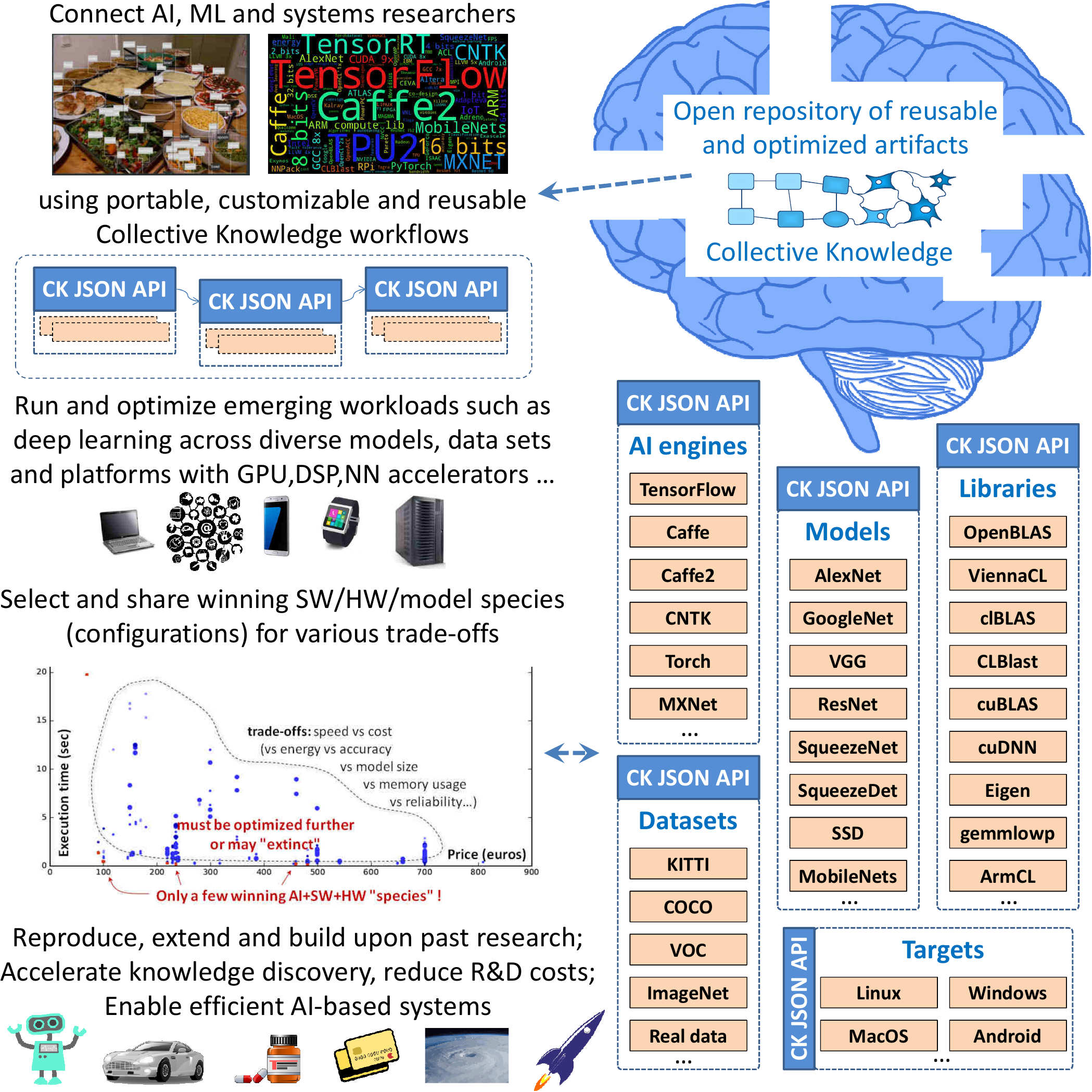} 
  \caption{
    Collective Knowledge framework as an open platform to support software/hardware/model
    co-design tournaments for Pareto-efficient deep learning
    and other emerging workloads in terms of speed, accuracy, 
    energy and various costs.
  }
  \label{fig:reproducible-tournaments}
\end{figure*}

We now plan to validate our Collective Knowledge approach
in the 1st reproducible ReQuEST tournament 
at the ACM ASPLOS'18 conference~\cite{request}
as presented in Figure~\ref{fig:reproducible-tournaments}.
ReQuEST is aimed at providing a scalable tournament framework, 
a common experimental methodology and an open repository for continuous evaluation 
and optimization of the quality vs.\ efficiency Pareto optimality of a wide range 
of real-world applications, libraries, and models across the whole 
hardware/software stack on complete platforms. 
ReQuEST also promote reproducibility of experimental results and reusability/customization 
of systems research artifacts by standardizing evaluation methodologies and facilitating 
the deployment of efficient solutions on heterogeneous platforms. 

   \begin{figure*}[!htbp]
     \centering
      \includegraphics[width=5in]
      {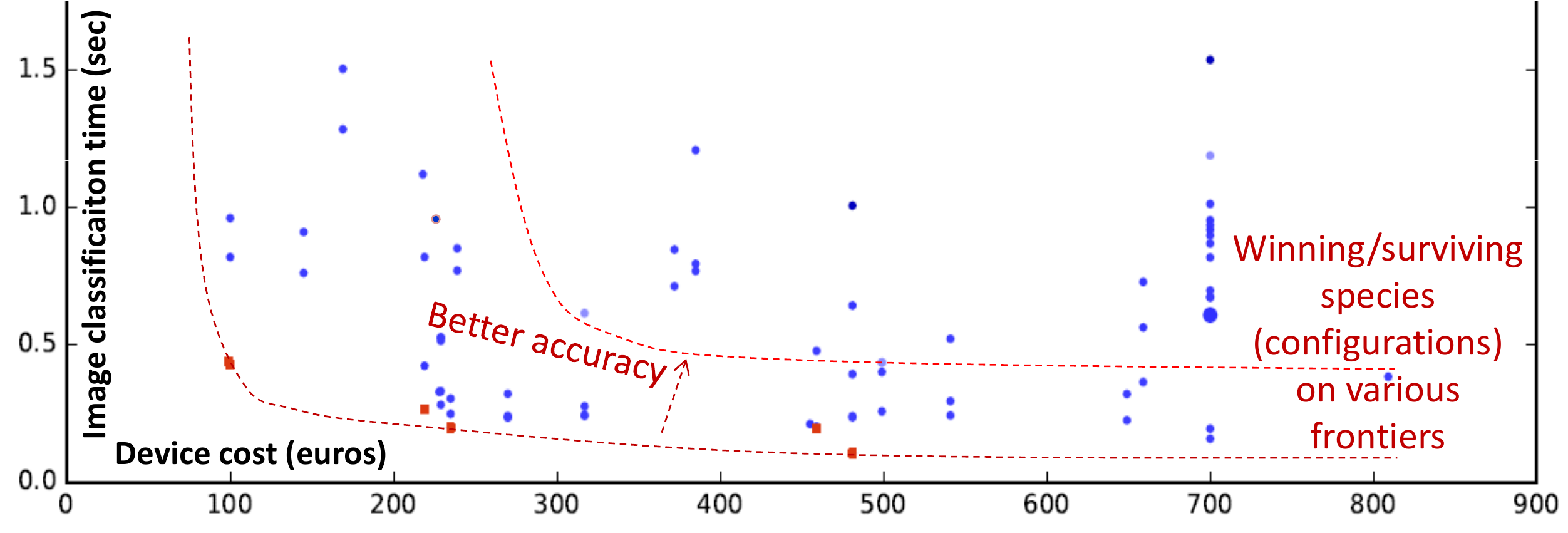} 
     \caption{
      An example of a live Collective Knowledge scoreboard to crowd-benchmark
      inference in terms of speed, accuracy and platform cost 
      across diverse deep learning frameworks, models, data sets, and 
      Android devices provided by volunteers. Red dots 
      are associated with the winning workflows (model/software/hardware)
      on different frontiers.
     }
     \vspace{-1em}
     \label{fig:dnn-crowdtuning-example}
   \end{figure*}

ReQuEST will use CK and our artifact evaluation methodology~\cite{ctuning-ae1} 
to provide unified evaluation and a live scoreboard of submissions. 
Figure~\ref{fig:dnn-crowdtuning-example} shows a proof-of-concept example of such a
scoreboard powered by CK to collaboratively benchmark inference (speed vs.\ platform cost) 
across diverse deep learning frameworks (TensorFlow, Caffe, MXNet, etc.), 
models (AlexNet, GoogleNet, SqueezeNet, ResNet, etc.), real user data sets, and mobile devices 
provided by volunteers (see the latest results at \href{http://cknowledge.org/repo}{cKnowledge.org/repo}).
Our goal is to teach students and researchers how to 
\begin{itemize}
  \item release research artifacts of their on-going or accomplished research
as portable and reusable components, standardize evaluation workflows, 
and facilitate deployment and tech transfer of state-of-the-art research,
  \item continuously optimize various algorithms
across diverse models, data sets and platforms in terms of speed, accuracy,
size, energy usage and other costs,
  \item build upon each others' work to develop the next generation 
of efficient software and hardware stack for emerging workloads.
\end{itemize}


\section{Conclusions and Future Work}
\label{sec:conclusions}
Researchers are now in a race to bring artificial intelligence to all possible
devices from IoT to supercomputers which will require 
much more efficient software and hardware then currently available.
At the same time, computer engineers have already been struggling for many years 
to develop efficient sub-components of computer systems including
algorithms, compilers and run-time systems.

The major issues including raising complexity, lack of a common experimental framework 
and lack of practical knowledge exchange between academia and industry.
Rather than innovating, researchers have to spend more and more time 
writing their own, ad-hoc and not easily customizable support tools 
to perform experiments such as multi-objective autotuning.

We presented our long-term educational initiative to teach
students and researchers how to solve the above problems 
using customizable workflow frameworks similar to other sciences.
We showed how to convert ad-hoc, multi-objective
and multi-dimensional autotuning into a portable and customizable workflow 
based on open-source Collective Knowledge workflow framework.
We then demonstrated how to use it to implement various scenarios
such as compiler flag autotuning of benchmarks and realistic workloads
across Raspberry Pi 3 devices in terms of speed and size.
We also demonstrated how to crowdsource such autotuning across different
devices provided by volunteers similar to SETI@home, collect the most efficient optimizations
in a reproducible way in a public repository of knowledge at ~\href{http://cknowledge.org/repo}{cKnowledge.org/repo}, 
apply various machine learning techniques including decision trees, the nearest neighbor classifier
and deep learning to predict the most efficient optimizations for previously
unseen workloads, and then continue improving models and features
as a community effort.
We now plan to develop an open web platform together with the community
to provide a user-friendly front-end to all presented workflows 
while hiding all complexity.

We use our methodology and open-source CK workflow framework and repository
to teach students how to exchange their research artifacts and results 
as reusable components with a a unified API and meta-information,
perform collaborative experiments, automate Artifact Evaluation
at journals and conferences~\cite{ctuning-ae1}, build upon each others' work,
make their research more reproducible and sustainable, 
and eventually accelerate transfer of their ideas to industry.
Students and researchers can later use such skills and unified artifacts
to participate in our open ReQuEST tournaments on reproducible and Pareto-efficient
co-design of the whole software and hardware stack for emerging workloads
such as deep learning and quantum computing in terms of
speed, accuracy, energy and costs~\cite{request}.
This, in turn, should help the community build an open repository of 
portable, reusable and customizable algorithms continuously optimized
across diverse platforms, models and data sets
to assemble efficient computer systems
and accelerate innovation.


\section{Acknowledgments}
\label{sec:ack}

We would like to thank Raspberry Pi foundation for initial financial support.
We are also grateful to dividiti and cTuning foundation colleagues, 
Flavio Vella, Marco Cianfriglia, Nikolay Chunosov, Daniil Efremov, Yuriy Kashnikov, 
Peter Green, Thierry Moreau and ReQuEST colleagues, and the Collective Knowledge community 
for evaluating Collective Knowledge concepts and providing useful feedback.



\newpage

\bibliographystyle{abbrv}
\bibliography{paper}

\newcommand{\noop}[1]{}
\begin{thebibliography}{100}

\bibitem{acm}
{Association for Computing Machinery (ACM)}.
\newblock \url{http://www.acm.org}.

\bibitem{ck-concepts}
{{Blog article}: CK concepts by Michel Steuwer}.
\newblock \url{http://michel.steuwer.info/About-CK}.

\bibitem{ck-slambench-repo}
{Collective Knowledge workflows for SLAMBench}.
\newblock \url{https://github.com/ctuning/reproduce-pamela-project}.

\bibitem{ctuning1}
{{cTuning.org}: public portal for collaborative and reproducible computer
  engineering}.
\newblock \url{http://cTuning.org}.

\bibitem{docker}
{Docker:} open source lightweight container technology that can run processes
  in isolation.
\newblock \url{http://www.docker.org}.

\bibitem{json-org}
Introducing {JSON}.
\newblock http://www.json.org.

\bibitem{milepost}
{MILEPOST project archive (MachIne Learning for Embedded PrOgramS
  opTimization)}.
\newblock \url{http://cTuning.org/project-milepost}.

\bibitem{live-ck-repo}
Open collective knowledge repository with shared optimization results from
  crowdsourced experiments across diverse platforms and data sets.
\newblock \url{http://cKnowledge.org/repo}.

\bibitem{prace}
{PRACE:} partnership for advanced computing in europe.
\newblock http://www.prace-project.eu.

\bibitem{live-ck-repo-rpi-gcc710}
Public optimization results when crowd-tuning gcc 7.1.0 across raspberry pi3
  devices.
\newblock
  \href{http://cKnowledge.org/repo/web.php?wcid=8289e0cf24346aa7:79bca2b76876b5c6}{link}.

\bibitem{ck-shared-repos}
{Public repositories with artifact and workflows in the Collective Knowledge
  format}.
\newblock \url{http://cKnowledge.org/shared-repos}.

\bibitem{all-milepost-features}
{Static Features available in MILEPOST GCC V2.1}.
\newblock
  \url{http://ctuning.org/wiki/index.php/CTools:MilepostGCC:StaticFeatures:MILEPOST_V2.1}.

\bibitem{doi}
{Digital Object Identifier or DOI} - a persistent identifier or handle used to
  uniquely identify objects, standardized by the iso.
\newblock \url{http://doi.org}, 2000.

\bibitem{imagenet-challenge}
{Imagenet challenge (ILSVRC)}: Imagenet large scale visual recognition
  challenge where software programs compete to correctly classify and detect
  objects and scenes.
\newblock \url{http://www.image-net.org}, 2010.

\bibitem{kaggle}
{Kaggle:} platform for predictive modelling and analytics competitions.
\newblock \url{https://www.kaggle.com}, 2010.

\bibitem{figshare}
{Figshare} - online digital repository where researchers can preserve and share
  their research outputs, including figures, datasets, images, and videos.
\newblock \url{http://figshare.com}, 2011.

\bibitem{zenodo}
{Zenodo} - research data repository.
\newblock \url{http://zenodo.org}, 2013.

\bibitem{trust2014}
Proceedings of the 1st workshop on reproducible research methodologies and new
  publication models in computer engineering (acm sigplan trust'14).
\newblock ACM, 2014.

\bibitem{tetracom}
{TETRACOM} - eu fp7 project to support technology transfer in computing
  systems.
\newblock \url{https://www.tetracom.eu}, 2014.

\bibitem{ctuning-ae1}
{Artifact Evaluation for Computer Systems Conferences including CGO,PPoPP,PACT
  and SuperComputing}: developing common experimental methodology and tools for
  reproducible and sustainable research.
\newblock \url{http://cTuning.org/ae}, 2014-cur.

\bibitem{ae}
Artifact evaluation for computer systems research.
\newblock \url{http://ctuning.org/ae}, 2014-cur.

\bibitem{ck-env}
Ck repository with multi-platform software and package manager implemented as
  ck modules which detect or install various software (compilers, libraries,
  tools).
\newblock \url{https://github.com/ctuning/ck-env}, 2015.

\bibitem{ck-clsmith}
Ck workflow for opencl crowd-fuzzing.
\newblock \url{https://github.com/ctuning/ck-clsmith}, 2015.

\bibitem{lpirc}
{LPIRC:} low-power image recognition challenge.
\newblock \url{https://rebootingcomputing.ieee.org/lpirc}, 2015.

\bibitem{ck}
{Collective Knowledge}: open-source, customizable and cross-platform workflow
  framework and repository for computer systems research.
\newblock \url{https://github.com/ctuning/ck}, 2016.

\bibitem{oss-fuzz}
Continuous fuzzing of open source software.
\newblock \url{https://github.com/google/oss-fuzz}, 2016.

\bibitem{microsoft-fuzz}
Microsoft security risk detection.
\newblock \url{https://www.microsoft.com/en-us/security-risk-detection/}, 2016.

\bibitem{cgo17-artifact}
{Artifacts and experimental workflows in the Collective Knowledge Format for
  the CGO'17 paper "Software Prefetching for Indirect Memory Accesses"}.
\newblock \url{https://github.com/SamAinsworth/reproduce-cgo2017-paper}, 2017.

\bibitem{hipeac_roadmap2017}
The {HiPEAC} vision on high-performance and embedded architecture and
  compilation (2012-2020).
\newblock http://www.hipeac.net/roadmap, 2017.

\bibitem{openbenchmarking}
{Open benchmarking: automated testing \& benchmarking on an open platform}.
\newblock \url{http://openbenchmarking.org}, 2017.

\bibitem{request}
{ReQuEST}: open tournaments on collaborative, reproducible and pareto-efficient
  software/hardware co-design of emerging workloads such as deep learning using
  collective knowledge technology.
\newblock \url{http://cKnowledge.org/request}, 2017.

\bibitem{europar97x}
B.~Aarts and et.al.
\newblock {OCEANS}: Optimizing compilers for embedded applications.
\newblock In {\em Proc. Euro-Par 97}, volume 1300 of {\em Lecture Notes in
  Computer Science}, pages 1351--1356, 1997.

\bibitem{DBLP:journals/corr/AbadiABBCCCDDDG16}
M.~Abadi, A.~Agarwal, P.~Barham, E.~Brevdo, Z.~Chen, C.~Citro, G.~S. Corrado,
  A.~Davis, J.~Dean, M.~Devin, S.~Ghemawat, I.~J. Goodfellow, A.~Harp,
  G.~Irving, M.~Isard, Y.~Jia, R.~J{\'{o}}zefowicz, L.~Kaiser, M.~Kudlur,
  J.~Levenberg, D.~Man{\'{e}}, R.~Monga, S.~Moore, D.~G. Murray, C.~Olah,
  M.~Schuster, J.~Shlens, B.~Steiner, I.~Sutskever, K.~Talwar, P.~A. Tucker,
  V.~Vanhoucke, V.~Vasudevan, F.~B. Vi{\'{e}}gas, O.~Vinyals, P.~Warden,
  M.~Wattenberg, M.~Wicke, Y.~Yu, and X.~Zheng.
\newblock Tensorflow: Large-scale machine learning on heterogeneous distributed
  systems.
\newblock {\em CoRR}, abs/1603.04467, 2016.

\bibitem{DBLP:conf/supercomputer/AbdelfattahHTD16}
A.~Abdelfattah, A.~Haidar, S.~Tomov, and J.~J. Dongarra.
\newblock Performance, design, and autotuning of batched {GEMM} for gpus.
\newblock In {\em High Performance Computing - 31st International Conference,
  {ISC} High Performance 2016, Frankfurt, Germany, June 19-23, 2016,
  Proceedings}, pages 21--38, 2016.

\bibitem{ABCP06}
F.~Agakov, E.~Bonilla, J.Cavazos, B.Franke, G.~Fursin, M.~O'Boyle, J.~Thomson,
  M.~Toussaint, and C.~Williams.
\newblock Using machine learning to focus iterative optimization.
\newblock In {\em Proceedings of the International Symposium on Code Generation
  and Optimization (CGO)}, 2006.

\bibitem{Ainsworth:2017:SPI:3049832.3049865}
S.~Ainsworth and T.~M. Jones.
\newblock Software prefetching for indirect memory accesses.
\newblock In {\em Proceedings of the 2017 International Symposium on Code
  Generation and Optimization}, CGO '17, pages 305--317, Piscataway, NJ, USA,
  2017. IEEE Press.

\bibitem{Anderson:2002:SEP:581571.581573}
D.~P. Anderson, J.~Cobb, E.~Korpela, M.~Lebofsky, and D.~Werthimer.
\newblock Seti@home: An experiment in public-resource computing.
\newblock {\em Commun. ACM}, 45(11):56--61, Nov. 2002.

\bibitem{Ansel:2009:PLC:1542476.1542481}
J.~Ansel, C.~Chan, Y.~L. Wong, M.~Olszewski, Q.~Zhao, A.~Edelman, and
  S.~Amarasinghe.
\newblock Petabricks: a language and compiler for algorithmic choice.
\newblock In {\em Proceedings of the 2009 ACM SIGPLAN conference on Programming
  language design and implementation}, PLDI '09, pages 38--49, New York, NY,
  USA, 2009. ACM.

\bibitem{ansel:pact:2014}
J.~Ansel, S.~Kamil, K.~Veeramachaneni, J.~Ragan-Kelley, J.~Bosboom, U.-M.
  O'Reilly, and S.~Amarasinghe.
\newblock Opentuner: An extensible framework for program autotuning.
\newblock In {\em International Conference on Parallel Architectures and
  Compilation Techniques}, Edmonton, Canada, August 2014.

\bibitem{citeulike:1671417}
K.~Asanovic, R.~Bodik, B.~C. Catanzaro, J.~J. Gebis, P.~Husbands, K.~Keutzer,
  D.~A. Patterson, W.~L. Plishker, J.~Shalf, S.~W. Williams, and K.~A. Yelick.
\newblock {The landscape of parallel computing research: a view from Berkeley}.
\newblock Technical Report UCB/EECS-2006-183, Electrical Engineering and
  Computer Sciences, University of California at Berkeley, Dec. 2006.

\bibitem{Ashouri:2017:MMC:3132652.3124452}
A.~H. Ashouri, A.~Bignoli, G.~Palermo, C.~Silvano, S.~Kulkarni, and J.~Cavazos.
\newblock Micomp: Mitigating the compiler phase-ordering problem using
  optimization sub-sequences and machine learning.
\newblock {\em ACM Trans. Archit. Code Optim.}, 14(3):29:1--29:28, Sept. 2017.

\bibitem{ashouri2016cobayn}
A.~H. Ashouri, G.~Mariani, G.~Palermo, E.~Park, J.~Cavazos, and C.~Silvano.
\newblock Cobayn: Compiler autotuning framework using bayesian networks.
\newblock {\em ACM Transactions on Architecture and Code Optimization (TACO)},
  13(2):21, 2016.

\bibitem{BCCP2008}
Bailey and et.al.
\newblock Peri auto-tuning.
\newblock {\em Journal of Physics: Conference Series (SciDAC 2008)}, 125:1--6,
  2008.

\bibitem{1742-6596-125-1-012089}
D.~Bailey, J.~Chame, C.~Chen, J.~Dongarra, M.~Hall, J.~Hollingsworth,
  P.~Hovland, S.~Moore, K.~Seymour, J.~Shin, A.~Tiwari, S.~Williams, and
  H.~You.
\newblock {PERI auto-tuning}.
\newblock {\em Journal of Physics: Conference Series}, 125(1):012089, 2008.

\bibitem{citeulike:873540}
C.~M. Bishop.
\newblock {\em {Pattern Recognition and Machine Learning (Information Science
  and Statistics)}}.
\newblock Springer, 1st ed. 2006. corr. 2nd printing 2011 edition, Oct. 2007.

\bibitem{CGJ1997}
B.~Calder, D.~Grunwald, M.~Jones, D.~Lindsay, J.~Martin, M.~Mozer, and B.~Zorn.
\newblock Evidence-based static branch prediction using machine learning.
\newblock {\em ACM Transactions on Programming Languages and Systems (TOPLAS)},
  1997.

\bibitem{CFAP2007}
J.~Cavazos, G.~Fursin, F.~Agakov, E.~Bonilla, M.~O'Boyle, and O.~Temam.
\newblock Rapidly selecting good compiler optimizations using performance
  counters.
\newblock In {\em Proceedings of the International Symposium on Code Generation
  and Optimization (CGO)}, March 2007.

\bibitem{childers2016artifact}
B.~R. Childers, G.~Fursin, S.~Krishnamurthi, and A.~Zeller.
\newblock {Artifact Evaluation for Publications (Dagstuhl Perspectives Workshop
  15452)}.
\newblock 5(11), 2016.

\bibitem{CSS99}
K.~Cooper, P.~Schielke, and D.~Subramanian.
\newblock Optimizing for reduced code space using genetic algorithms.
\newblock In {\em Proceedings of the Conference on Languages, Compilers, and
  Tools for Embedded Systems (LCTES)}, pages 1--9, 1999.

\bibitem{Tapus:2002:AHT:762761.762771}
C.~\c{T}\u{a}pu\c{s}, I.-H. Chung, and J.~K. Hollingsworth.
\newblock Active harmony: towards automated performance tuning.
\newblock In {\em Proceedings of the 2002 ACM/IEEE conference on
  Supercomputing}, Supercomputing '02, pages 1--11, Los Alamitos, CA, USA,
  2002. IEEE Computer Society Press.

\bibitem{DBLP:conf/IEEEpact/CumminsP0L17}
C.~Cummins, P.~Petoumenos, Z.~Wang, and H.~Leather.
\newblock End-to-end deep learning of optimization heuristics.
\newblock In {\em 26th International Conference on Parallel Architectures and
  Compilation Techniques, {PACT} 2017, Portland, OR, USA, September 9-13,
  2017}, pages 219--232, 2017.

\bibitem{Dongarra:2011:IES:1943326.1943339}
J.~Dongarra~et.al.
\newblock The international exascale software project roadmap.
\newblock {\em Int. J. High Perform. Comput. Appl.}, 25(1):3--60, Feb. 2011.

\bibitem{DJBP2009}
C.~Dubach, T.~M. Jones, E.~V. Bonilla, G.~Fursin, and M.~F. O'Boyle.
\newblock Portable compiler optimization across embedded programs and
  microarchitectures using machine learning.
\newblock In {\em Proceedings of the IEEE/ACM International Symposium on
  Microarchitecture (MICRO)}, December 2009.

\bibitem{Duran:1981:RRT:800078.802530}
J.~W. Duran and S.~Ntafos.
\newblock A report on random testing.
\newblock In {\em Proceedings of the 5th International Conference on Software
  Engineering}, ICSE '81, pages 179--183, Piscataway, NJ, USA, 1981. IEEE
  Press.

\bibitem{Flick:2015:PCA:2807591.2807619}
P.~Flick, C.~Jain, T.~Pan, and S.~Aluru.
\newblock A parallel connectivity algorithm for de bruijn graphs in metagenomic
  applications.
\newblock In {\em Proceedings of the International Conference for High
  Performance Computing, Networking, Storage and Analysis}, SC '15, pages
  15:1--15:11, New York, NY, USA, 2015. ACM.

\bibitem{FOTP2005}
B.~Franke, M.~O'Boyle, J.~Thomson, and G.~Fursin.
\newblock Probabilistic source-level optimisation of embedded programs.
\newblock In {\em Proceedings of the Conference on Languages, Compilers, and
  Tools for Embedded Systems (LCTES)}, 2005.

\bibitem{Fur2004}
G.~Fursin.
\newblock {\em Iterative Compilation and Performance Prediction for Numerical
  Applications}.
\newblock PhD thesis, University of Edinburgh, United Kingdom, 2004.

\bibitem{Fur2009}
G.~Fursin.
\newblock {Collective Tuning Initiative}: automating and accelerating
  development and optimization of computing systems.
\newblock In {\em Proceedings of the GCC Developers' Summit}, June 2009.

\bibitem{FCOP2007}
G.~Fursin, J.~Cavazos, M.~O'Boyle, and O.~Temam.
\newblock {MiDataSets}: Creating the conditions for a more realistic evaluation
  of iterative optimization.
\newblock In {\em Proceedings of the International Conference on High
  Performance Embedded Architectures \& Compilers (HiPEAC 2007)}, January 2007.

\bibitem{new_pub_model}
G.~Fursin and C.~Dubach.
\newblock Community-driven reviewing and validation of publications.
\newblock In {\em Proceedings of the 1st Workshop on Reproducible Research
  Methodologies and New Publication Models in Computer Engineering (ACM SIGPLAN
  TRUST'14)}. ACM, 2014.

\bibitem{29db2248aba45e59:a31e374796869125}
G.~Fursin, Y.~Kashnikov, A.~W. Memon, Z.~Chamski, O.~Temam, M.~Namolaru,
  E.~Yom-Tov, B.~Mendelson, A.~Zaks, E.~Courtois, F.~Bodin, P.~Barnard,
  E.~Ashton, E.~Bonilla, J.~Thomson, C.~Williams, and M.~F.~P. O'Boyle.
\newblock Milepost gcc: Machine learning enabled self-tuning compiler.
\newblock {\em International Journal of Parallel Programming (IJPP)},
  39:296--327, 2011.
\newblock 10.1007/s10766-010-0161-2.

\bibitem{ck-date16}
G.~Fursin, A.~Lokhmotov, and E.~Plowman.
\newblock {Collective Knowledge}: towards {R\&D} sustainability.
\newblock In {\em Proceedings of the Conference on Design, Automation and Test
  in Europe (DATE'16)}, March 2016.

\bibitem{fursin_lokhmotov_upton_2018}
G.~Fursin, A.~Lokhmotov, and E.~Upton.
\newblock Collective knowledge repository with reproducible experimental
  results from collaborative program autotuning on raspberry pi (program
  reactions to most efficient compiler optimizations).
\newblock https://doi.org/10.6084/m9.{f}{i}gshare.5789007.v2, Jan 2018.

\bibitem{cm:29db2248aba45e59:cd11e3a188574d80}
G.~Fursin, A.~Memon, C.~Guillon, and A.~Lokhmotov.
\newblock {Collective Mind, Part II}: Towards performance- and cost-aware
  software engineering as a natural science.
\newblock In {\em 18th International Workshop on Compilers for Parallel
  Computing (CPC'15)}, January 2015.

\bibitem{fursin:hal-01054763}
G.~Fursin, R.~Miceli, A.~Lokhmotov, M.~Gerndt, M.~Baboulin, D.~Malony, Allen,
  Z.~Chamski, D.~Novillo, and D.~D. Vento.
\newblock {Collective Mind}: Towards practical and collaborative auto-tuning.
\newblock {\em Scientific Programming}, 22(4):309--329, July 2014.

\bibitem{FOK02}
G.~Fursin, M.~O'Boyle, and P.~Knijnenburg.
\newblock Evaluating iterative compilation.
\newblock In {\em Proceedings of the Workshop on Languages and Compilers for
  Parallel Computers (LCPC)}, pages 305--315, 2002.

\bibitem{FOTP04}
G.~Fursin, M.~O'Boyle, O.~Temam, and G.~Watts.
\newblock Fast and accurate method for determining a lower bound on execution
  time.
\newblock {\em Concurrency: Practice and Experience}, 16(2-3):271--292, 2004.

\bibitem{Gamblin:2015:SPM:2807591.2807623}
T.~Gamblin, M.~LeGendre, M.~R. Collette, G.~L. Lee, A.~Moody, B.~R.
  de~Supinski, and S.~Futral.
\newblock The spack package manager: Bringing order to hpc software chaos.
\newblock In {\em Proceedings of the International Conference for High
  Performance Computing, Networking, Storage and Analysis}, SC '15, pages
  40:1--40:12, New York, NY, USA, 2015. ACM.

\bibitem{Grauer-Gray2012-hn}
S.~Grauer-Gray, L.~Xu, R.~Searles, S.~Ayalasomayajula, and J.~Cavazos.
\newblock Auto-tuning a high-level language targeted to {GPU} codes.
\newblock In {\em Innovative Parallel Computing ({InPar)}, 2012}, pages 1--10,
  May 2012.

\bibitem{DBLP:conf/cgo/GreweWO13}
D.~Grewe, Z.~Wang, and M.~F.~P. O'Boyle.
\newblock Portable mapping of data parallel programs to opencl for
  heterogeneous systems.
\newblock In {\em Proceedings of the 2013 {IEEE/ACM} International Symposium on
  Code Generation and Optimization, {CGO} 2013, Shenzhen, China, February
  23-27, 2013}, pages 22:1--22:10, 2013.

\bibitem{Hall:2009:CRN:1461928.1461946}
M.~Hall, D.~Padua, and K.~Pingali.
\newblock Compiler research: The next 50 years.
\newblock {\em Commun. ACM}, 52(2):60--67, Feb. 2009.

\bibitem{DBLP:conf/ipps/HartonoNS09}
A.~Hartono, B.~Norris, and P.~Sadayappan.
\newblock Annotation-based empirical performance tuning using orio.
\newblock In {\em 23rd {IEEE} International Symposium on Parallel and
  Distributed Processing, {IPDPS} 2009, Rome, Italy, May 23-29, 2009}, pages
  1--11, 2009.

\bibitem{HE2008}
K.~Hoste and L.~Eeckhout.
\newblock Cole: Compiler optimization level exploration.
\newblock In {\em Proceedings of the International Symposium on Code Generation
  and Optimization (CGO)}, 2008.

\bibitem{DBLP:conf/sc/HosteTGW12}
K.~Hoste, J.~Timmerman, A.~Georges, and S.~D. Weirdt.
\newblock Easybuild: Building software with ease.
\newblock In {\em 2012 {SC} Companion: High Performance Computing, Networking
  Storage and Analysis, Salt Lake City, UT, USA, November 10-16, 2012}, pages
  572--582, 2012.

\bibitem{JGVP2009}
V.~Jimenez, I.~Gelado, L.~Vilanova, M.~Gil, G.~Fursin, and N.~Navarro.
\newblock Predictive runtime code scheduling for heterogeneous architectures.
\newblock In {\em Proceedings of the International Conference on High
  Performance Embedded Architectures \& Compilers (HiPEAC 2009)}, January 2009.

\bibitem{Khan:2013:SAC:2400682.2400690}
M.~Khan, P.~Basu, G.~Rudy, M.~Hall, C.~Chen, and J.~Chame.
\newblock A script-based autotuning compiler system to generate
  high-performance cuda code.
\newblock {\em ACM Trans. Archit. Code Optim.}, 9(4):31:1--31:25, Jan. 2013.

\bibitem{KKO2000}
T.~Kisuki, P.~Knijnenburg, and M.~O'Boyle.
\newblock Combined selection of tile sizes and unroll factors using iterative
  compilation.
\newblock In {\em Proceedings of the International Conference on Parallel
  Architectures and Compilation Techniques (PACT)}, pages 237--246, 2000.

\bibitem{vista}
P.~Kulkarni, W.~Zhao, H.~Moon, K.~Cho, D.~Whalley, J.~Davidson, M.~Bailey,
  Y.~Paek, and K.~Gallivan.
\newblock Finding effective optimization phase sequences.
\newblock In {\em Proceedings of the Conference on Languages, Compilers, and
  Tools for Embedded Systems (LCTES)}, pages 12--23, 2003.

\bibitem{DBLP:journals/corr/LascuD15}
A.~Lascu and A.~F. Donaldson.
\newblock Integrating a large-scale testing campaign in the {CK} framework.
\newblock {\em CoRR}, abs/1511.02725, 2015.

\bibitem{la2004}
C.~Lattner and V.~Adve.
\newblock {LLVM}: A compilation framework for lifelong program analysis \&
  transformation.
\newblock In {\em Proceedings of the 2004 International Symposium on Code
  Generation and Optimization (CGO'04)}, Palo Alto, California, March 2004.

\bibitem{DBLP:journals/taco/LeatherBO14}
H.~Leather, E.~V. Bonilla, and M.~F.~P. O'Boyle.
\newblock Automatic feature generation for machine learning-based optimising
  compilation.
\newblock {\em {TACO}}, 11(1):14:1--14:32, 2014.

\bibitem{Lidbury:2015:MCF:2737924.2737986}
C.~Lidbury, A.~Lascu, N.~Chong, and A.~F. Donaldson.
\newblock Many-core compiler fuzzing.
\newblock In {\em Proceedings of the 36th ACM SIGPLAN Conference on Programming
  Language Design and Implementation}, PLDI '15, pages 65--76, New York, NY,
  USA, 2015. ACM.

\bibitem{5160988}
Y.~Liu, E.~Z. Zhang, and X.~Shen.
\newblock A cross-input adaptive framework for gpu program optimizations.
\newblock In {\em 2009 IEEE International Symposium on Parallel Distributed
  Processing}, pages 1--10, May 2009.

\bibitem{LCYP04}
J.~Lu, H.~Chen, P.-C. Yew, and W.-C. Hsu.
\newblock Design and implementation of a lightweight dynamic optimization
  system.
\newblock In {\em Journal of Instruction-Level Parallelism}, volume~6, 2004.

\bibitem{LCWP2009}
L.~Luo, Y.~Chen, C.~Wu, S.~Long, and G.~Fursin.
\newblock Finding representative sets of optimizations for adaptive
  multiversioning applications.
\newblock In {\em 3rd Workshop on Statistical and Machine Learning Approaches
  Applied to Architectures and Compilation (SMART'09), colocated with HiPEAC'09
  conference}, January 2009.

\bibitem{Manotas:2014:SSE:2568225.2568297}
I.~Manotas, L.~Pollock, and J.~Clause.
\newblock Seeds: A software engineer's energy-optimization decision support
  framework.
\newblock In {\em Proceedings of the 36th International Conference on Software
  Engineering}, ICSE 2014, pages 503--514, New York, NY, USA, 2014. ACM.

\bibitem{Marin:2004:CPP:1012888.1005691}
G.~Marin and J.~Mellor-Crummey.
\newblock Cross-architecture performance predictions for scientific
  applications using parameterized models.
\newblock {\em SIGMETRICS Perform. Eval. Rev.}, 32(1):2--13, June 2004.

\bibitem{Mars:2010:CAE:1772954.1772991}
J.~Mars, N.~Vachharajani, R.~Hundt, and M.~L. Soffa.
\newblock Contention aware execution: Online contention detection and response.
\newblock In {\em Proceedings of the 8th Annual IEEE/ACM International
  Symposium on Code Generation and Optimization}, CGO '10, pages 257--265, New
  York, NY, USA, 2010. ACM.

\bibitem{fftw}
F.~Matteo and S.~Johnson.
\newblock {FFTW}: An adaptive software architecture for the {FFT}.
\newblock In {\em Proceedings of the {IEEE} International Conference on
  Acoustics, Speech, and Signal Processing}, volume~3, pages 1381--1384,
  Seattle, {WA}, May 1998.

\bibitem{Miceli:2012:APA:2451764.2451792}
R.~Miceli~et.al.
\newblock Autotune: A plugin-driven approach to the automatic tuning of
  parallel applications.
\newblock In {\em Proceedings of the 11th International Conference on Applied
  Parallel and Scientific Computing}, PARA'12, pages 328--342, Berlin,
  Heidelberg, 2013. Springer-Verlag.

\bibitem{Monsifrot}
A.~Monsifrot, F.~Bodin, and R.~Quiniou.
\newblock A machine learning approach to automatic production of compiler
  heuristics.
\newblock In {\em Proceedings of the International Conference on Artificial
  Intelligence: Methodology, Systems, Applications}, LNCS 2443, pages 41--50,
  2002.

\bibitem{DBLP:conf/cc/MooreC13}
R.~W. Moore and B.~R. Childers.
\newblock Automatic generation of program affinity policies using machine
  learning.
\newblock In {\em CC}, pages 184--203, 2013.

\bibitem{slambench_paper}
L.~Nardi, B.~Bodin, M.~Z. Zia, J.~Mawer, A.~Nisbet, P.~H.~J. Kelly, A.~J.
  Davison, M.~Luj\'an, M.~F.~P. O'Boyle, G.~Riley, N.~Topham, and S.~Furber.
\newblock Introducing {SLAMBench}, a performance and accuracy benchmarking
  methodology for {SLAM}.
\newblock In {\em {IEEE Intl. Conf. on Robotics and Automation (ICRA)}}, May
  2015.
\newblock arXiv:1410.2167.

\bibitem{Nis1998}
A.~Nisbet.
\newblock Iterative feedback directed parallelisation using genetic algorithms.
\newblock In {\em Proceedings of the Workshop on Profile and Feedback Directed
  Compilation in conjunction with International Conference on Parallel
  Architectures and Compilation Technique (PACT)}, 1998.

\bibitem{doi:10.1093/bioinformatics/bth361}
T.~Oinn, M.~Addis, J.~Ferris, D.~Marvin, M.~Senger, M.~Greenwood, T.~Carver,
  K.~Glover, M.~R. Pocock, A.~Wipat, and P.~Li.
\newblock Taverna: a tool for the composition and enactment of bioinformatics
  workflows.
\newblock {\em Bioinformatics}, 20(17):3045--3054, 2004.

\bibitem{PE2006}
Z.~Pan and R.~Eigenmann.
\newblock Fast and effective orchestration of compiler optimizations for
  automatic performance tuning.
\newblock In {\em Proceedings of the International Symposium on Code Generation
  and Optimization (CGO)}, pages 319--332, 2006.

\bibitem{DBLP:journals/ijpp/ParkCPBCS13}
E.~Park, J.~Cavazos, L.-N. Pouchet, C.~Bastoul, A.~Cohen, and P.~Sadayappan.
\newblock Predictive modeling in a polyhedral optimization space.
\newblock {\em International Journal of Parallel Programming}, 41(5):704--750,
  2013.

\bibitem{scikit-learn}
F.~Pedregosa, G.~Varoquaux, A.~Gramfort, V.~Michel, B.~Thirion, O.~Grisel,
  M.~Blondel, P.~Prettenhofer, R.~Weiss, V.~Dubourg, J.~Vanderplas, A.~Passos,
  D.~Cournapeau, M.~Brucher, M.~Perrot, and E.~Duchesnay.
\newblock Scikit-learn: Machine learning in {P}ython.
\newblock {\em Journal of Machine Learning Research}, 12:2825--2830, 2011.

\bibitem{DBLP:journals/ijhpca/PuschelMSXJPVJ04}
M.~P{\"{u}}schel, J.~M.~F. Moura, B.~Singer, J.~Xiong, J.~R. Johnson, D.~A.
  Padua, M.~M. Veloso, and R.~W. Johnson.
\newblock Spiral: {A} generator for platform-adapted libraries of signal
  processing alogorithms.
\newblock {\em {IJHPCA}}, 18(1):21--45, 2004.

\bibitem{Ren:2010:GPC:1849301.1849332}
G.~Ren, E.~Tune, T.~Moseley, Y.~Shi, S.~Rus, and R.~Hundt.
\newblock Google-wide profiling: A continuous profiling infrastructure for data
  centers.
\newblock {\em IEEE Micro}, 30(4):65--79, July 2010.

\bibitem{DBLP:journals/corr/RussakovskyDSKSMHKKBBF14}
O.~Russakovsky, J.~Deng, H.~Su, J.~Krause, S.~Satheesh, S.~Ma, Z.~Huang,
  A.~Karpathy, A.~Khosla, M.~S. Bernstein, A.~C. Berg, and F.~Li.
\newblock Imagenet large scale visual recognition challenge.
\newblock {\em CoRR}, abs/1409.0575, 2014.

\bibitem{sammutencyclopedia}
C.~Sammut and G.~Webb.
\newblock {\em Encyclopedia of Machine Learning and Data Mining}.
\newblock Springer reference. Springer Science + Business Media.

\bibitem{DBLP:conf/cf/ShenVSAS13}
J.~Shen, A.~L. Varbanescu, H.~J. Sips, M.~Arntzen, and D.~G. Simons.
\newblock Glinda: a framework for accelerating imbalanced applications on
  heterogeneous platforms.
\newblock In {\em Conf. Computing Frontiers}, page~14, 2013.

\bibitem{Shende:2006:TPP:1125980.1125982}
S.~S. Shende and A.~D. Malony.
\newblock The tau parallel performance system.
\newblock {\em Int. J. High Perform. Comput. Appl.}, 20(2):287--311, May 2006.

\bibitem{spiral}
B.~Singer and M.~Veloso.
\newblock Learning to predict performance from formula modeling and training
  data.
\newblock In {\em Proceedings of the Conference on Machine Learning}, 2000.

\bibitem{Smith:2005:AVM:1069588.1069632}
J.~E. Smith and R.~Nair.
\newblock The architecture of virtual machines.
\newblock {\em Computer}, 38(5):32--38, May 2005.

\bibitem{SA2005}
M.~Stephenson and S.~Amarasinghe.
\newblock Predicting unroll factors using supervised classification.
\newblock In {\em Proceedings of the International Symposium on Code Generation
  and Optimization (CGO)}. IEEE Computer Society, 2005.

\bibitem{SAMP2003}
M.~Stephenson, S.~Amarasinghe, M.~Martin, and U.-M. O'Reilly.
\newblock Meta optimization: Improving compiler heuristics with machine
  learning.
\newblock In {\em Proceedings of the ACM SIGPLAN Conference on Programming
  Language Design and Implementation (PLDI'03)}, pages 77--90, June 2003.

\bibitem{Takanen:2008:FSS:1404500}
A.~Takanen, J.~DeMott, and C.~Miller.
\newblock {\em Fuzzing for Software Security Testing and Quality Assurance}.
\newblock Artech House, Inc., Norwood, MA, USA, 1 edition, 2008.

\bibitem{tnld10}
S.~Tomov, R.~Nath, H.~Ltaief, and J.~Dongarra.
\newblock Dense linear algebra solvers for multicore with {GPU} accelerators.
\newblock In {\em Proc. of the IEEE IPDPS'10}, pages 1--8, Atlanta, GA, April
  19-23 2010. IEEE Computer Society.
\newblock {DOI:~10.1109/IPDPSW.2010.5470941}.

\bibitem{DBLP:conf/sc/TsaiLKD16}
Y.~M. Tsai, P.~Luszczek, J.~Kurzak, and J.~J. Dongarra.
\newblock Performance-portable autotuning of opencl kernels for convolutional
  layers of deep neural networks.
\newblock In {\em 2nd Workshop on Machine Learning in {HPC} Environments,
  MLHPC@SC, Salt Lake City, UT, USA, November 14, 2016}, pages 9--18, 2016.

\bibitem{VE00}
M.~Voss and R.~Eigenmann.
\newblock {ADAPT: A}utomated de-coupled adaptive program transformation.
\newblock In {\em Proceedings of International Conference on Parallel
  Processing}, 2000.

\bibitem{atlas}
R.~Whaley and J.~Dongarra.
\newblock Automatically tuned linear algebra software.
\newblock In {\em Proceedings of the Conference on High Performance Networking
  and Computing}, 1998.

\bibitem{ck-acm}
D.~Wilkinson, B.~Childers, R.~Bernard, W.~Graves, and J.~Davidson.
\newblock Acm pilot demo 1 - collective knowledge: Packaging and sharing.
  version 3.
\newblock 2017.

\bibitem{Yang:2011:FUB:1993498.1993532}
X.~Yang, Y.~Chen, E.~Eide, and J.~Regehr.
\newblock Finding and understanding bugs in c compilers.
\newblock In {\em Proceedings of the 32Nd ACM SIGPLAN Conference on Programming
  Language Design and Implementation}, PLDI '11, pages 283--294, New York, NY,
  USA, 2011. ACM.

\bibitem{soffa2005}
M.~Zhao, B.~R. Childers, and M.~L. Soffa.
\newblock A model-based framework: an approach for profit-driven optimization.
\newblock In {\em Third Annual {IEEE}/{ACM} Interational Conference on Code
  Generation and Optimization}, pages 317--327, 2005.

\end{thebibliography}

\appendix

\clearpage
\newpage
\pagebreak

\section{Artifact Appendix}
\label{artifact_appendix}

\textbf{Submission guidelines:}\begin{center}{\it \href{http://ctuning.org/ae/submission-20161020.html}{cTuning.org/ae/submission-20161020.html}}\end{center}

This is an example of an Artifact Appendix which we introduced 
at the computer systems conferences including CGO, PPoPP, PACT and SuperComputing
to gradually unify artifact evaluation, sharing and reuse~\cite{ctuning-ae1,childers2016artifact,new_pub_model,Fur2009}.
We briefly describe how to install and use our autotuning workflow, visualize optimization results and reproduce them.
We also shared all scripts which we used to generate data and
graphs in all sections from this report though we did not yet 
have time to thoroughly document them.
In fact, we plan to gradually document them and standardize APIs 
of shared CK modules with the help of the community and motivated students.

\subsection{Abstract}

We provided the whole Collective Knowledge workflow with all dependencies
for collaborative, customizable, multi-dimensional and multi-objective autotuning 
of realistic workloads on Raspberry Pi 3 and other devices.

Current optimization results are available for GCC 7.1.0 (\href{http://cknowledge.org/repo/web.php?wcid=8289e0cf24346aa7:79bca2b76876b5c6}{link})
and for GCC 4.9.2 (\href{http://cknowledge.org/repo/web.php?wcid=8289e0cf24346aa7:d24a4fde9f120e10}{link}).
They are also available as a \href{https://github.com/ctuning/ck-rpi-optimization-results}{CK repository} 
and can be replayed on another platform via CK.

\subsection{Description}

\subsubsection{Check-list (artifact meta information)}

{\small
\begin{itemize}
  \item {\bf Algorithm:} -
  \item {\bf Program:} shared programs from the \href{https://github.com/ctuning/ctuning-programs}{CK ctuning-programs repository}
  \item {\bf Compilation:} any GCC
  \item {\bf Transformations:} compiler flag optimizations
  \item {\bf Binary:} will be produced during autotuning
  \item {\bf Data set:} real inputs from the \href{https://github.com/ctuning/ctuning-datasets-min}{CK ctuning-datasets-min repository}
  \item {\bf Run-time environment:} Raspbian (or any other)
  \item {\bf Hardware:} Raspberry Pi 3 (or any other)
  \item {\bf Run-time state:} will be monitored by CK (CPU frequency)
  \item {\bf Execution:} empirical measurements of the execution time of autotuned workloads via CK workflow
  \item {\bf Output:} best combinations of GCC compiler flags that improve execution time and code size 
  \item {\bf Experiment workflow:} autotuning, crowd-tuning  and collaborative machine learning workflow implemented using CK framework
  \item {\bf Experiment customization:} standard customization via CK API: select compiler, programs and data sets for autotuning, crowd-tuning and predictive modeling
  \item {\bf Publicly available?:} yes - CK autotuning and machine learning workflow (available under BSD 3-clause license) and all related artifacts are shared as reusable and customizable components via GitHub. 
\end{itemize}
}

\subsubsection{How software can be obtained (if available)}

You can obtain CK repositories with optimization results, shared programs and data sets, workflow for autotuning and crowd-tuning as following:

\begin{flushleft}
\texttt{\$ sudo pip install ck \newline
\$ ck pull repo:ck-rpi-optimization-results}
\end{flushleft}

Note that you may need around 1GB of free space. You can install 2 additional CK repositories~\cite{fursin_lokhmotov_upton_2018} 
from the public FigShare repository as following (need ~3GB of free space):

\begin{flushleft}
\texttt{\$ ck add repo:ck-rpi-optimization-results-reactions --zip=https://ndownloader.figshare.com/files/10218435 --quiet \newline
\$ ck add repo:ck-rpi-optimization-results-reactions2 --zip=https://ndownloader.figshare.com/files/10218441 --quiet \newline
\$ ck ls experiment:rpi3-* \newline}
\end{flushleft}

These repositories are so large because they contain all experiments from this report in a reproducible way
(we also plan to considerably reduce this size by removing duplicate information in the future). 
But if you want to prepare and run your own repositories you will likely need less than 100MB. 
See this artifact in the CK from the ACM CGO'17 paper~\cite{Ainsworth:2017:SPI:3049832.3049865} as example: 
\href{https://github.com/SamAinsworth/reproduce-cgo2017-paper}{github.com/SamAinsworth/reproduce-cgo2017-paper}

\subsubsection{Hardware dependencies}

Tested on Raspberry Pi Model B 3 devices with 4-core BCM2709 processor but should work on any platform.

\subsubsection{Software dependencies}

\begin{itemize}
  \item Raspbian GNU/Linux 8 (jessie)
  \item Collective Knowledge Framework~\cite{ck,ck-date16}
  \item Python 2.7+ or 3.4+
  \item Git client
  \item GCC 4.9.2 or GCC 7.1.0
\end{itemize}

\subsubsection{Data sets}

A minimal set of inputs for cTuning benchmarks available from the \href{https://github.com/ctuning/ctuning-datasets-min}{CK ctuning-datasets-min repository}.

\subsection{Installation}

Installation is performed using CK with the help of integrated cross-platform package manager~\cite{ck-env}:

\begin{flushleft}
\texttt{\$ sudo pip install ck \newline
\$ ck pull repo:ck-rpi-optimization-results \newline
\$ ck compile program:zlib --speed \newline}
\end{flushleft}

CK will automatically detect required software which is already installed on your platform, 
install missing packages, and prepare autotuning workflow for execution.

Note that CK allows multiple versions of different software to natively co-exist.
Therefore, you can install several versions of GCC which will be automatically
detected by CK and their environment prepared accordingly.
For example, you can install (build) GCC 7.1.0 on RPi 3 via CK as following:

\begin{flushleft}
\texttt{\$ ck pull repo:ck-dev-compilers \newline
\$ ck install package:compiler-gcc-any-src-linux-no-deps --env.PARALLEL\_BUILDS=1 --env.GCC\_COMPILE\_CFLAGS=-O0 --env.GCC\_COMPILE\_CXXFLAGS=-O0 --env.EXTRA\_CFG\_GCC=--disable-bootstrap --env.RPI3=YES --force\_version=7.1.0 \newline
\$ ck show env --tags=gcc}
\end{flushleft}

Note that you may need to install extra dependencies including

\begin{flushleft}
\texttt{\$ sudo apt-get install texinfo build-essential libgmp-dev libmpfr-dev libisl-dev libcloog-isl-dev libmpc-dev}
\end{flushleft}

You may also want to increase a swap size on RPi 3 to speed up GCC building. 
You can change "CONF\_SWAPSIZE=100" in /etc/dphys-swapfile to "CONF\_SWAPSIZE=1000". 
But do not forget to change it back after successful build to avoid damaging your SD card.

\subsection{Experiment workflow}

\textbf{Autotuning example}

You can run zlib autotuning via CK as following:

\begin{flushleft}
\texttt{\$ ck autotune program:zlib --iterations=150 --repetitions=3 --scenario=9d88674c45b94971 --cmd\_key=decode --record\_uoa=my-first-experiment}
\end{flushleft}

CK will automatically detect available compilers, will ask user to select data set, 
and will evaluate 150 combinations of random compiler flags 
(repeating each experiment 3 times for statistical analysis 
of empirical variation of results).

Experimental results will be aggregated in a CK entry "experiment:my-first-experiment" a local CK repository:

\begin{flushleft}
\texttt{\$ ck find experiment:my-first-experiment}
\end{flushleft}

You can plot graph (execution time vs binary size) or view results in a web browser as following:

\begin{flushleft}
\texttt{\$ ck plot graph:my-first-experiment \newline
\$ ck browser experiment:my-first-experiment}
\end{flushleft}

You can compile and run zlib program via CK as following:
\begin{flushleft}
\texttt{\$ ck compile program:zlib --flags="some flags" \newline
\$ ck run program:zlib}
\end{flushleft}

Finally, you can participate in GCC crowd-tuning as following:
\begin{flushleft}
\texttt{\$ ck crowdsource program.optimization --gcc}
\end{flushleft}

\subsection{Evaluation and expected result}

You can find all scripts to perform experiments from this article as following:

\begin{flushleft}
\texttt{\$ ck ls ck-rpi-optimization-results:script:* | sort}
\end{flushleft}

You can then go to each individual entry and see related scripts:
\begin{flushleft}
\texttt{\$ ls `ck find script:rpi3-susan-autotune`}
\end{flushleft}

You can find all experimental results in the following entries:
\begin{flushleft}
\texttt{\$ ck ls ck-rpi-optimization-results:experiment:* | sort}
\end{flushleft}

You can then browse all results in your web browser as following:
\begin{flushleft}
\texttt{\$ ck browser experiment:rpi3-zlib-decode-gcc4-150b-rnd-frontier}
\end{flushleft}

You can find information about how to replay each autotuning iteration there, for example:
\begin{flushleft}
\texttt{\$ ck replay experiment:b0f31c56475aa510 --point=46049203405c5347}
\end{flushleft}

CK should normally show expected and new results while reporting any unexpected 
behavior (if difference is more than some threshold such as 5\%).

\subsection{Experimental methodology}
\label{experimental_methodology}

One of the most important points of using Collective Knowledge framework
is to take advantage of the experimental methodology for computer systems research
continuously improved by the community.
For this purpose, we instrument programs using a small xOpenME library~\cite{fursin:hal-01054763}
which allows us to monitor behavior of some code regions and dump 
final statistics to a JSON file in the CK format.
CK will then repeat each autotuning iteration N times,
apply statistical analysis on all exposed characteristics, report min, max and mean values,
and calculate expected value based on a histogram of all results (if supported by used Python)
as shown in Figure~\ref{fig:reproducibility} (a).

We then calculate improvements of a given optimization over reference one (-O3)
using minimal and expected execution times, and record differences.
If the difference is more than 5\%, we mark such experient is noise and untrustable 
to be analyzed and improved later by the community.
If several system states are detected as shown in Figure~\ref{fig:reproducibility} (b),
CK will not be able to reproduce them - it then means that the common CK experimental workflow
should also be improved for this hardware and environment to be able to distinguish such
states (such as CPU and GPU frequency due to DVFS for example).

\subsection{Notes}
\label{notes}

We did not have time to thoroughly document experiments from sections~\ref{sec:crowdfuzzing}+ of this report.
However we shared all CK modules, workflows and scripts we used in this report 
in the following CK entries:

\begin{flushleft}
\texttt{\$ ck ls script:rpi3-*} \newline
\texttt{\$ ck ls converting-ad-hoc-works-to-ck-*} \newline
\end{flushleft}

Scripts from Sections~\ref{sec:autotuning} and \ref{sec:flag_autotuning}
to invoke portable and customizable CK autotuning workflow:

\begin{flushleft}
\texttt{\$ ck find script:rpi3-susan-autotune} \newline
\texttt{\$ ck find script:rpi3-susan-graphs} \newline
\texttt{\$ ck find script:rpi3-susan-reduce} \newline
\texttt{\$ ck find script:rpi3-all-autotune} \newline
\end{flushleft}

Scripts from Section~\ref{sec:crowdtuning}:

\begin{flushleft}
\texttt{\$ ck find script:rpi3-all-autotune} \newline
\texttt{\$ ck find script:rpi3-crowdtune} \newline
\end{flushleft}

Scripts from Section~\ref{sec:collaborative}:

\begin{flushleft}
\texttt{\$ ck find script:rpi3-zlib-decode-autotune} \newline
\texttt{\$ ck find script:rpi3-zlib-decode-graphs} \newline
\texttt{\$ ck find script:rpi3-zlib-decode-reduce} \newline
\texttt{\$ ck find script:rpi3-zlib-encode-autotune} \newline
\texttt{\$ ck find script:rpi3-zlib-encode-graphs} \newline
\texttt{\$ ck find script:rpi3-zlib-encode-reduce} \newline
\end{flushleft}

Scripts from Section~\ref{sec:crowdfuzzing}:

\begin{flushleft}
\texttt{\$ ck find script:rpi3-susan-fuzz-bugs} \newline
\end{flushleft}

Scripts from Sections~\ref{sec:crowdmodeling} and \ref{sec:features}:

\begin{flushleft}
\texttt{\$ ck find script:rpi3-crowdmodel} \newline
\end{flushleft}

Scripts from Section~\ref{sec:datasets}: //input-aware

\begin{flushleft}
\texttt{\$ ck find script:rpi3-all-autotune-multiple-datasets} \newline
\texttt{\$ ck find script:rpi3-input-aware-autotune-blas} \newline
\end{flushleft}

Scripts from Section~\ref{sec:competitions}:

\begin{flushleft}
\texttt{\$ ck find script:converting-ad-hoc-works-to-ck-slambench-autotuning} \newline
\end{flushleft}

\subsection{Conclusion}

We hope that our customizable autotuning and machine learning workflow 
can teach students, scientists and engineers learn how to collaboratively
co-design Pareto-efficient software and hardware stack for emerging workloads.
Please feel free to send us updates and patches to fix, help us improve or extend
our artifacts with documentation, and keep in touch with our community via
CK mailing list:~\href{https://groups.google.com/d/forum/collective-knowledge}{groups.google.com/d/forum/collective-knowledge}!

\end{document}